\documentclass{jfm}
\usepackage{graphicx,epsf,subfigure}
\usepackage{epstopdf, epsfig}
\usepackage{natbib}
\usepackage{multirow}
\usepackage{amsmath}
\usepackage{amsfonts,amssymb}
\usepackage{hyperref}

\shorttitle{Inertial instabilities in ellipsoids}
\shortauthor{J. Vidal and D. C\'ebron}

\title{Inviscid instabilities in rotating ellipsoids on eccentric Kepler orbits}

\author{J\'er\'emie Vidal
  \corresp{\email{jeremie.vidal@univ-grenoble-alpes.fr}}
  \and David C\'ebron}

\affiliation{Universit\'e Grenoble Alpes, CNRS, ISTerre, F-38000 Grenoble}

\begin{document}

\maketitle

\begin{abstract}
	We consider the hydrodynamic stability of homogeneous, incompressible and rotating ellipsoidal fluid masses. 
	The latter are the simplest models of fluid celestial bodies with internal rotation and subjected to tidal forces.
	The classical problem is the stability of Roche-Riemann ellipsoids moving on circular Kepler orbits. 
	However previous stability studies have to be reassessed.
	Indeed they only consider global perturbations of large wavelength or local perturbations of short wavelength.
	Moreover many planets and stars undergo orbital motions on eccentric Kepler orbits, implying time-dependent ellipsoidal semi-axes. This time dependence has never been taken into account in hydrodynamic stability studies.
	In this work we overcome these stringent assumptions. We extend the hydrodynamic stability analysis of rotating ellipsoids to the case of eccentric orbits.
	We have developed two open source and versatile numerical codes to perform global and local inviscid stability analyses.
	They give sufficient conditions for instability. 
    The global method, based on an exact and closed Galerkin basis, handles rigorously global ellipsoidal perturbations of unprecedented complexity. 
	Tidally driven and libration-driven elliptical instabilities are first recovered and unified within a single framework.
	Then we show that new global fluid instabilities can be triggered in ellipsoids by tidal effects due to eccentric Kepler orbits. Their existence is confirmed by a local analysis and direct numerical simulations of the fully nonlinear and viscous problem.
	Thus a non-zero orbital eccentricity may have a strong destabilising effect in celestial fluid bodies, which may lead to space-filling turbulence in most of the parameters range.
\end{abstract}

\section{Introduction}
	\subsection{Physical context}
As a result of gravitational tidal forces generated by their orbital partners, most planets and moons have time-dependent spin rates and ellipsoidal shapes \citep[e.g.][]{chandrasekhar1969ellipsoidal}, which disturb their rotational dynamics. 
It bears the name of mechanical or harmonic forcing \citep{le2015flows,le2016flows}, such as tides, librations or precession.
Librations are oscillations of the figure axes of a synchronised body with respect to a given mean rotation axis. 
Precession refers to the case whereby the instantaneous rotation vector rotates itself about a secondary axis that is fixed in an inertial frame of reference \citep{poincare1910precession}.
Observations of mechanical forcings of a celestial body can be used to infer its internal structure \citep[e.g.][]{dehant2015precession}.

Mechanical forcings also play an important role in the dynamics of planetary and stellar fluid interiors, extracting a part of the available rotational energy to sustain large-scale flows \citep{tilgner2015rotational}.
Many orbiting celestial bodies have orbits sufficiently close to their hosts such that strong tidal interactions are expected. Tides create a tidal bulge, leading to angular momentum exchange between the orbital motion and the spinning bodies, and they also dissipate energy through the induced fluid flows. 
Therefore tides may play an important role in the (internal and orbital) dynamics of binary systems and orbiting extra-solar planets \citep[e.g.][]{ogilvie2004tidal,cebron2012elliptical}. For instance we expect tidal interactions to be responsible for the spin synchronisation and of the circularisation of the orbits in binary systems \citep[e.g.][]{hut1981tidal,hut1982tidal,rieutord2004evolution}. However these problems are not yet fully resolved. Many studies are devoted to understand the mechanisms of tidal dissipation in such systems \citep[e.g.][]{rieutord2010viscous,ogilvie2007tidal}, but uncertainties remain.

Mechanical forcings may also sustain dynamos, such as tidal dynamos \citep{barker2013non,cebron2014tidally}.  \citet{malkus1963precessional,malkus1968precession,malkus1989experimental} first pointed out the relevance of harmonic forcings to drive planetary core flows, suggesting that the Earth's magnetic field is maintained by luni-solar precession.
Using energy and power considerations, \citet{kerswell1996upper} showed that turbulent precession-driven flows are sufficiently vigorous to potentially sustain a dynamo.
Numerical dynamos driven by precession \citep{tilgner2005precession,tilgner2007kinematic,wu2009dynamo,goepfert2016dynamos,barker2016turbulence} have been found.

	\subsection{Inertial instabilities}
The flow stability in ellipsoids is a long standing issue. It dates back to the stability study of self-gravitating ellipsoids. More than a century ago, \citet{riemann1860untersuchungen} considered the stability of ellipsoidal flows with a linear dependence in Cartesian space coordinates.
\citet{hough1895oscillations}, \citet{sloudsky1895rotation} and \citet{poincare1910precession} also assumed that flows depend linearly on Cartesian space coordinates. This simplifies the mathematical complexity of the problem, because flows reduce to time-dependent uniform vorticity flows governed by ordinary differential equations.
They are the order zero response of a rotating fluid enclosed in a rigid ellipsoid undergoing mechanical forcing \citep{roberts2011flows}. It was first predicted by theoretical studies on precessing flows \citep{bondi1953dynamical,stewartson1963motion,roberts1965motion,busse1968steady} and later confirmed by experiments \citep{pais2001precession,noir2003experimental,cebron2010tilt} and simulations \citep{lorenzani2001fluid,tilgner2001fluid,noir2013precession} in the laminar regime. However a basic flow of uniform vorticity is actually established only if it is dynamically stable, i.e. if no inviscid perturbation grows upon the basic state \citep{kerswell1993instability}. Otherwise, the basic flow is dynamically unstable and is prone to inertial instabilities, as it is the case for precessing flows \citep{kerswell1993instability,cebron2010tilt,wu2011high}.

The basic role of uniform vorticity flows in the hydrodynamic instabilities which are triggered in precessing flows suggests considering more generally the stability of uniform vorticity flows. Indeed such flows are also observed for tidal \citep{cebron2010systematic,cebron2012elliptical,cebron2013elliptical,grannan2016tidally} and libration forcings \citep{zhang2012asymptotic,cebron2012libration,grannan2014experimental,vantieghem2015latitudinal,favier2015generation}.
Both tidal and librating basic flows are prone to the elliptical instability \citep{kerswell2002elliptical}, which was discovered in various contexts \citep[][]{gledzer1978finite,bayly1986three,gledzer1992instability,bayly1986three,pierrehumbert1986universal,waleffe1990three}.
The elliptical instability may play a fundamental role in astrophysics. Indeed tidally driven basic flows, associated with the equilibrium tide \citep{zahn1966marees,goupil2008tidal}, are not an efficient source of dissipation for small enough molecular viscosity. Yet, the elliptical instability may be a viable alternative as a strong source of dissipation \citep{cebron2010systematic,le2010tidal,barker2016nonlinear}.
The libration-driven elliptical instability also occurs in synchronised moons \citep{kerswell1998tidal,cebron2012libration,vantieghem2015latitudinal}.
Finally the elliptical instability is the first ingredient to explain the observed transition towards turbulence in experiments and simulations \citep{grannan2014experimental,favier2015generation,grannan2016tidally,le2017inertial}.

	\subsection{Motivations}
Previously cited theoretical works have studied inertial instabilities (i) in containers departing very weakly from spheres; (ii) for a subset of simple mechanical forcings; (iii) for rigid ellipsoidal containers. However, (i) laboratory experiments and simulations depart strongly from spherical containers to overcome viscous effects and celestial bodies have mainly triaxial shapes; (ii) celestial bodies are subject to a combination of mechanical forcings; (iii) celestial bodies may deform in time to adjust to time-dependent gravitational constraints along their orbits.
To relax these three assumptions, we have developed two open source numerical codes to perform the local and global linear stability analyses of various mechanically driven basic flows. Both methods give only sufficient conditions for instability.
The local method, first introduced by \citet{bayly1986three,pierrehumbert1986universal} and later developed by \citet{lifschitz1991local,friedlander1991instability}, assumes short-wavelength perturbations which are insensitive to the fluid boundary. The global method takes into account the ellipsoidal geometry of the fluid boundary. 
It relies on a Galerkin expansion of the perturbations onto a basis which satisfies the boundary conditions.
Finding an appropriate basis is a difficult task. Furthermore, bases in complex geometries often require advanced numerical schemes \citep{theofilis2011global}.
Extending the works of \citet{gledzer1978finite,gledzer1992instability,lebovitz1989stability,wu2011high}, we use a polynomial basis made of Cartesian monomials generated for any polynomial degree.

As a result of the complexity of the tidal response in the fluid layers of rotating planets and stars, we consider a simplified model that captures the most important physical elements.  
The problem of tidal flows in ellipsoidal homogeneous bodies orbiting on eccentric orbits was tackled by \citet{nduka1971roche}, but calculations were very incomplete. Indeed the latter study only solved the ellipsoidal shapes. Their physical relevance remains elusive, since the fluid instabilities that can grow upon the basic state were not considered.
The hydrodynamic stability of ellipsoidal fluid masses has been tackled by \citet{lebovitz1996short,lebovitz1996new} for isolated fluid masses and recently by \citet{Barker11062016,barker2016nonlinear} for circular orbital motions.

The present paper is a step in the direction of completing this picture from a fluid dynamics point of view. It is a first attempt to understand the case of eccentric orbits. We relax the assumption of ellipsoidal equilibrium to perform the stability analysis of basic flows in arbitrary triaxial ellipsoids orbiting on eccentric Kepler orbits.

The paper is organised as follows. In \S\ref{sec:base_flow} we present the orbitally driven basic flow of uniform vorticity.
In \S\ref{sec:stability}, we describe the stability analysis methods.
Then in \S\ref{sec:results} we survey the hydrodynamic instabilities driven by orbital motions. In addition to tidally driven and libration-driven elliptical instabilities which are recovered, we find new orbitally driven elliptical instabilities (ODEI) associated with the eccentric Kepler orbits.
We discuss the physical mechanism responsible for these new instabilities in \S\ref{sec:physics} and we end the paper with a conclusion in \S\ref{sec:ccl}.

\section{Modelling of the basic state}
\label{sec:base_flow}
	\subsection{Orbital forcing}
\begin{figure}
	\centering
	\begin{tabular}{cc}
		\subfigure[]{\includegraphics[width=0.49\textwidth]{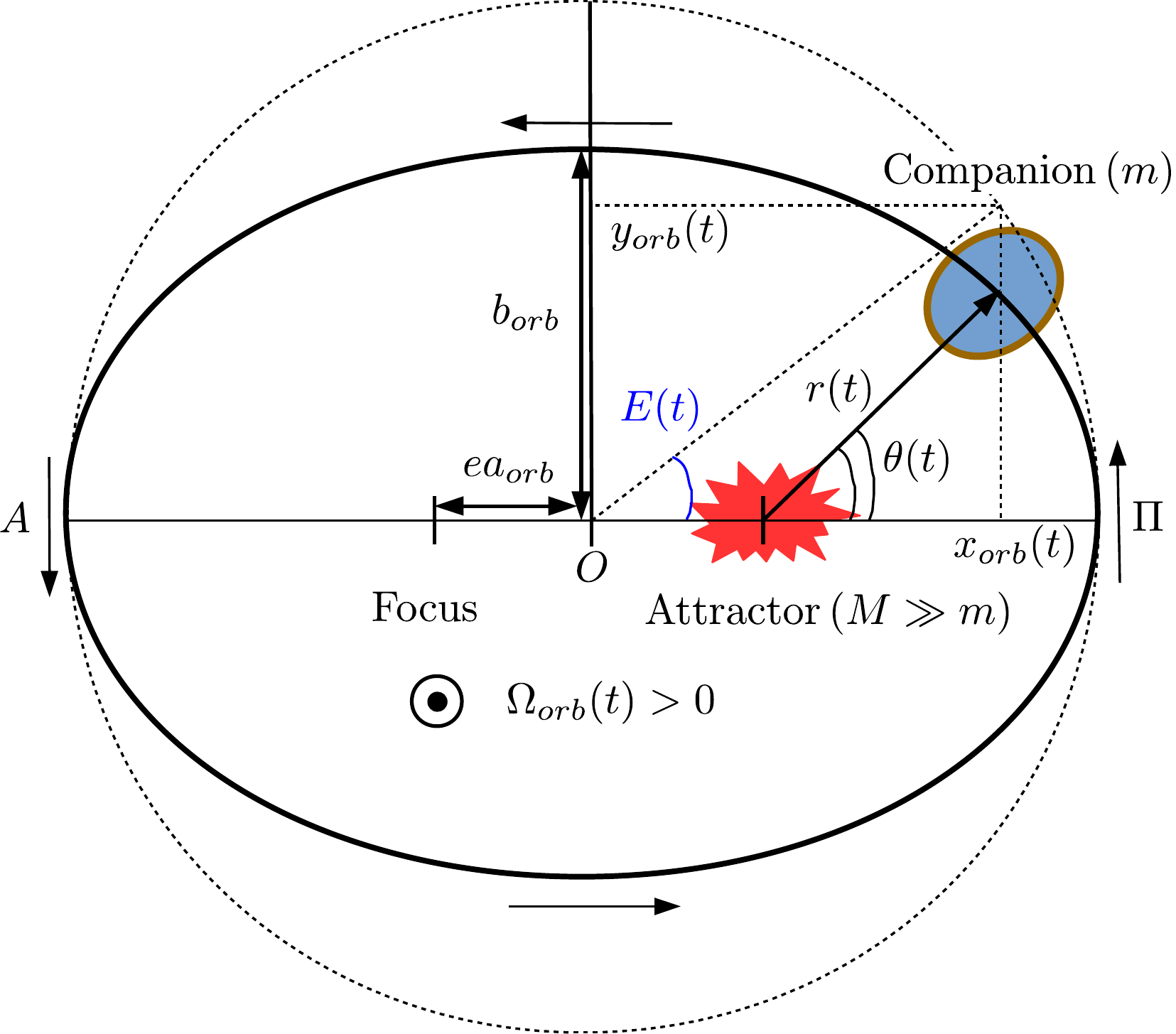}} &
		\subfigure[]{\includegraphics[width=0.49\textwidth]{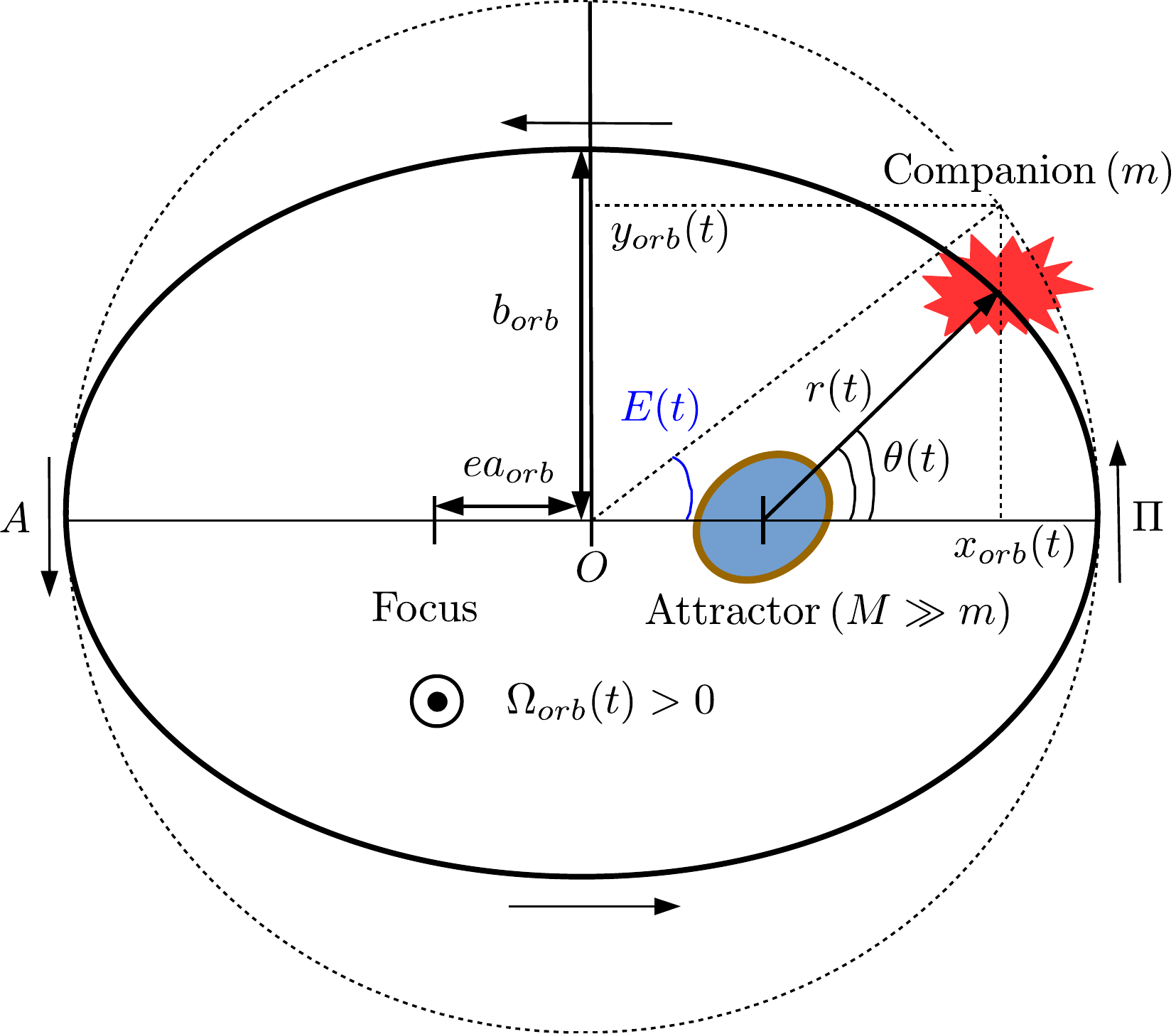}}
	\end{tabular}
	\caption{Eccentric Kepler orbit for our two-body problem. The companion body rotates around an attractor at an orbital angular velocity orthogonal to the orbital plane of amplitude $\Omega_{orb} (t) > 0$. The eccentric Kepler orbit, of eccentricity $e$ (thick black line) and of geometric centre $O$, has semi-axes $a_{orb}$ and $b_{orb}$. The perihelion (resp. aphelion) point of the orbit is $\Pi$ (resp. $A$). The dashed black line is the circumscribed circle of radius $a_{orb}$. The orbital position vector of the orbiting companion body, relating the centre-of-mass of the attractor to the one of the companion body, is $r(t)$. We denote the true anomaly $\theta(t)$ and the mean anomaly $E(t)$. Using Cartesian coordinates centred on the attractor, the position of the orbiting body is $x_{orb} = a_{orb} (\cos E - e)$ and $y_{orb} = a_{orb} \sqrt{1-e^2} \sin E$. (a) A fluid ellipsoidal companion body orbiting around a point-source mass attractor. (b) A companion point-source mass orbiting around a fluid ellipsoidal attractor.}
	\label{Fig_ODEI_draw}
\end{figure}

We are interested in the orbital problem of a companion body of mass $m$ (e.g. a moon, a gaseous planet or a low mass star), which moves on an eccentric Kepler orbit of eccentricity $e$ around an attractor of mass $M \gg m$.
The centre-of-mass of the attractor coincides with the centre-of-mass of the two-body system, which is also a focus of the eccentric Kepler orbit described by the companion body.

Depending on the astrophysical configuration, we consider that either the companion or the attractor is a tidally deformed homogeneous fluid body. The latter has rotating internal motions of time-dependent uniform vorticity $\boldsymbol{\omega}(t)$. The fluid is incompressible, of uniform density $\rho$ and kinematic viscosity $\nu$. The other celestial body is then modelled as a rigid point-source mass. The two situations are sketched in figure \ref{Fig_ODEI_draw}.
Under these circumstances the fluid body experiences its self-gravitating acceleration, the disturbing tidal acceleration and the centrifugal, Coriolis and Poincar\'e accelerations. 
Following \citet{aizenman1968equilibrium,chandrasekhar1969ellipsoidal,nduka1971roche}, we keep only the quadratic terms in the expansion of the tidal potential generated by a point-source mass. Then a mathematically exact description of the fluid boundary is achieved by considering a triaxial ellipsoid, denoting the principal semi-axes $(a (t), b(t), c(t))$. They depend on time $t$ because of the time-dependent gravitational force exerted along the eccentric orbit. The fluid ellipsoid is characterised by its equatorial and polar ellipticities
\begin{equation}
	\beta_{ab} (t)= \frac{|a^2-b^2|}{a^2+b^2} < 1, \ \, \ \beta_{ac}(t) = \frac{a^{2}-c^{2}}{a^{2} + c^{2}}. 
	\label{Eq_betaAB}
\end{equation}
The limit $\beta_{ac} \to 1$ corresponds to the limit case of a disk ($c=0$), whereas $\beta_{ac} \to -1$ corresponds to an infinite cylinder ($c \to \infty$). 

To describe the orbital and fluid motions, we introduce two reference frames. We define an inertial frame with fixed axes and whose origin is the centre-of-mass of the attractor.
The horizontal plane defines the orbital plane and the vertical axis $\boldsymbol{\widehat{z}}$ is parallel to the direction of the orbital angular velocity of the companion body (of amplitude $\Omega_{orb}(t)$). It is the natural frame for describing the orbital motions.
A tractable frame to describe the fluid motions is the rotating body frame, whose origin is the centre-of-mass of the fluid ellipsoid (either the attractor or the companion). 
Its main axes coincide with the directions of its principal ellipsoidal axes $(a (t), b(t), c(t))$. The body frame is rotating at the angular velocity $\boldsymbol{\Omega}^{\mathcal{B}}(t)$ with respect to the inertial frame. The general problem with an arbitrary orientation of $\boldsymbol{\omega}(t)$ with respect to $\Omega_{orb}(t) \widehat{\boldsymbol{z}}$ is of great mathematical complexity \citep[see the general equations (44) to (52) of][]{nduka1971roche}. It could be solved but it is beyond the scope of the present paper. 
Instead following \citet{aizenman1968equilibrium,chandrasekhar1969ellipsoidal,nduka1971roche}, we assume a null obliquity such that $\boldsymbol{\omega} (t) = \omega(t) \widehat{\boldsymbol{z}}$ and $\boldsymbol{\Omega}^{\mathcal{B}} (t)$ is along $\widehat{\boldsymbol{z}}$.

To make the problem dimensionless, we choose $L=\sqrt{(a_{0}^2+b_{0}^2)/2}$ as the length scale, where $(a_{0}, b_{0})$ are characteristic hydrostatic semi-axes of the fluid ellipsoid, and $\Omega_s^{-1}$ as the time scale, where $\Omega_s$ is the steady internal fluid spin rate in the inertial frame. For clarity, the dimensionless variables will be also noted as their dimensional counterparts.

The time dependencies of $\Omega_{orb} (t)$ and $\beta_{ab}(t)$ are given by the orbital dynamics (figure \ref{Fig_ODEI_draw}).
We introduce the dimensionless mean orbiting angular velocity $\Omega_0$ of the body along the elliptical orbit.
The orbit has main orbital semi-axes ($a_{orb}, b_{orb}$).
Following \citet{murray1999solar}, an elliptical orbit is described by Kepler's equation at a given time $t$, 
\begin{equation}
	E(t) - e \sin E(t) = \Omega_0 t,
	\label{Eq_Kepler}
\end{equation}
with $E(t)$ the eccentric anomaly.
The orbital rotation rate on the elliptical orbit is
\begin{equation}
	\Omega_{orb} (t) = \frac{\mathrm{d} \theta}{\mathrm{d} t} = \Omega_0 \frac{[ 1 + e \cos \theta(t)]^2}{\left ( 1 - e^2 \right )^{{3}/{2}}},
	\label{Eq_ODEI_Worb}
\end{equation}
where $\theta(t)$ is the true anomaly defined by
\begin{equation}
	\theta(t) = 2 \arctan \left [ \sqrt{\frac{1+e}{1-e}} \tan \left (\frac{E(t)}{2} \right ) \right ].
	\label{Eq_Trueanomaly}
\end{equation}
The orbital position $r(t)$, describing the position of the centre-of-mass of the companion with respect to the attractor, is
\begin{equation}
	r(t) = a_{orb} \frac{1-e^2}{1 + e \cos \theta(t)} = a_{orb} \left [ 1 - e \cos E(t)  \right ].
	\label{eq:orbitalpositionvector}
\end{equation}

The fluid ellipsoid may have a relative orientation with respect to the orbital position vector (\ref{eq:orbitalpositionvector}). However the relative orientation is extremely small in the null obliquity case \citep{nduka1971roche}. So we assume that the tidal bulge is always aligned with the orbital position vector (instantaneous bulge response), i.e. $\boldsymbol{\Omega}^{\mathcal{B}} (t) = \Omega_{orb} (t) \widehat{\boldsymbol{z}}$.
We estimate at first order the equatorial ellipticity (\ref{Eq_betaAB}) of the fluid ellipsoid with an hydrostatic balance. Following \citet{cebron2012elliptical} it reads
\begin{equation}
	\beta_{ab} (t) = \frac{3}{2} (1 + k_2) \mathcal{M} \left ( \frac{D}{r(t)} \right )^3 = \beta_0\left ( \frac{1+e\cos \theta(t) }{1-e^2} \right )^3 < 1,
	\label{Eq_ODEI_beta}
\end{equation}
with $\beta_0$ a characteristic equatorial ellipticity, $\mathcal{M}$ the ratio between the mass of the celestial body responsible for the disturbing tidal potential and the mass of the fluid ellipsoid, $D$ the mean spherical radius of the fluid ellipsoid and $k_2$ the potential Love number. The latter can be computed with the Clairaut-Radau theory \citep[e.g.][]{van2008librations}. A typical value is $k_2 = 3/2$ for an incompressible homogeneous body in hydrostatic equilibrium \citep{greff2005analytical}.
To take into account all the possible triaxial geometries, the polar ellipticity $\beta_{ac}(t)$ is a free parameter.

\begin{figure}
	\centering
	\includegraphics[width=0.95\textwidth]{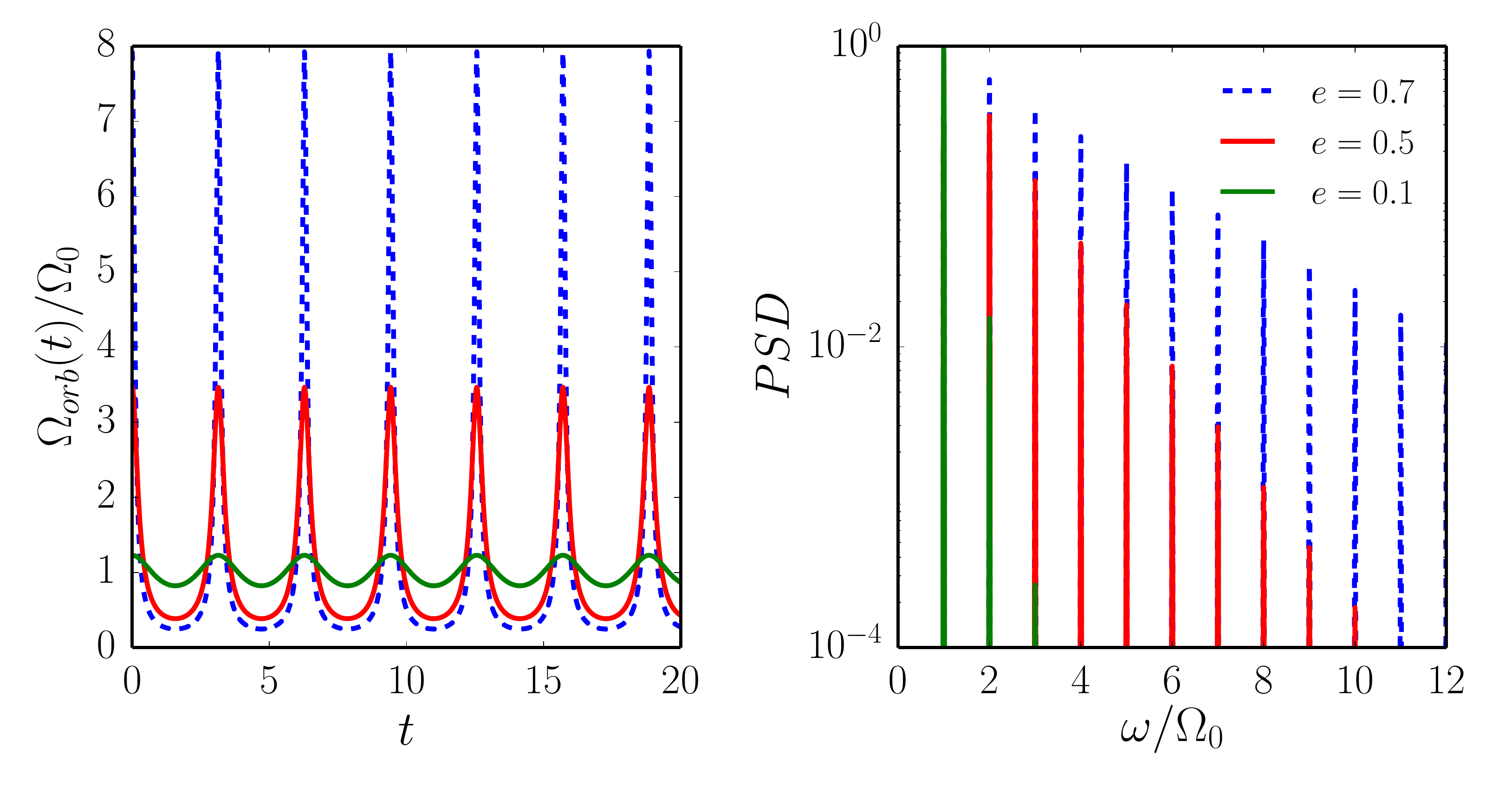}
	\caption{(Left) Normalised orbital spin rate $\Omega_{orb}(t)/\Omega_0$ given by formula (\ref{Eq_ODEI_Worb}) for various eccentricities $e$. Kepler's equation (\ref{Eq_Kepler}) is solved with an iterative Newton's algorithm at each time step. (Right) Associated power spectral density (PSD) in function of the normalised angular frequency $\omega / \Omega_0$ of the Fourier transform.}
	\label{Fig_ODEI_Keplersol}
\end{figure}

Note that $\beta_0$ is the ellipticity of a body with the same mass $m$ but moving on a circular orbit of radius $a_{orb}$ (dashed circle in figure \ref{Fig_ODEI_draw}). It is not the time averaged value of $\beta_{ab}(t)$. The ellipticity $\beta_{0}$ refers to the static (tidal) bulge or equilibrium tide \citep{zahn1966marees}.
The fluctuations in time superimposed on this equilibrium tide are the dynamical tides \citep{zahn1975dynamical}, which are excited by the periodic terms of the disturbing tidal potential.
From formula (\ref{Eq_ODEI_beta}), minimum and maximum values $(\beta_{\min}, \beta_{\max})$ of the ellipticity at the aphelion point $A$ ($\theta=\pi$) and the perihelion point $\Pi$ ($\theta=0$) are
\begin{equation}
	\beta_{\min} = \beta_0 (1+e)^{-3}, \ \, \ \beta_{\max} = \beta_0 (1-e)^{-3}.
    \label{eq:betaminmax_e}
\end{equation}
Because the ellipticity is bounded ($\beta_{ab}(t) < 1$), the upper bound of the maximum allowable eccentricity for a given ellipticity $\beta_0$, denoted $e_{\max}$, is
\begin{equation}
	e_{\max} = 1 - \beta_0^{1/3}.
	\label{eq:emax_beta0}
\end{equation}
However from a physical ground, the maximum allowable ellipticity is governed by a balance between the internal cohesion model of the fluid body (e.g. self-gravitation, molecular\dots) and the disturbing tidal and centrifugal accelerations. For homogeneous self-gravitating ellipsoids, the maximum allowable ellipticity is given by the Roche limit \citep{aizenman1968equilibrium}. 
The classical Roche problem considers an homogeneous self-gravitating ellipsoidal body moving on a circular orbit.
However as long as the eccentric orbit remains outside the Roche limiting circle, the variations of the ellipsoidal figure are small \citep{nduka1971roche}. Therefore in our framework we estimate a lower bound of the orbital eccentricity $e_{R}$ as the eccentricity of the first orbit crossing the Roche limiting circle. It reads
\begin{equation}
	e_R = 1 - \left ( \frac{\beta_0}{\beta^*} \right )^{1/3},
    \label{eq:eroche}
\end{equation}
where $\beta^* = 0.59$ is the lower bound of the equatorial ellipticity of unstable homogeneous ellipsoids moving on circular orbits in the Roche limit \citep[estimated from  point B in figure 3 of][]{aizenman1968equilibrium}.
When $0 \leq  e \leq e_R$ the ellipsoidal configuration is assumed to be stable, whereas when $e_R \leq e < e_{\max}$ some ellipsoidal configurations could be unstable (and hence physically unrealistic) for self-gravitating bodies.

For a circular orbit ($e=0$), the orbital rotation rate is steady $\Omega_{orb}(t) = \Omega_0$ and $\beta_{ab} (t) = \beta_0$.
For an eccentric orbit ($e\neq0$), we determine $\Omega_{orb}(t)$ by solving Kepler's equation (\ref{Eq_Kepler}) numerically using an iterative Newton's algorithm (starting with $E=0$ as initial guess at $t=0$).
We show in figure \ref{Fig_ODEI_Keplersol} the normalised ratio $\Omega_{orb}(t) / \Omega_0$ and its associated power spectral density for different eccentricities.
$\Omega_{orb}(t)$ has a fundamental angular frequency $\omega=\Omega_0$ but, as $e$ is increased, more and more harmonics are required to properly describe the time dependence of $\Omega_{orb}(t)$.

Finally it is worth noting that the case $\Omega_0 = 1$ corresponds to a synchronised fluid body, since the dimensional averaged orbital rate and the averaged fluid spin rate are equal.
When $\Omega_0 \neq 1$ the fluid body is not synchronised. A mean differential rotation exists between the elliptical deformation and the fluid spin rate over one spin period. 

	\subsection{Fluid equations}
	\label{sec:odei0}
In the body reference frame rotating at the dimensionless angular velocity $\boldsymbol{\Omega}^{\mathcal{B}} (t) = \Omega_{orb} (t) \, \widehat{\boldsymbol{z}}$, the time-dependent fluid boundary is ellipsoidal at any time. It is described by the equation
\begin{equation}
	\left ( \frac{x}{a(t)} \right )^2 + \left ( \frac{y}{b(t)} \right )^2 + \left ( \frac{z}{c(t)} \right )^2 = 1.
	\label{Eq_Forcing_Body}
\end{equation}
Because the fluid is incompressible, we restrict ourselves (without loss of generality) to the case $a(t) b(t) c(t) = 1$, such that fluid ellipsoid has a dimensionless constant volume of $4\pi/3$. To take into account all possible triaxial ellipsoids, we define the semi-axes as 
\begin{equation}
	a = R \sqrt{1 + \beta_{ab}(t)}, \, \ \, b = R \sqrt{1 - \beta_{ab}(t)} \, \ \, c = 1/(ab),
\end{equation}
where $R$ is a free parameter governing the polar ellipticity $\beta_{ac} (t)$. Note that many celestial bodies are flattened at their poles. The flattening condition, valid at each time, leads to the sufficient condition $R \geq R_m$ for a given orbit, with 
\begin{equation}
	R_{m}(e) = \left [ ( 1-\beta_{\max}(e)^{2} ) (1-\beta_{\max}(e)) \right ]^{-1/6}
	\label{Eq_RM_ODEI}
\end{equation}
and $\beta_{\max}(e)$ defined by the expression (\ref{eq:betaminmax_e}).

For a fluid mechanics study, the knowledge of the axes $(a(t), b(t), c(t))$, of the forcing $\boldsymbol{\Omega}^{\mathcal{B}} (t)$ and of the internal vorticity $\omega (t)$ is sufficient to fully determine the internal dynamics. The basic flow driven by the orbital motions in the fluid ellipsoid, expressed in the body frame, is
\begin{equation}
	\boldsymbol{U} (\boldsymbol{r}, t) = \left[ 1 - \Omega_{orb} (t) \right] \left ( -[1 + \beta_{ab}(t)] y \,\widehat{\boldsymbol{x}} \,+\, [1 - \beta_{ab}(t)] x \,\widehat{\boldsymbol{y}} \right),
	\label{Eq_BF_Orbit}
\end{equation}
with $\boldsymbol{r} = (x,y,z)^T$ the position vector in the body frame, $(\boldsymbol{\widehat{x}}, \boldsymbol{\widehat{y}}, \boldsymbol{\widehat{z}})$ the unit Cartesian basis vectors in that frame and $\beta_{ab}(t)$ the time-dependent equatorial ellipticity defined by formula (\ref{Eq_ODEI_beta}).
It is an incompressible ($\nabla \cdot \boldsymbol{U} = 0$) and laminar flow of uniform vorticity $\omega(t) = 2 \left[ 1 - \Omega_{orb} (t) \right]$ in the body frame. It is an exact solution of the dimensionless and nonlinear Navier-Stokes equation in the body frame
\begin{equation}
	\frac{\partial \boldsymbol{U}}{\partial t} + ( \boldsymbol{U} \cdot \nabla ) \boldsymbol{U} + 2 \boldsymbol{\Omega}^{\mathcal{B}}(t) \times \boldsymbol{U} = - \nabla P + E_k \nabla^{2} \boldsymbol{U} + \boldsymbol{r} \times \frac{\mathrm{d} \boldsymbol{\Omega}^{\mathcal{B}}}{\mathrm{d}t},
	\label{Eq_BF_NS}
\end{equation}
with $E_k = \nu/(\Omega_{s}L^{2})$ the dimensionless Ekman number and $P$ the modified pressure. Equation (\ref{Eq_BF_NS}) is supplemented with the impermeability condition $\boldsymbol{U} \cdot \boldsymbol{n} = 0$, at the boundary with $\boldsymbol{n}$ the unit vector normal to the boundary (\ref{Eq_Forcing_Body}). However the viscous boundary condition (either no-slip or stress-free) is violated. 
Moreover, the exact kinematic boundary condition is in fact $\boldsymbol{U} \cdot \nabla F + \partial F/\partial t=0$, with $F (\boldsymbol{r},t) = 1-(x/a(t))^2 - (y/b(t))^2 - (z/c(t))^2$. Neglecting the heterogeneous term $\partial F/\partial t$ in the boundary condition is relevant when typically $||\boldsymbol{U}|| \gg || \partial F/\partial t || / || \nabla F||$, i.e. when $||\boldsymbol{U}|| \gg e \beta_0 \Omega_0$. The latter condition is the astrophysically relevant limit. Indeed, celestial fluid bodies are typically characterised by $e \ll 1$ and $\beta_0 \ll 1$. At leading order, the basic flow is made of a solid body rotation of order $O(1)$ and a tidal correction of order $O(\beta_0)$. Corrections of the basic flow only appear at the next order  $O(e\beta_0) \ll 1$ and are neglected in the following.

In the literature, the stability of basic flows (\ref{Eq_BF_Orbit}) has only been studied for steady values of $\beta_{ab}$. We relax here this assumption.
To investigate whether the basic flow (\ref{Eq_BF_Orbit}) is stable against small perturbations, we perform a linear stability analysis. We expand the total velocity field into the sum of the basic flow $\boldsymbol{U} (\boldsymbol{r},t)$ (\ref{Eq_BF_Orbit}) and a perturbation $\boldsymbol{u}(\boldsymbol{r},t)$.
The inviscid ($E_k \ll 1$), linearised governing equations for the perturbation in the body frame are
\begin{subequations}
	\label{Eq_IE}
	\begin{align}
		\frac{\partial \boldsymbol{u}}{\partial t} + (\boldsymbol{U} \cdot \nabla) \boldsymbol{u} + (\boldsymbol{u} \cdot \nabla) \boldsymbol{U} + 2 \, \boldsymbol{\Omega}^{\mathcal{B}} (t) \times \boldsymbol{u} &= - \nabla \pi, 	\label{Eq_IE_Momentum} \\
		\nabla \cdot \boldsymbol{u} &= 0, \ \, \ \boldsymbol{u} \cdot \boldsymbol{n} = 0	\label{Eq_IE_Div+Impermeability}
	\end{align}
\end{subequations}
with $\pi$ the modified pressure perturbation. 
By consistency with the basic flow, we also neglect $\partial F/ \partial t$ in the boundary condition for the velocity perturbation. However, because the stability problem is linear, our results are not affected. Indeed the heterogeneous term $\partial F/\partial t$ in the boundary condition could only generate additional instabilities. Consequently we emphasise that our stability study gives sufficient conditions for instability.
The basic flow $\boldsymbol{U}(\boldsymbol{r},t)$ is linearly unstable if the amplitude $||\boldsymbol{u} (\boldsymbol{r},t)||$ grows without bound with time. 

\section{Inviscid stability analysis methods}
\label{sec:stability}
	\subsection{Global method in triaxial ellipsoids}
To remove the pressure term in equation (\ref{Eq_IE_Momentum}), we take the curl of equation (\ref{Eq_IE_Momentum}) and obtain the governing equation for the vorticity of the perturbation $\boldsymbol{\zeta} = (\nabla \times \boldsymbol{u})$
\begin{equation}
	\frac{\partial \boldsymbol{\zeta}}{\partial t} + \left ( \boldsymbol{U} \cdot \nabla \right ) \boldsymbol{\zeta} + \left ( \boldsymbol{u} \cdot \nabla \right ) \boldsymbol{\omega} - \left ( \boldsymbol{\zeta} \cdot \nabla \right ) \boldsymbol{U} = \left ( \boldsymbol{\omega} + 2\boldsymbol{\Omega}^{\mathcal{B}} \right ) \cdot \nabla \boldsymbol{u}.
	\label{Eq_IE_Vorticity}
\end{equation}
As originally devised by \citet{gledzer1978finite,gledzer1992instability}, if $\boldsymbol{u}$ is a Cartesian polynomial of maximum degree $n$ in the Cartesian coordinates $(x,y,z)$, then each term in equation (\ref{Eq_IE_Vorticity}) is a polynomial in the Cartesian coordinates of maximum degree $n-1$.
It suggests to look for perturbations $\boldsymbol{u}$ which belong to a finite-dimensional vector space of Cartesian polynomials.

We consider the finite-dimensional vector space $\boldsymbol{\mathcal{V}}_{n}$, such that an element $\boldsymbol{v} \in \boldsymbol{\mathcal{V}}_{n}$ is of maximum degree $n$ and satisfies (at any time) $\boldsymbol{v} \cdot \boldsymbol{n} = 0$ at the ellipsoidal boundary (\ref{Eq_Forcing_Body}) and $\nabla \cdot \boldsymbol{v} = 0$. Elements of $\boldsymbol{\mathcal{V}}_{n}$  represent vortical perturbations that are tangential to the ellipsoidal boundary at any time. The dimension of $\boldsymbol{\mathcal{V}}_{n}$ is \citep{lebovitz1989stability,backus2017completeness,ivers2017enumeration}
\begin{equation}
	N_{\mathcal{V}} = n (n+1) (2n+7)/6.
	\label{Eq_DimGPall}
\end{equation}
Finding an appropriate basis is a difficult task. Any polynomial basis of $\boldsymbol{\mathcal{V}}_{n}$ is a complete basis for velocity fields defined over triaxial ellipsoids and meeting the impermeable boundary condition \citep{lebovitz1989stability,backus2016completeness,ivers2017enumeration}, i.e. any velocity field can be projected in theory onto $\boldsymbol{\mathcal{V}}_{n}$ in the limit $n\to \infty$. Ellipsoidal harmonics, which are the eigenfunctions of the Laplace operator in ellipsoidal coordinates, are known to form a complete basis \citep{dassios2012ellipsoidal}. Unfortunately, ellipsoidal harmonics have neither explicit expressions nor known recurrence relationships to generate them.
 
For a given degree $n$, alternative bases have been proposed. \citet{vantieghem2014inertial,backus2017completeness,ivers2017enumeration} established that inertial modes, i.e. the eigenmodes of rotating fluids restored by the Coriolis force, form a basis of $\boldsymbol{\mathcal{V}}_{n}$ in ellipsoids rotating at steady angular velocities. Hence, the global method describes the dynamics of the perturbation $\boldsymbol{u}$ in terms of a superposition of inertial (and geostrophic) modes. This approach has been considered by \citet{kerswell1993instability,kerswell1998tidal,zhang2010fluid,zhang2012asymptotic,zhang2013non,zhang2014precessing}. The latter studies look at the effects of various mechanical forcings in steady spheroidal containers ($a=b$), using the explicit formula of spheroidal inertial modes \citep{greenspan1968,zhang2004inertial}. 
On the other hand, \citet{vantieghem2015latitudinal} study global instabilities driven by latitudinal libration in triaxial ellipsoids, but only considering the inertial modes of degree $n\leq3$ \citep{vantieghem2014inertial}.
The explicit spatial dependence of inertial modes is not available in triaxial ellipsoids for higher degrees \citep{vantieghem2014inertial,backus2017completeness}, and also in spheroidal containers as soon as the time dependence of the figure axes (\ref{Eq_Forcing_Body}) is taken into account. 

Instead we build an algebraic polynomial basis of $\boldsymbol{\mathcal{V}}_{n}$ in time-dependent ellipsoids, denoted $\left \{ \boldsymbol{v}_{i} (\boldsymbol{r},t) \right \}$. The basis does not required to satisfy any dynamical equation, such that it holds at any time provided that the boundary (\ref{Eq_Forcing_Body}) is ellipsoidal in the body frame. It is thus an alternative to ellipsoidal harmonics to perform spectral computations in ellipsoids. This basis has two main advantages over other ellipsoidal harmonics: (i) the Cartesian coordinate system is easier to tackle than the ellipsoidal one; (ii) the basis is explicit and can be generated for any polynomial degree $n$.

We consider linearly independent Cartesian monomials $x^{i}y^{j}z^{k}$ of degree $i+i+k  \leq n-1$. The number of such independent monomials is $N_2 = n (n+1) (n+2)/6$. Among them, there are $N_1 = n (n+1)/2$ monomials independent of $z$, denoted $g_i (\boldsymbol{r})$. The other monomials, denoted $h_i (\boldsymbol{r})$, contain $z$ as factor. We index the set of these polynomials as \citep{lebovitz1989stability,Barker11062016}
\begin{subequations}
	\label{Eq_ghi}
	\begin{align}
		\left \{ g_i (\boldsymbol{r}) \right \} &= \left \{ 1,x,y,x^2,xy,y^2,\dots, x^{n-1}, y^{n-1} \right \},& i \in [1, N_1], \\
		\left \{ h_i (\boldsymbol{r}) \right \} &= \left \{ z, xz, yz, z^2, \dots, z^{n-1} \right \}, & i \in [N_1 + 1, N_2].
	\end{align}
\end{subequations}
Next we consider the linearly independent basis elements at each time
\begin{subequations}
	\label{Eq_Basis_Vn}
	\begin{align}
		\boldsymbol{v}_{i} (\boldsymbol{r},t) &= \nabla [g_i (\boldsymbol{r}) F (\boldsymbol{r},t) ] \times \boldsymbol{\widehat{x}}, & i \in [1,N_2], \\
		\boldsymbol{v}_{N_2 + i} (\boldsymbol{r},t) &= \nabla [g_i (\boldsymbol{r}) F (\boldsymbol{r},t)] \times \boldsymbol{\widehat{y}}, & i \in [1,N_2], \\
		\boldsymbol{v}_{2N_2 + i} (\boldsymbol{r},t) &= \nabla [h_i (\boldsymbol{r}) F (\boldsymbol{r},t)] \times \boldsymbol{\widehat{z}}, & i \in [1,N_1],
	\end{align}
\end{subequations}
with $F (\boldsymbol{r},t) = 1 -  ( {x}/{a(t)} )^{2} - ( {y}/{b(t)} )^{2} - ( {z}/{c(t)} )^{2}$.
Note that the total number of basis elements (\ref{Eq_DimGPall}) satisfies $N_{\mathcal{V}} = N_{1} + 2N_{2}$. The polynomial set (\ref{Eq_ghi}) ensures that basis elements (\ref{Eq_Basis_Vn}) are linearly independent. It is worth noting that the basis (\ref{Eq_Basis_Vn}) is neither orthogonal nor normalised, which is not necessary to build the stability equations. They can be \emph{a posteriori} orthonormalised with the modified Gram-Schmidt algorithm. The basis (\ref{Eq_Basis_Vn}) is explicit and can be built analytically for any degree $n$.

We have also implemented another algorithm to build the basis of $\boldsymbol{\mathcal{V}}_{n}$ for arbitrary $n$. It relies on spherical harmonics, after transforming the triaxial ellipsoid into a sphere with the Poincar\'e transform \citep{poincare1910precession,wu2011high}. The method is described in Appendix \ref{Appendix_GP_Vn}. However the algorithm is less efficient than the above procedure, since the basis is built numerically.

We consider perturbations $\boldsymbol{u} (\boldsymbol{r},t) \in \boldsymbol{\mathcal{V}}_{n}$ at any time and expand them as linear combinations of the $N_{\mathcal{V}}$ basis elements
\begin{equation}
	\boldsymbol{u} (\boldsymbol{r},t) = \sum \limits_{i=1}^{N_{\mathcal{V}}} \alpha_{i} (t) \, \boldsymbol{v}_{i} (\boldsymbol{r},t),
	\label{Eq_IE_GPexpand}
\end{equation}
where $\left \{ \alpha_{i} (t) \right \}$ is a set of arbitrary time-dependent coefficients. In the expansion (\ref{Eq_IE_GPexpand}) we emphasise that the basis polynomial elements are also time dependent, since the ellipsoid in the body frame has time-dependent axes. We substitute the expansion (\ref{Eq_IE_GPexpand}) into the stability equation (\ref{Eq_IE}) and project the resulting equation on the polynomial basis (\ref{Eq_Basis_Vn}), using the (real) inner product defined by the integral over the ellipsoidal volume
\begin{equation}
	\langle \boldsymbol{v_i} , \boldsymbol{v_j} \rangle (t) = \int \limits_{\mathcal{V}}  \boldsymbol{v_i} (\boldsymbol{r},t) \cdot \boldsymbol{v_j} (\boldsymbol{r},t) \, \mathrm{d}x \, \mathrm{d}y \, \mathrm{d}z.
	\label{Eq_Inner_Product}
\end{equation}
Then the stability problem (\ref{Eq_IE}) yields a finite number of ordinary differential equations
\begin{equation}
	\sum_{i=1}^{N_{\mathcal{V}}} N_{ij} \frac{\mathrm{d} \alpha_{j}}{\mathrm{d} t} + \sum_{i=1}^{N_{\mathcal{V}}} L_{ij} \alpha_{j} (t) = \sum_{i=1}^{N_{\mathcal{V}}} M_{ij} \alpha_{j} (t),
	\label{Eq_Stab_Sum}
\end{equation}
where $N_{ij}$, $L_{ij}$ and $M_{ij}$ are the time-dependent elements of squares matrices $\boldsymbol{N}$, $\boldsymbol{L}$ and $\boldsymbol{M}$ of size $N_{\mathcal{V}} \times N_{\mathcal{V}}$. Explicitly these elements are given by
\begin{subequations}
	\label{Eq_NijMij}
	\begin{align}
		N_{ij} &= \langle \boldsymbol{v_i} , \boldsymbol{v_j} \rangle (t), \ \, L_{ij} (t) = \langle \boldsymbol{v_i} , \mathrm{d}\boldsymbol{v_j}/\mathrm{d}t \rangle (t),\\
		M_{ij} &= - \langle \boldsymbol{v_i} , (\boldsymbol{U} \cdot \nabla) \boldsymbol{v_j} + (\boldsymbol{v_j} \cdot \nabla ) \boldsymbol{U} + 2 \boldsymbol{\Omega}^{\mathcal{B}} (t) \times \boldsymbol{v_j} \rangle (t). \label{Eq_Mij}
	\end{align}
\end{subequations}
Note that the pressure term does not contribute to (\ref{Eq_Mij}).
We compute the elements (\ref{Eq_NijMij}) explicitly using the formula \citep[misprint corrected from][]{lebovitz1989stability}
\begin{equation}
	\int \limits_{\mathcal{V}} x^{i} y^{j} z^{k} \, \mathrm{d}x \, \mathrm{d}y \, \mathrm{d}z = 
	\begin{cases}
		8 \pi [a(t)]^{2 \gamma_{1} + 1} [b(t)]^{2 \gamma_{2} + 1} [c(t)]^{2 \gamma_{3} + 1} \frac{(\gamma+1)! (2\gamma)!}{(2\gamma+3)! \gamma!}  & \text{if} \ i,j,k \ \text{all even}, \\
		0 & \text{if} \ i,j \ \text{or} \ k \ \text{odd},
	\end{cases}
	\label{Eq_IntVol_Lebovitz}
\end{equation}
with $2\gamma_{1} = i$, $2\gamma_{2} = j$, $2\gamma_{3} = k$, $\gamma = \gamma_{1} + \gamma_{2} + \gamma_{3}$ and $\gamma! = \gamma_{1}! \gamma_{2}! \gamma_{3}!$. Stability equations (\ref{Eq_Stab_Sum}) are written in canonical matrix form
\begin{equation}
	\frac{\mathrm{d}\boldsymbol{\alpha}}{\mathrm{d} t} = \boldsymbol{N}^{-1} \left ( \boldsymbol{M} - \boldsymbol{L} \right ) \boldsymbol{\alpha} = \boldsymbol{\mathcal{J}} \boldsymbol{\alpha},
	\label{Eq_IE_StabGP2}
\end{equation}
with the unknown vector $\boldsymbol{\alpha} (t) = \left ( \alpha_{1} (t), \alpha_{2} (t), \dots \right )^{T}$ and $\boldsymbol{\mathcal{J}}$ is the time-dependent Jacobian matrix of the system. 

The basic flow $\boldsymbol{U}(\boldsymbol{r},t)$ is unstable if the perturbation $\boldsymbol{u}(\boldsymbol{r},t)$ has at least one modal coefficient $\alpha_{i}(t)$, governed by the stability equation (\ref{Eq_IE_StabGP2}), which grows without bound in time. The most unstable perturbation is associated with the fastest growth rate denoted $\sigma$.
Since the basic flow (\ref{Eq_BF_Orbit}) is periodic of period $T=2\pi/\Omega_0$, its stability can be determined using the Floquet theory. We compute the eigenvalues (Floquet exponents) $\left \{ \mu_i \right \}$ of the fundamental matrix $\boldsymbol{\Phi}(t)$ evaluated at time $T$. The fundamental matrix is solution of
\begin{equation}
	\frac{\mathrm{d} \boldsymbol{\Phi} }{\mathrm{d} t} = \boldsymbol{\mathcal{J}} \boldsymbol{\Phi}, \, \ \, \boldsymbol{\Phi} (0) = \boldsymbol{I},
	\label{Eq_StabFloquet}
\end{equation}
with $\boldsymbol{I}$ the identity matrix. Then the growth rates $\left \{ \sigma_i \right \}$ and the frequencies $\left \{ \omega_i \right \}$ associated with the flow perturbations are then
\begin{equation}
	\sigma_i = \frac{1}{T} \Re_e \left [ \ln \left ( \mu_i \right ) \right ] \, \ \, \omega_i = \frac{1}{T} \Im_m \left [ \ln \left ( \mu_i \right ) \right ],
\end{equation}
The fastest growth rate, associated with the most dangerous unstable flow, is $\sigma = \max_i \sigma_i$ and its associated angular frequency is $\omega$.
Note that in the special case of a circular orbit ($e=0$) the orbital forcing (\ref{Eq_ODEI_Worb}) is steady, such that the above Floquet analysis reduces to a classical eigenvalue stability analysis.

A key point of the global method is that the vector space $\boldsymbol{\mathcal{V}}_{n}$ is invariant under the action of the perturbation stability equation (\ref{Eq_IE}) \citep{kerswell1993instability,lebovitz1989stability,vantieghem2014inertial,backus2016completeness,ivers2017enumeration}. Thus the stability equation (\ref{Eq_IE}) exactly reduces to equation (\ref{Eq_IE_StabGP2}) for any finite value of the degree $n$ in the expansion (\ref{Eq_IE_GPexpand}). It is not an approximation and it is not necessary to replace $n$ by $\infty$ in the expansion (\ref{Eq_IE_GPexpand}), and then to truncate at finite $n$, to get the stability equation (\ref{Eq_IE_StabGP2}). It is a main difference with classical spectral Galerkin expansions in various geometries (e.g. spherical harmonics), even in the linear framework. For our purposes, it is another advantage of our properly chosen polynomial basis over the spectral basis of ellipsoidal harmonics.

Finally because the expansion (\ref{Eq_IE_GPexpand}) is exact, the global method at a given degree $n$ gives only exact sufficient conditions for inviscid instability (to the numerical precision of the numerical solver). 
New tongues of instability of the basic flow generally appear in the parameter space when $n$ increases. The largest growth rate also generally increases by considering larger and larger $n$, reaching progressively its asymptotic value. This phenomenon has already been noticed in previous global analyses performed at lower degrees $n\leq 7$ \citep[e.g.][]{kerswell1993instability,wu2011high,wu2013dynamo,vantieghem2015latitudinal,Barker11062016}. Indeed, more resonances are expected when $n$ increases. So we conclude that the global method at finite values of $n$ cannot prove the stability of the basic flow, but it gives sufficient conditions for instability. Finally, it is usually expected that the upper bounds of global growth rates, in the asymptotic limit $n \to \infty$, coincide with the growth rates of local perturbations of short wavelength (see \S\ref{subsec:wkb}). However no general mathematical proof is available to justify it.

	\subsection{Local method in unbounded fluids}
	\label{subsec:wkb}
To get a complementary physical understanding of fluid instabilities growing upon the basic flow (\ref{Eq_BF_Orbit}), we also perform a local (WKB) stability analysis. It probes the stability of any inviscid, three-dimensional and time-dependent basic flow in an unbounded fluid, considering localised plane wave perturbations of small wavelength which are advected along the basic flow \citep{lifschitz1991local,friedlander1991instability,friedlander2003localized}. Because the orbitally driven basic flow (\ref{Eq_BF_Orbit}) is linear in space coordinates, the short-wavelength perturbations exactly reduce to Kelvin waves \citep{bayly1986three,craik1986evolution,craik1989stability,waleffe1990three}
\begin{equation}
	\boldsymbol{u} (\boldsymbol{r},t) = \boldsymbol{a} (t) \exp [\mathrm{i} \, \boldsymbol{k}(t) \cdot \boldsymbol{r}],
	\label{Eq_Kelvin_Waves} 
\end{equation}
with $\boldsymbol{k}(t)$ the time-dependent wavenumber and $\boldsymbol{a}(t)$ the time-dependent amplitude of the velocity perturbation. Kelvin waves (\ref{Eq_Kelvin_Waves}) are exact inviscid, nonlinear and incompressible solutions upon the basic flow (\ref{Eq_BF_Orbit}) in the body frame if
\begin{subequations}
	\label{Eq_WKB}
	\begin{align}
		\frac{\mathrm{d} \boldsymbol{k}}{\mathrm{d} t} &= - \left (\nabla \boldsymbol{U} \right )^{T} (t) \, \boldsymbol{k}, \label{Eq_WKB_K} \\
		\frac{\mathrm{d} \boldsymbol{a}}{\mathrm{d} t} &= \left [ \left ( \frac{2 \boldsymbol{k} \boldsymbol{k}^{T}}{||\boldsymbol{k}||^{2}} - \boldsymbol{I} \right ) \nabla \boldsymbol{U} (t) + 2 \left ( \frac{\boldsymbol{k} \boldsymbol{k}^{T}}{||\boldsymbol{k}||^{2}} - \boldsymbol{I} \right )  \boldsymbol{\Omega}^{\mathcal{B}} (t) \times \right ] \boldsymbol{a}, \label{Eq_WKB_A}
	\end{align}
\end{subequations}
and the incompressibility condition $\boldsymbol{k}(t) \cdot \boldsymbol{a} (t) = 0$ hold.
The latter is always satisfied if it holds for the initial condition $(\boldsymbol{k}_{0}, \boldsymbol{a}_{0})$. 
The existence of an unbounded solution for $\boldsymbol{a}(t)$ is a sufficient condition for instability \citep{lifschitz1991local,friedlander2003localized}. 

Equations (\ref{Eq_WKB}) are independent of the magnitude of $\boldsymbol{k}_{0}$. So we restrict the initial wave vector to the spherical surface of unit radius $\boldsymbol{k}_0 = (\sin(\theta_0) \cos(\phi), \sin(\theta_0) \sin(\phi), \cos(\theta_0))^T$, where $\phi \in [0, 2\pi]$ is the longitude and $\theta_0 \in [0, \pi]$ is the colatitude between the spin axis $\boldsymbol{\widehat{z}}$ and the initial wave vector $\boldsymbol{k}_0$.
In practice, the equation (\ref{Eq_WKB_K}) is time stepped with a numerical solver from a range of initial wave vectors. Then we compute the maximum growth rate $\sigma$ of equation (\ref{Eq_WKB_A}) as the fastest growing solution from all possible initial wave vectors.

	\subsection{Numerical implementation}
\begin{figure}
	\centering
	\includegraphics[width=0.5\textwidth]{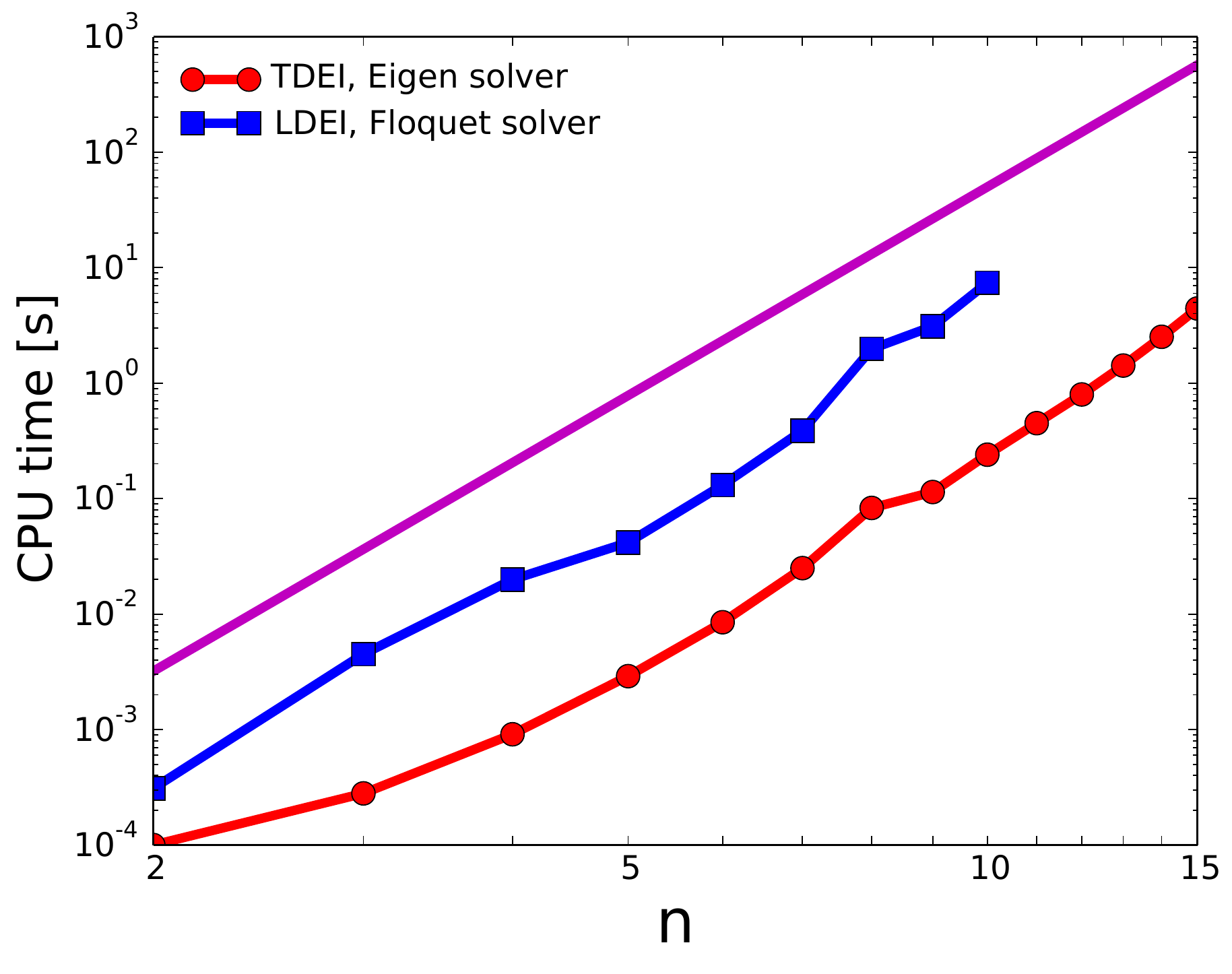}
	\caption{Characteristic CPU time to compute a growth rate for tidally driven and libration-driven instabilities presented below in section \ref{sec:results} with the different numerical solvers. To smooth out the variability of computation time between different parameters, we compute a stability map of 100 points in the plane $(\beta_{ac}, \Omega_0)$ for tides and in the plane $(\beta_0, 2e)$ for longitudinal-libration to extract an average time for one iteration. Circle symbols stand for tidal forcing and square ones for longitudinal libration. The magenta solid line shows the power law $\propto n^6$, in good agreement with the numerical scaling.}
	\label{Fig_CPUtime}
\end{figure}

We have developed for the global stability analysis the SIREN code (Stability with IneRtial EigeNmodes), freely available at \url{https://bitbucket.org/vidalje/siren}. It handles any mechanical forcing and basic flow of uniform vorticity. 
The matrices $\boldsymbol{N}$, $\boldsymbol{L}$ and $\boldsymbol{M}$ are first computed symbolically with Sympy (\url{http://www.sympy.org/}), a computer algebra system (CAS) for Python, which is used to manipulate the Cartesian polynomials $x^iy^jz^k$ in a symbolic way.
Then they are converted to Fortran subroutines with the Sympy fcode function and finally wrapped with f2py \citep{peterson2009f2py} for fast numerical evaluation inside Python using Numpy \citep{van2011numpy}.
The Jacobian matrix $\boldsymbol{\mathcal{J}}$ is computed numerically, because we cannot compute the symbolic inverse $\boldsymbol{N}^{-1}$ for arbitrary $n$.
Because of the difficulty to build the Jacobian matrix for an arbitrary forcing, previous global studies have only considered less than 200 basis elements ($n \leq 7$) \citep{kerswell1993instability,lebovitz1996new,wu2011high,vantieghem2015latitudinal,barker2016nonlinear,Barker11062016}. In practice, we have built and solved numerically the stability system (\ref{Eq_IE_StabGP2}) for degrees as large as $n=25$, yielding more than 6000 basis elements.

For the local stability analysis we have also developed the SWAN code (Short-Wavelength stability Analysis), freely available at  \url{https://bitbucket.org/vidalje/swan}. It gives sufficient conditions for inviscid instability of any basic flow (not necessarily of uniform vorticity) expressed in Cartesian coordinates. The stability equations (\ref{Eq_WKB}) are built using Sympy, then converted to a Fortran subroutine with the Sympy fcode function and finally wrapped with f2py for fast numerical evaluations with Numpy.

Both numerical codes use an explicit Runge-Kutta time step solver with adaptive step size (available in the Python library Scipy) to integrate the stability differential equations. Performing a survey in parameter space is an embarrassingly parallel problem, and our implementation takes full advantage of this situation using mpi4py (\url{http://mpi4py.scipy.org/}).

To validate our codes, we have first considered the precession of a steady spheroid ($a=b \neq c$). This benchmark is described in Appendix \ref{app:prec}. We perfectly recover previous studies \citep{kerswell1993instability,wu2011high}. We also get a very good agreement between local and global analyses, because we can reach large enough degrees $n$ with the SIREN code. Then we assess the performance of our SIREN code in figure \ref{Fig_CPUtime}, which shows the evolution of CPU time with $n$ for the tidally driven and libration-driven flows considered below (see \S\ref{sec:tdei} and \S\ref{sec:ldei}). 
We observe that the CPU time scales as $n^6$, in agreement with formula (\ref{Eq_DimGPall}). Indeed the number of basis elements scales as $n^3$ and so the number of elements in matrices $\boldsymbol{N}$,  $\boldsymbol{L}$ and $\boldsymbol{M}$ is of order $n^6$. 
As expected the eigenvalue solver is faster than the Floquet solver.

\section{Orbitally driven elliptical instabilities}
\label{sec:results}
In this section, we perform the stability analysis of the orbitally driven basic flow (\ref{Eq_BF_Orbit}). First we consider two particular cases of orbital forcing, namely the tidal forcing in non-synchronised bodies on circular orbits in \S\ref{sec:tdei} and the libration forcing in synchronised bodies on eccentric orbits in \S\ref{sec:ldei}. Then we survey in the whole parameter space the stability of ellipsoids moving along eccentric orbits in \S\ref{sec:odei}. In this case, time variations of the ellipsoidal axes can play a significant role and drive new vigorous instabilities, called orbitally driven elliptical instabilities (ODEI).

	\subsection{Tidally driven elliptical instability on circular orbits}
	\label{sec:tdei}
\begin{figure}
	\centering
	\begin{tabular}{cc}
		\subfigure[Local analysis]{\includegraphics[width=0.49\textwidth]{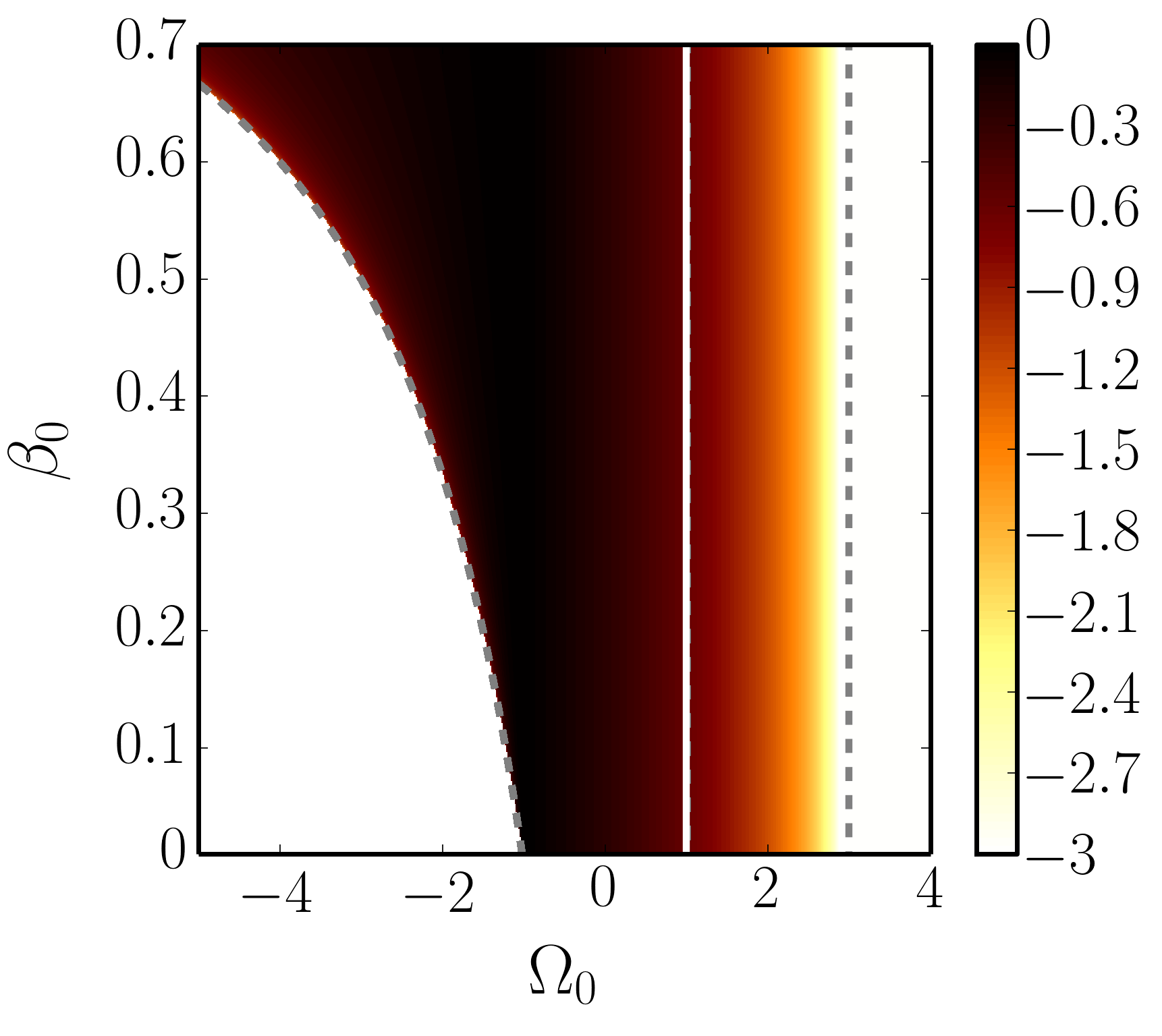}} &
		\subfigure[Global analysis $n=15$]{\includegraphics[width=0.49\textwidth]{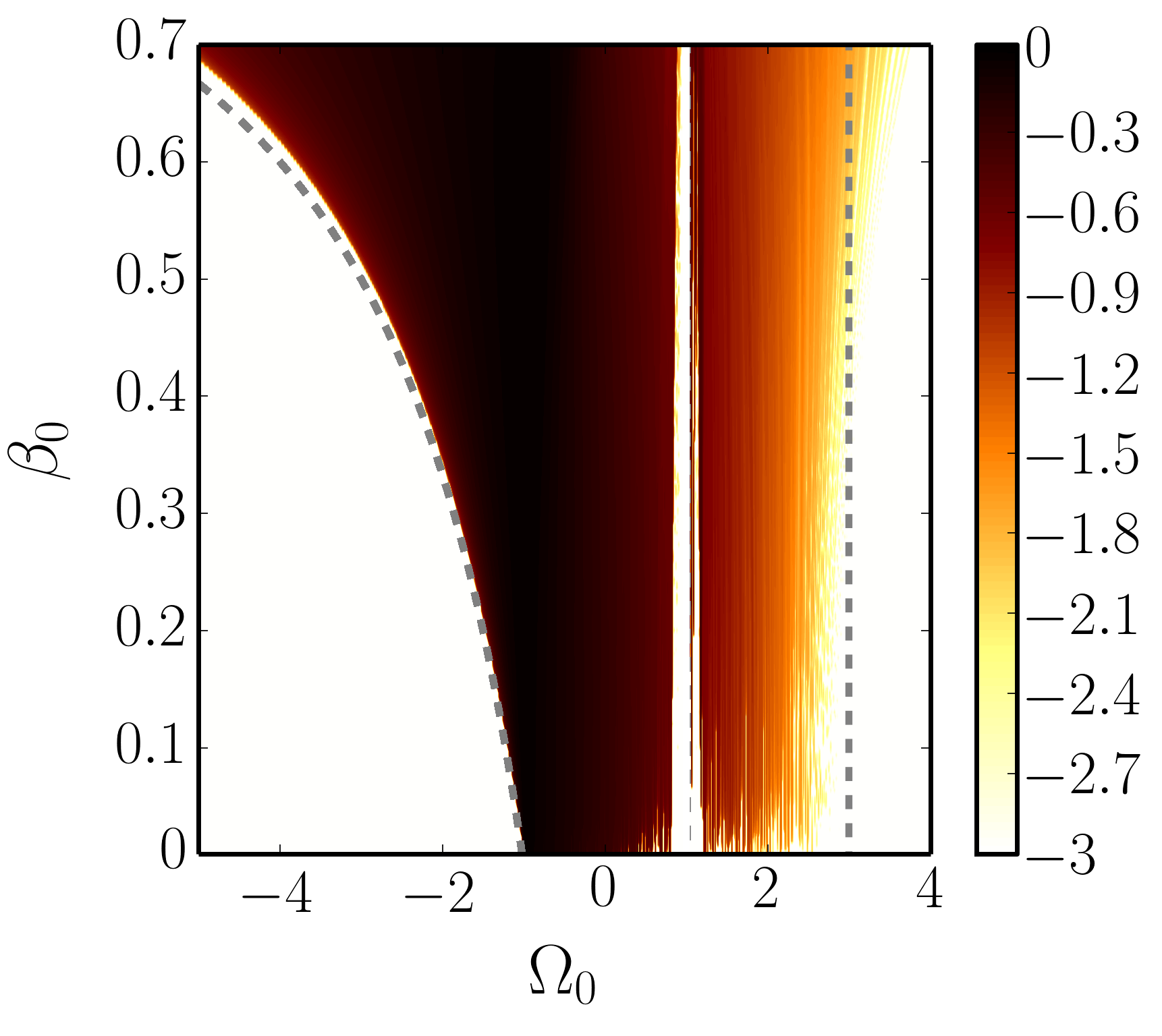}}
	\end{tabular}
	\caption{Areas of instability of the tidally driven flow in the $(\beta_{0}, \Omega_{0}$) plane. Color map shows $\log_{10} (\sigma / [\beta_{0} |1 - \Omega_0|] )$. Triaxial geometry $a=\sqrt{1+\beta_{0}}$, $b = \sqrt{1-\beta_{0}}$ and $c=1/(a b)$ such that the triaxial container has a constant dimensionless volume $4\pi/3$. On the vertical white line $\Omega_0=1$ the TDEI does not exist. Black dashed lines $\Omega_0 = (1~+~\beta_0)/(1~-~\beta_0)$ and $\Omega_0=3$ are the bounds of the forbidden zone FZ$_{\beta_0}$.}
	\label{Fig_TDEI2}
\end{figure}

We focus here on the effect of the equilibrium tide on a circular orbit ($e=0$). The fluid ellipsoid has steady semi-axes ($a, b, c$) and rotates at the steady orbital rate $\Omega_0$.
Basic flow (\ref{Eq_BF_Orbit}) thus reduces to the tidally driven basic flow
\begin{equation}
	\boldsymbol{U} (\boldsymbol{r}) = (1 - \Omega_{0}) \left [ -(1 + \beta_{0})y \,\widehat{\boldsymbol{x}} \,+\, (1 - \beta_{0}) x \,\widehat{\boldsymbol{y}} \right ].
	\label{eq:TDEI1}
\end{equation}
This flow can be unstable if $\Omega_{0} \neq 1$, leading to the classical tidally driven elliptical instability (TDEI).

On one hand, the TDEI has been widely studied with a local analysis in unbounded domains \citep{bayly1986three,craik1989stability,waleffe1990three,cebron2012elliptical}. Note that in the asymptotic limit $\beta_{0} \to 0$, equations (\ref{Eq_WKB}) can be solved analytically using a multiple-scale analysis in $\beta_{0}$ to get a theoretical growth rate (see appendix \ref{app:tdei_ledizes}). \citet{le2000three} shows that the TDEI exists in the range  $(\beta_0~+~1)/(\beta_0~-~1)<\Omega_0<3$.
Outside this range, the flow is stable and lies in the classical forbidden zone for $\beta_0 \ll 1$, hereafter denoted FZ$_{\beta_{0}}$. 

On the other hand, the global stability analysis of tidal basic flow (\ref{eq:TDEI1}) has been mainly performed for weakly deformed spheroids \citep{lacaze2004elliptical} or cylinders \citep{malkus1989experimental,eloy2003elliptic}.
Triaxial ellipsoids have also been considered \citep{gledzer1978finite,gledzer1992instability,kerswell2002elliptical,roberts2011flows,Barker11062016,barker2016nonlinear}, but only disturbed by perturbations of small polynomial degrees ($n \leq 7$).
Such large-wavelength instabilities do not compare well with the aforementioned local stability analyses. 
Using our framework, we can reach much larger polynomial degrees $n$. A comparison between local and numerical analyses is given in figure \ref{Fig_TDEI2}. As expected, the local results are upper bounds of global results (with expected matches for large enough polynomial degrees). A more in-depth discussion is given in appendix \ref{app:tdei_ledizes}. 

\begin{figure}
	\centering
	\begin{tabular}{cc}
		\subfigure[$\beta_{0}=0.15$]{\includegraphics[width=0.49\textwidth]{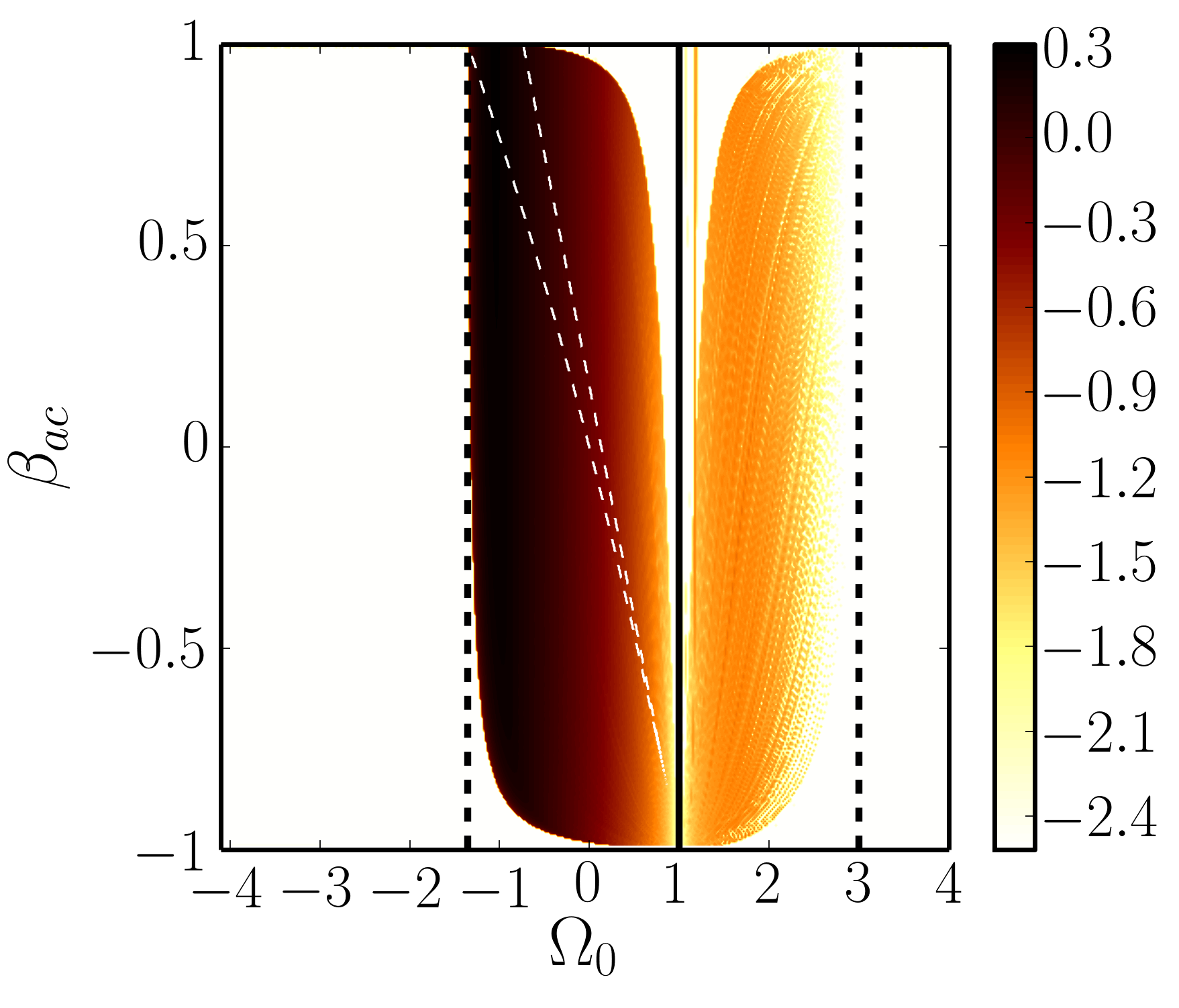}} &
		\subfigure[$\beta_{0}=0.6$]{\includegraphics[width=0.49\textwidth]{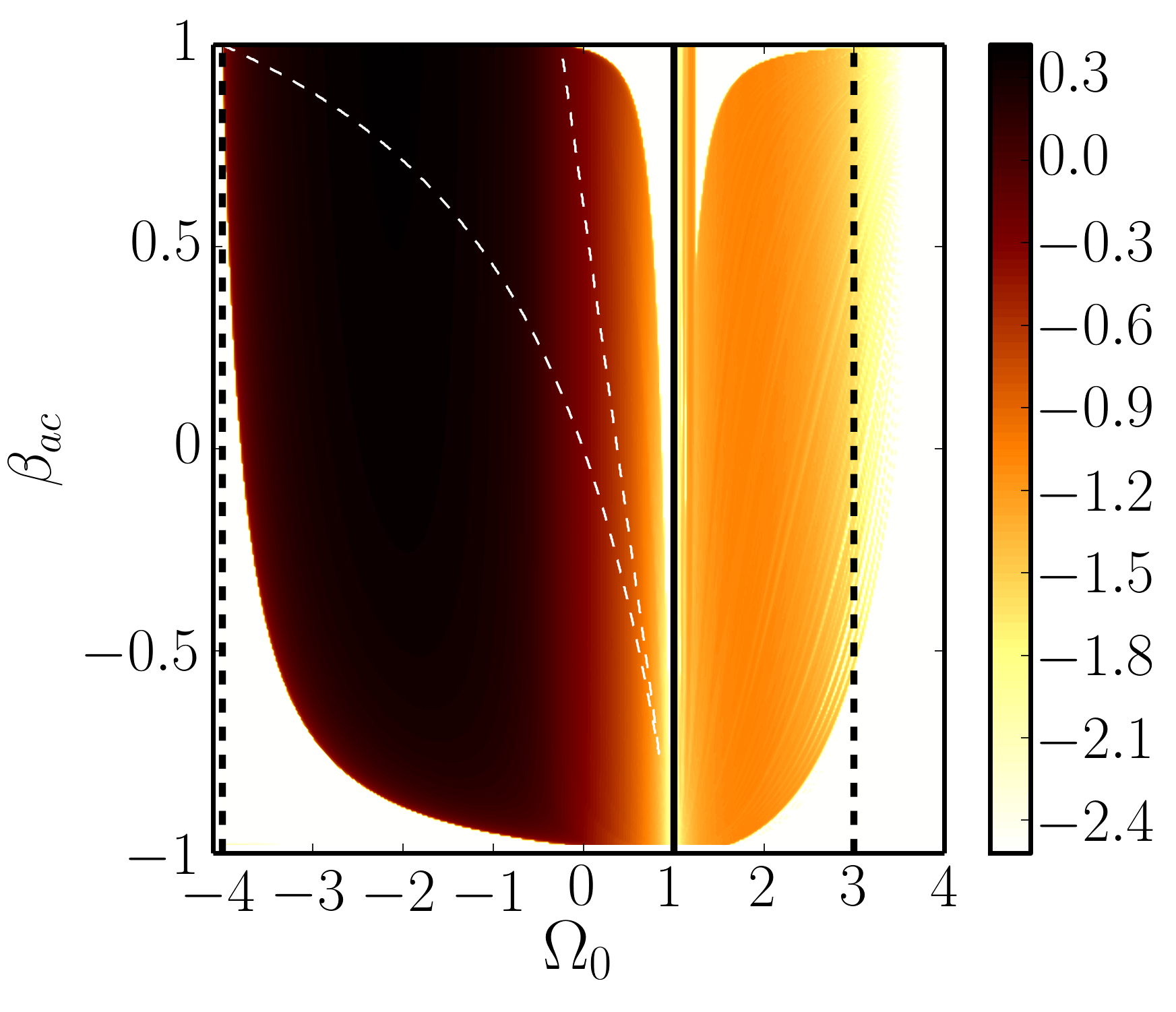}} \\
	\end{tabular}
	\caption{(a) \& (b) Areas of instability of the tidally driven flow in the $(\beta_{ac}, \Omega_{0}$) plane at degree $n=15$. Color map shows $\log_{10} \left ( \sigma / \beta_{0} \right )$. White areas correspond to marginally stable regions. Triaxial geometry $a=\sqrt{1+\beta_0}$ and $b = \sqrt{1-\beta_0}$ and $c = a \sqrt{(1-\beta_{ac})/(1+\beta_{ac})}$. Vertical dashed blacks lines represent the lower and upper bounds of the forbidden zone FZ$_{\beta_{0}}$. The solid black line indicates the synchronized case $\Omega_{0} = 1$ (no instability). White dashed lines correspond to the isoline $\sigma/\beta_0=0.01$ for the stability problem reduced to degree $n=1$, such that the spin-over instability is excited in between \cite[in this case, $\sigma$ is analytically known, see e.g.][]{roberts2011flows}.}
	\label{Fig_TDEI1}
\end{figure}

Global stability results at maximum degree $n=15$ are shown in figure \ref{Fig_TDEI1}, where the ratio $\sigma / \beta_{0}$ of the instability is computed for two equatorial ellipticities ($\beta_{0} = 0.15$ and $\beta_{0} = 0.6$). We always find $\sigma = 0$ when $\Omega_{0} = 1$ as expected, because $\boldsymbol{U} = \boldsymbol{0}$ from the expression (\ref{eq:TDEI1}). When $\beta_0$ increases, the zone of instability extends but it is still outside of the forbidden zone FZ$_{\beta_{0}}$.

We observe that ellipsoids spinning in the retrograde direction ($\Omega_{0} < 0$) are more unstable than the prograde ones ($\Omega_{0} > 0$). By varying the polynomial degree, we note that the TDEI for prograde rotation ($\Omega_0 > 0$) appears at larger $n$ than the TDEI for retrograde rotation ($\Omega_0 < 0$). As an example the spin-over mode \citep{kerswell2002elliptical}, associated with the linear basis ($n=1$), appears only for $\Omega_0 < 0$. Similarly the largest $\sigma$ is reached at smaller $n$ for retrograde rotation than for prograde rotation (not shown). 
We also observe an effect of $\beta_{ac}$ at large values of $|\beta_{ac}|$, not predicted by the local analysis (insensitive to $\beta_{ac}$). There, higher polynomial degrees may be needed to reach the asymptotic limit of the local analysis to completely fill in the map with new global unstable tongues.

\begin{figure}
	\centering
	\begin{tabular}{cc}
    	\subfigure[$\beta_{ac} = 0.5, \, \Omega_0 = -1, \, \omega = 0$]{\includegraphics[width=0.47\textwidth]{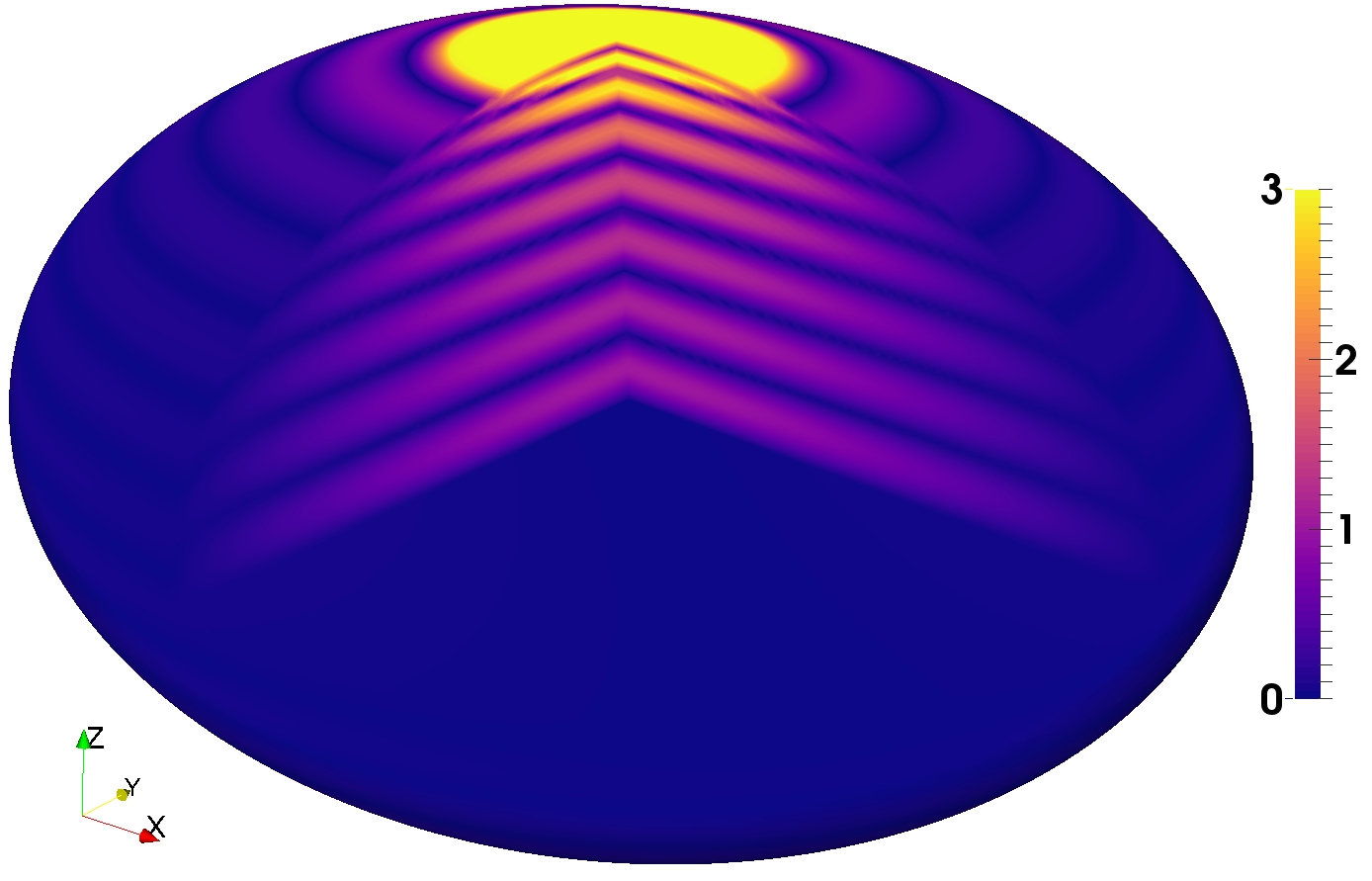}} &
		\subfigure[$\beta_{ac} = 0.5, \, \Omega_0 = -0.5, \, \omega = 0$]{\includegraphics[width=0.47\textwidth]{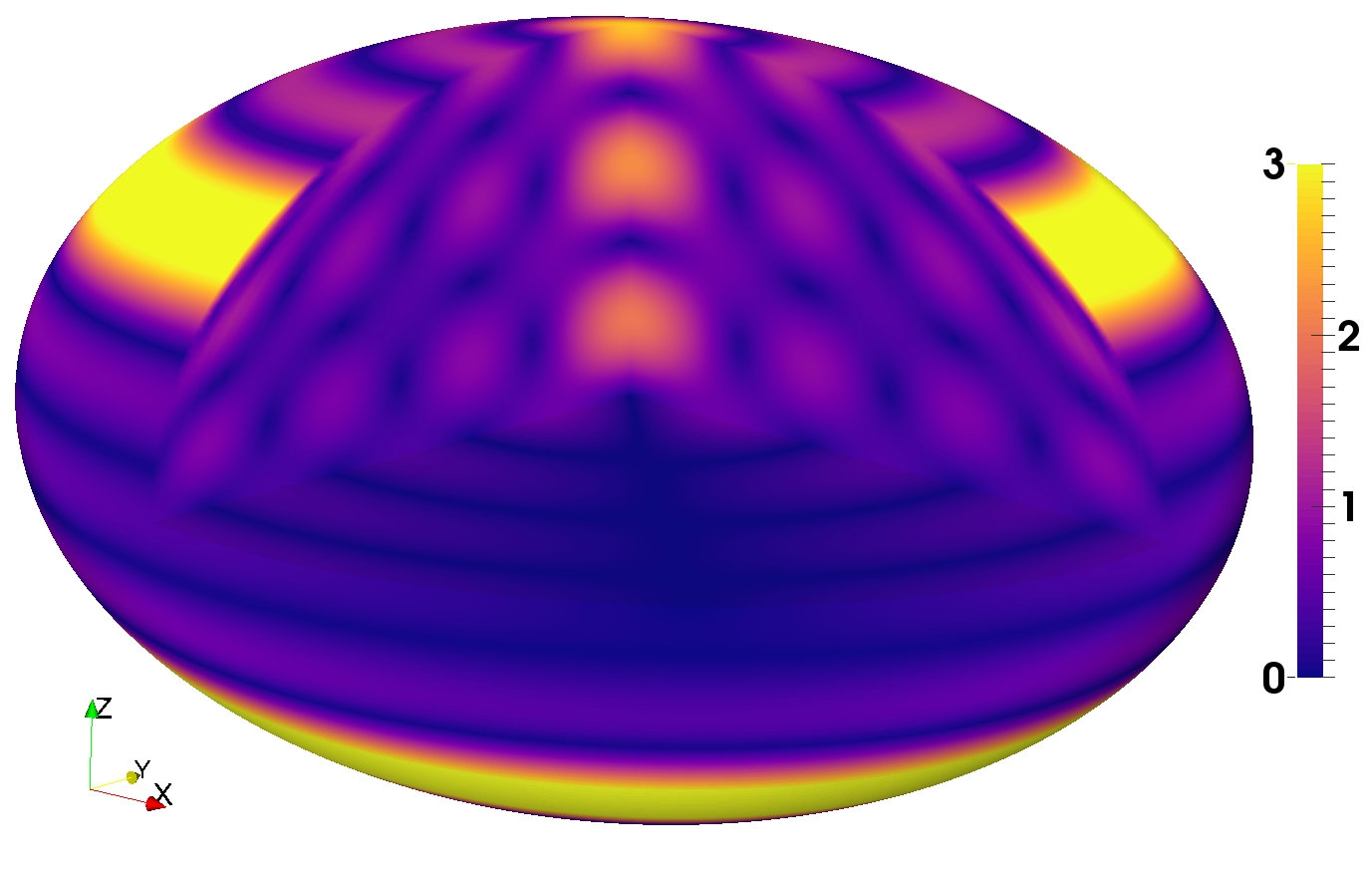}} \\
         \subfigure[$\beta_{ac} = 0.5, \, \Omega_0 = -0.1, \, \omega=3.27$]{\includegraphics[width=0.47\textwidth]{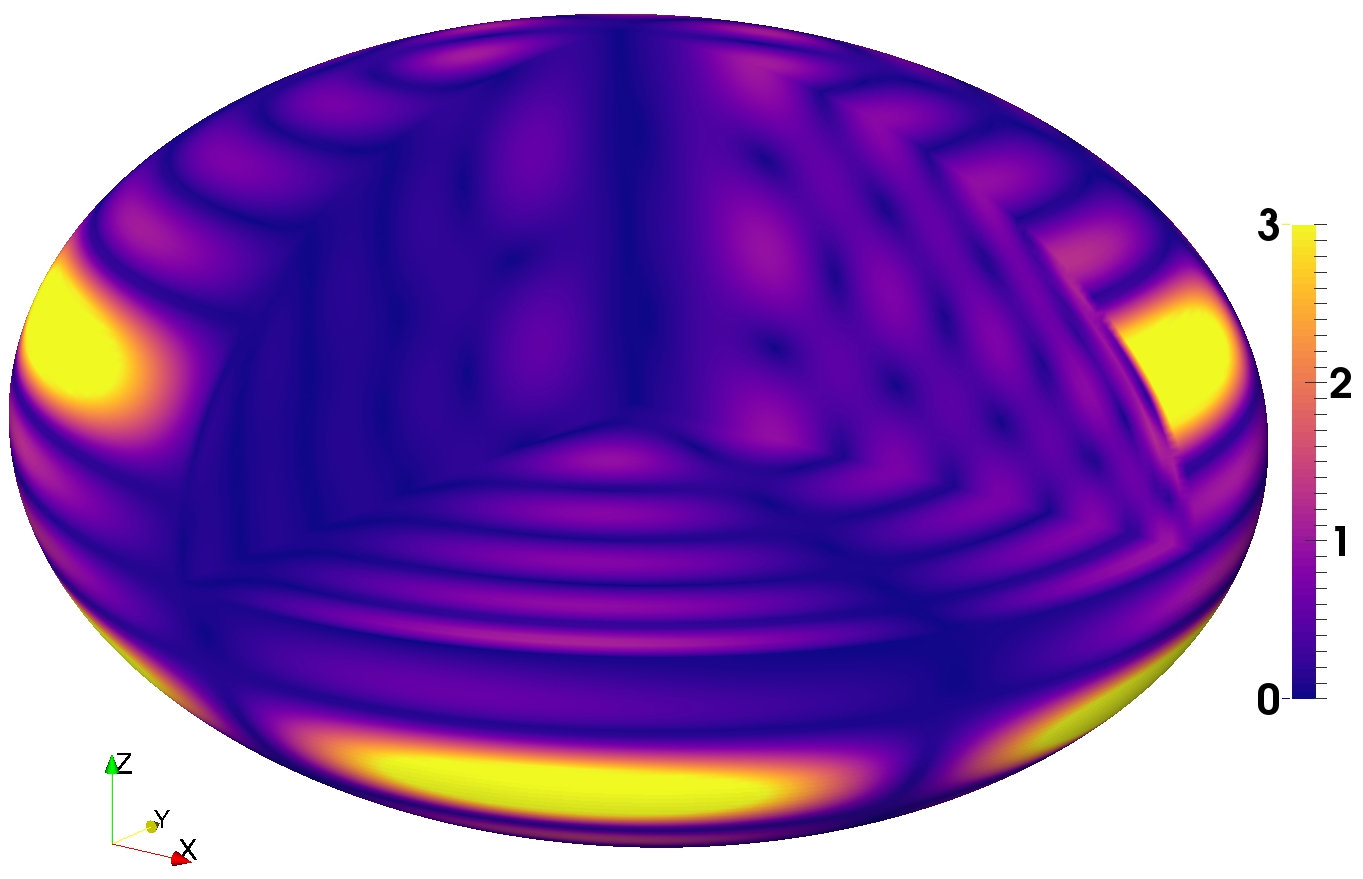}} &
		\subfigure[$\beta_{ac} = 0.5, \, \Omega_0 = 2, \, \omega=0$]{\includegraphics[width=0.47\textwidth]{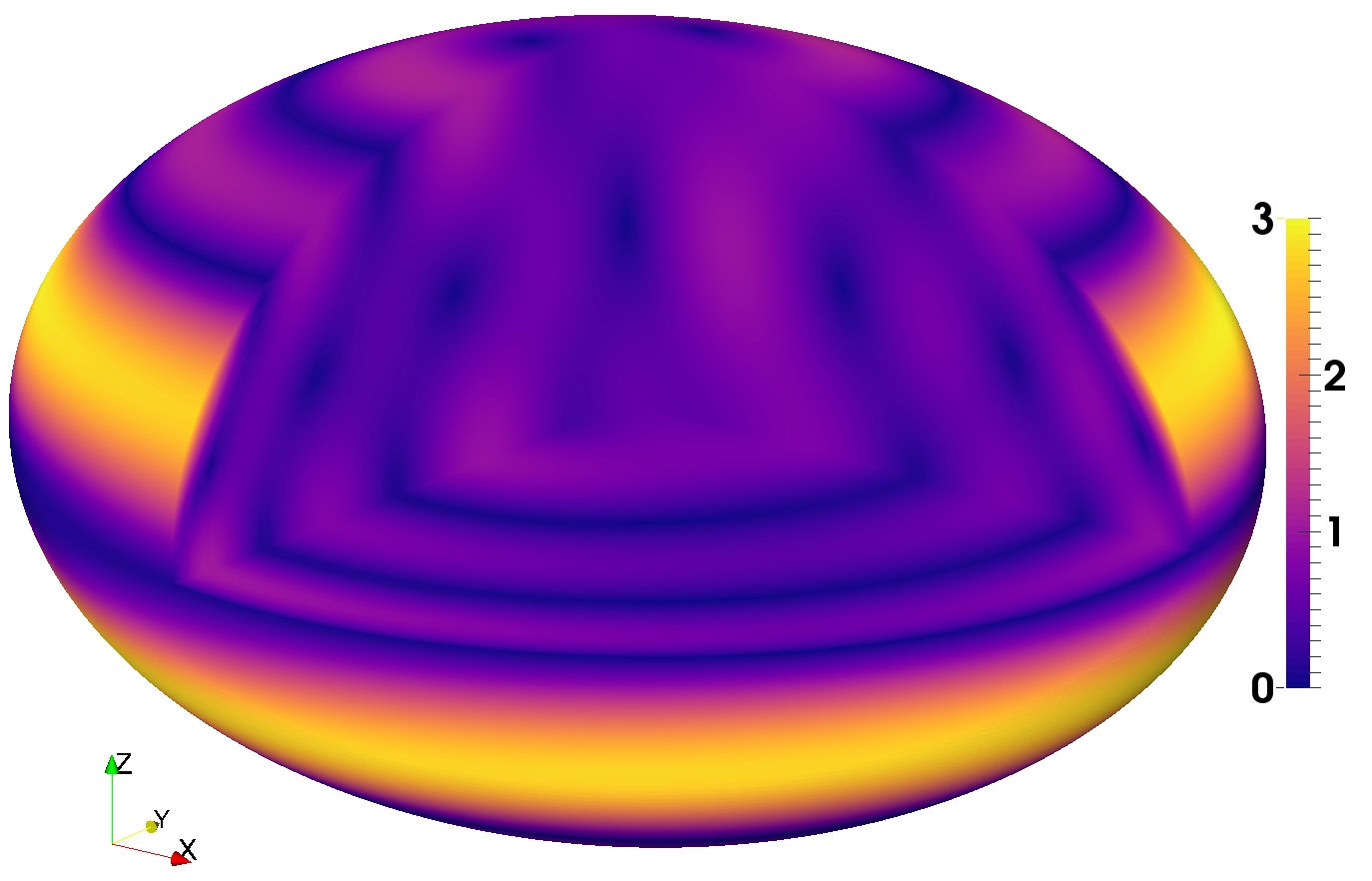}} \\
 \end{tabular}
	\caption{Three-dimensional renderings of the most unstable flows associated with figure \ref{Fig_TDEI1}. Degree $n=15$ and amplitude of equilibrium tide $\beta_0 = 0.15$. Velocity magnitude $||\boldsymbol{u}||$ is shown in meridional/equatorial planes and at the ellipsoidal surface. The colour map is saturated for $|| \boldsymbol{u}|| \geq 3$.}
	\label{Fig_TDEI_Flow}
\end{figure}

In figure \ref{Fig_TDEI_Flow} we show the most dangerous unstable flows of the TDEI, as varying $\Omega_0$ when $\beta_0=0.15$.
When $\beta_0=0.6$ the spatial structures of the flows, at the same values of $\Omega_0$, are similar (not shown).
In all these flows, the motions seem to be concentrated in conical layers tilted from the fluid rotation axis. 
Between these layers, the flow has low or zero amplitude.
Some flows also exhibit one or several nodes in azimuth.

\begin{figure}
	\centering
	\begin{tabular}{cc}
    	\subfigure[$n=10, \sigma=0.297, \omega=0$]{\includegraphics[width=0.49\textwidth]{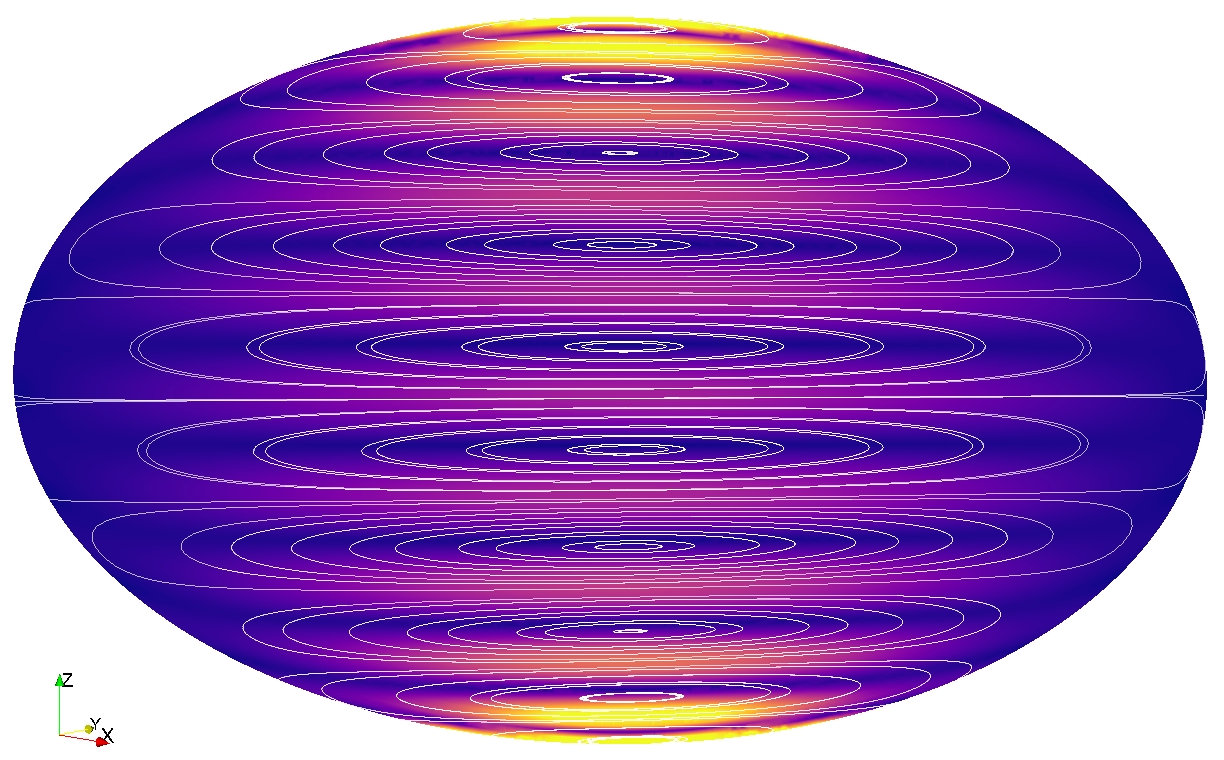}} &
		\subfigure[$n=15, \sigma=0.299, \omega=0$]{\includegraphics[width=0.49\textwidth]{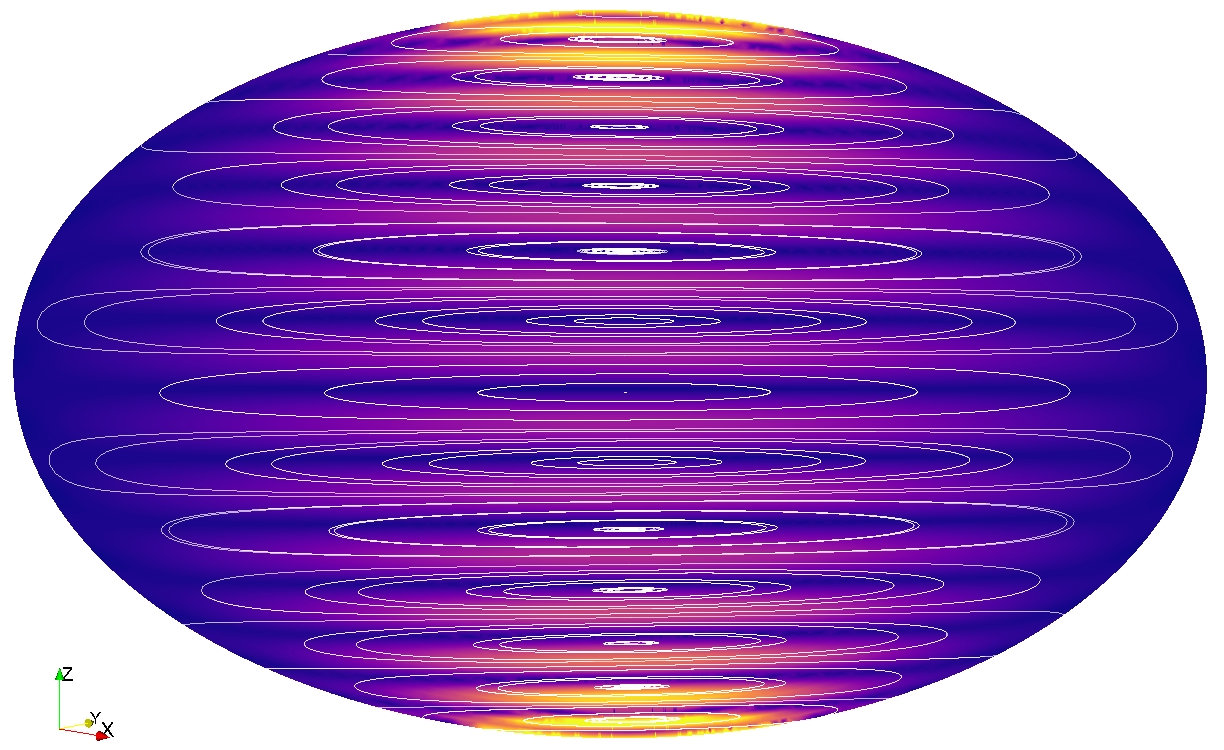}}
   \end{tabular}
	\caption{"Stack of pancakes"-like instability for $\beta_{0}=0.15, \beta_{ac} = 0.5$ and $\Omega_0 = -1$. Magnitude $||\boldsymbol{u}||$ and velocity streamlines in the meridional plane at $45$ degrees from the long axis where the stretching is maximum.} 
	\label{Fig_TDEI_Flow_Pancake}
\end{figure}

In figure \ref{Fig_TDEI_Flow} (a) computed at $\Omega_0 = -1$, the flow has the particular structure of a stack of pancakes (SoP). Note that the modal angular frequencyof SoP is $\omega=0$ in the body frame, as predicted theoretically by \citet{lebovitz1996short,Barker11062016}. SoP structures have also been observed in experiments \citep{grannan2014experimental} and in nonlinear direct simulations \citep{favier2015generation,barker2016nonlinear}.
For this instability, each pancake moves horizontally in the direction opposite to its neighbours (horizontal epicyclic motions), independently of all other heights, in a plane at $45^\circ$ from the main equatorial axis where the stretching is maximum \citep{waleffe1990three}. SoP structures are illustrated by the streamlines in figure \ref{Fig_TDEI_Flow_Pancake}.
High amplitudes are located near the poles.
Note also that the number of pancakes increases as $n$ increases, as suggested by figure \ref{Fig_TDEI_Flow_Pancake}. However this number seems to be insensitive to the amplitude of the equilibrium tide $\beta_{0}$ and to $\beta_{ac}$ (not shown).
Such a small-scale flow will undoubtedly lead to turbulence if it reaches high enough amplitudes.

When $\Omega_0 = -0.5$ the unstable flow shown in figure \ref{Fig_TDEI_Flow} (b) for $n=15$ is identical for degrees $n=10$ and $n=6$. So we expect the observed flow to be the most unstable flow when $\Omega_0 = -0.5$.
It is mostly an equatorially symmetric mode, with high intensities located on the rotation axis and within a surface band at mid-latitudes.
Finally flows in figures \ref{Fig_TDEI_Flow} (c) and (d) share a common structure which could evolve with $n$.

	\subsection{Libration-driven elliptical instability}
	\label{sec:ldei}
We investigate here the stability of a synchronised fluid body ($\Omega_{0} = 1$) on an eccentric orbit ($0 < e \leq 1$).
The associated forcing, called longitudinal librations, leads to the libration-driven elliptical instability (LDEI).
We distinguish the following two limit cases of longitudinal librations.

If the fluid is enclosed in a solid container with a strong enough rigidity (e.g. a silicate mantle), the entire body rigidly rotates with a fixed shape. Dynamical tides can be neglected with respect to the equilibrium tide, such that $\beta_{ab} (t) = \beta_0$. The forcing bears the name of physical librations. A differential rotation exists between the fluid spin rate and the equilibrium tide, rotating at leading order at the angular velocity (equal to the orbital angular velocity in our model)
 \begin{equation}
	\boldsymbol{\Omega}^\mathcal{B} (t) = \left ( 1 + \epsilon \sin t \right ) \widehat{\boldsymbol{z}}
	\label{Eq_LongLibPhysic_Forcing}
\end{equation}
with $\epsilon \leq 2e$ the libration amplitude. This amplitude depends on the rheology of the celestial body.
The LDEI driven by physical librations has been studied amongst others by \citet{cebron2012libration,noir2012experimental,grannan2014experimental} and \citet{favier2015generation}.
Note that the physical libration forcing (\ref{Eq_LongLibPhysic_Forcing}) could actually contain multiple frequencies due to the presence of several attracting bodies \citep{rambaux2011moon}. 
In the limit $\beta_{0} \ll 1$, the local growth rate of this physical LDEI is \citep{herreman2009effects,cebron2012libration,cebron2014libration}
\begin{equation}
	\sigma_\text{wkb} = \frac{17}{64} \epsilon \beta_{0}.
    \label{Eq_LDEI_WKB}
\end{equation}
According to the local formula (\ref{Eq_LDEI_WKB}) the LDEI is triggered for any non-zero $\epsilon$ and $\beta_0$.

On the other hand, if the fluid spin rate is low enough for the shape of the fluid body to have time to adapt to the gravitational constraints or is enclosed within a solid container with a small enough rigidity (e.g. a thin icy shell), the ellipsoidal cavity always points toward the attractor. Then the container has a time-dependent equatorial ellipticity given by the expression (\ref{Eq_ODEI_beta}). A differential rotation exists between the fluid spin rate and the dynamical tides (superimposed on the equilibrium tide). 
In the inviscid framework of this work, we call this forcing optical librations \cite[because the amplitude of optical librations is $2e$, see e.g.][]{murray1999solar}. In the limit $e \to 0$, this forcing simply associates a prescribed time evolution of $(a(t),b(t),c(t))$ to the forcing (\ref{Eq_LongLibPhysic_Forcing}), rather than considering a constant ellipsoidal shape.
At this first order in $e$, the time dependence of the dynamical tides is monochromatic, in agreement with numerical results of figure \ref{Fig_ODEI_Keplersol} at small $e$.
Physical librations with maximum amplitude $\epsilon = 2e$ are recovered if we neglect the dynamical tides, yielding $\beta_{ab}(t) = \beta_0$.

\begin{figure}
	\centering
    \begin{tabular}{cc}
		\subfigure[]{\includegraphics[width=0.49\textwidth]{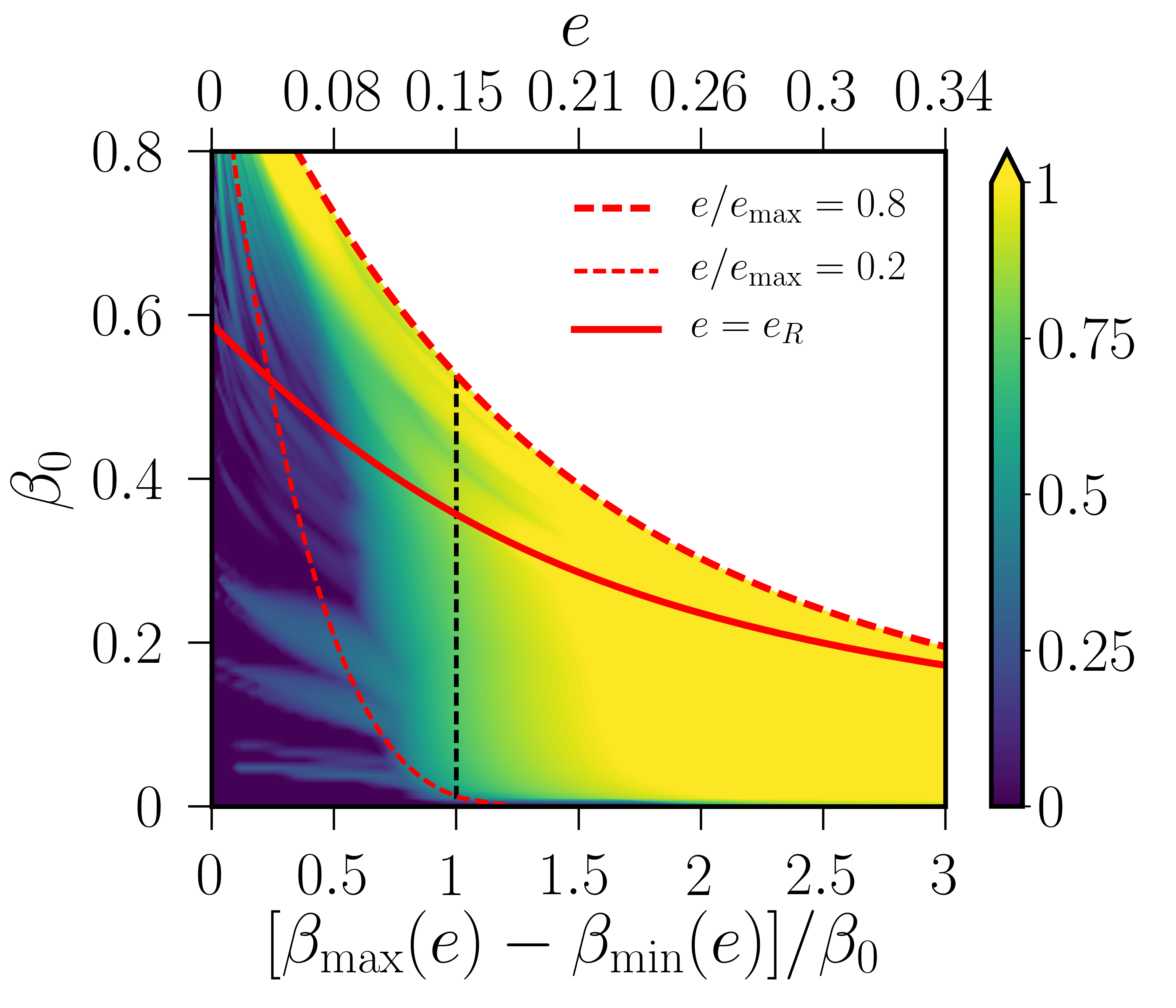}} & \subfigure[ $n=10, \omega=0.357$]{\includegraphics[width=0.45\textwidth]{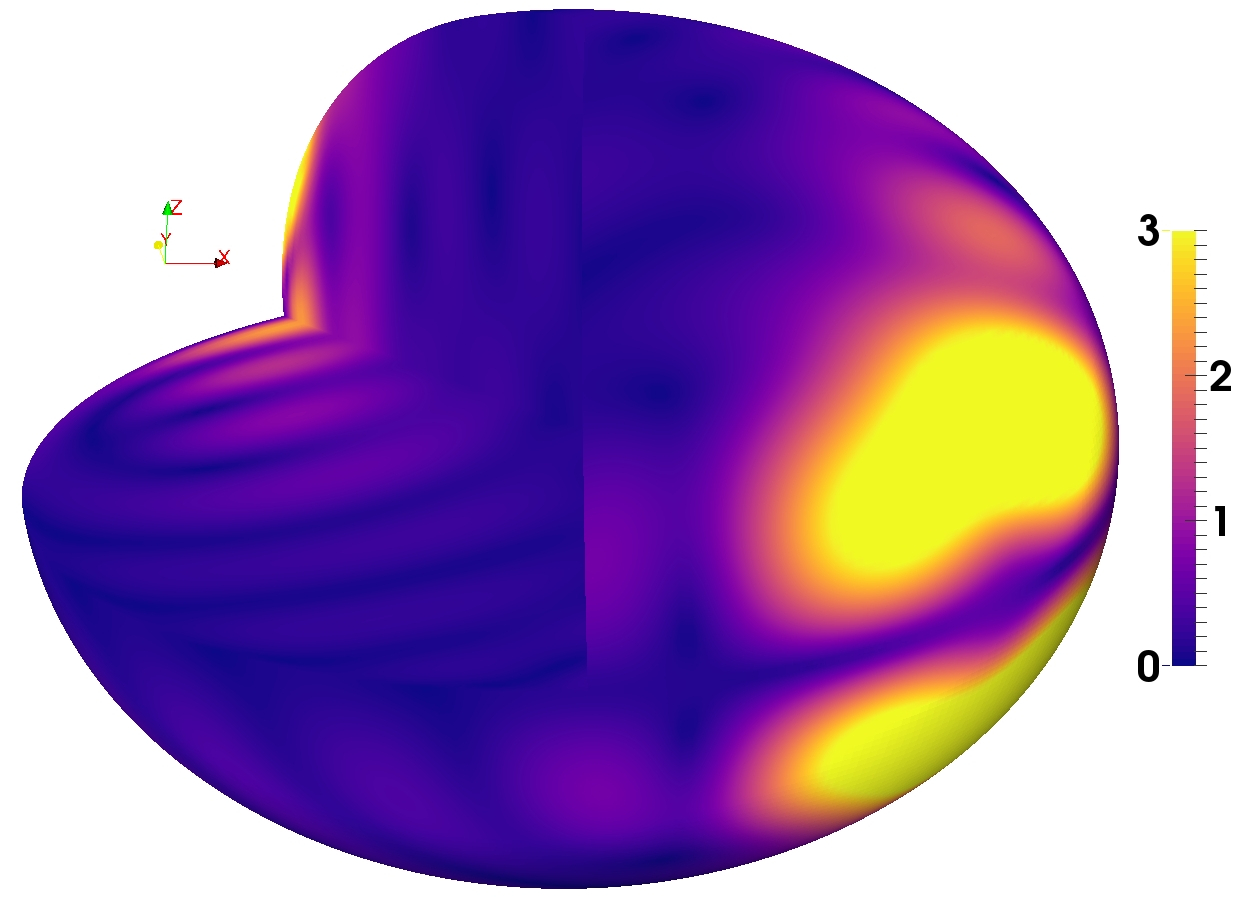}} \\
    \end{tabular}
	\caption{Libration-driven elliptical instability ($\Omega_0 = 1$) for eccentric Kepler orbits. (a) Ratio $\log_{10}(\sigma / \sigma_\text{wkb})$ with $\sigma_\text{wkb}$ the local growth rate of LDEI given by the formula (\ref{Eq_LDEI_WKB}). Color bar saturated at $\log_{10}(\sigma / \sigma_\text{wkb}) = 1$. Larger values ($\log_{10}(\sigma / \sigma_\text{wkb}) \leq 2$) occur for small enough $\beta_0$ when $[\beta_{\max}(e)-\beta_{\min}(e)]/\beta_0 \geq 3$ (not shown). Solid red line shows the Roche limiting circular orbit. Ellipsoids moving along orbits located above the Roche limit could be unstable against free-surface perturbations (not considered in this work). (b) Flow magnitude $||\boldsymbol{u}||$ for $\beta_0=0.05, \, e/e_{\max}=0.4$. Flow computed at $\theta(t) = \pi/2$ on the orbit (see figure \ref{Fig_ODEI_draw}).}
	\label{Fig_ODEI_LDEI}
\end{figure}

We consider the general optical LDEI, taking into account the exact orbital motion (\ref{Eq_ODEI_Worb}) and associated dynamical tides.
We survey in figure \ref{Fig_ODEI_LDEI} (a) the optical LDEI on eccentric orbits, varying the equilibrium tide $\beta_0$ and the eccentricity $e$ from the circular case to $e/e_{\max} = 0.8$.
Two distinctive behaviours occur. The transition is associated with a physical change in the main tidal effect.
To compare the effects of dynamical and equilibrium tides, we introduce the normalised ratio
\begin{equation}
	\Delta \beta_{ab} / \beta_0 = [\beta_{\max}(e)-\beta_{\min}(e)]/\beta_0 = (1-e)^{-3} - (1+e)^{-3},
	\label{eq:dbetaAB}
\end{equation}
with $\beta_{\max}(e)$ (respectively $\beta_{\min}(e)$) the maximum (respectively minimum) equatorial ellipticity for a given eccentricity as defined in expressions (\ref{eq:betaminmax_e}).
Physically when $\Delta \beta_{ab} / \beta_0 \ll 1$ the equilibrium tide $\beta_0$ is of prime importance compared with dynamical tides. 
For weakly eccentric orbits ($e \to 0$), the growth rates of the optical LDEI coincide with the ones of the physical LDEI predicted by formula (\ref{Eq_LDEI_WKB}), as shown by the unit ratio $\sigma / \sigma_\text{wkb} = 1$.
However we observe that new tongues of instability, with normalised growth rates $\sigma / \sigma_\text{wkb} \geq 1$, appear when $\Delta \beta_{ab} / \beta_0 \leq 1$ ($e \leq 0.15$). These new tongues are not predicted by the local formula (\ref{Eq_LDEI_WKB}). This phenomenon is already visible at large $\beta_0$ in figure \ref{Fig_ODEI_LDEI} (a), computed at $n=10$. For smaller values of $\beta_0$, several tongues also appear but higher degrees ($n\geq 15)$ are required to catch them all (not shown).
Thus the LDEI can be more vigorous than predicted before, even in the range of small eccentricities relevant in geo and astrophysics, with growth rates as high as $10 \, \sigma_{\text{wkb}}$.  

On the other hand when $\Delta \beta_{ab} / \beta_0 \geq 1$, the effects of dynamical tides overcome the ones of the equilibrium tide. The eccentricity of the orbit ($e\geq 0.15$) plays now a fundamental role in the tidal effects and the fluid body tends to forget its equilibrium tide. In figure \ref{Fig_ODEI_LDEI} (b) we show the most dangerous unstable flow for an equilibrium tide of amplitude $\beta_0 = 0.05$ and on an orbit of eccentricity $e/e_{\max}=0.4$.
Violent instabilities occur with growth rates $\sigma / \sigma_\text{wkb} \geq 10$ figure \ref{Fig_ODEI_LDEI} (a), whatever the value of $\beta_0$, and can even reach extreme values $10 \leq \sigma / \sigma_\text{wkb} \leq 100$ when $\Delta \beta_{ab} / \beta_0 \geq 3$ (not shown in (a)).  This latter effect occurs for small enough $\beta_0$ and highly eccentric orbits ($e \geq 0.34$), a situation relevant for planets.
 
	\subsection{Survey of the orbitally driven elliptical instability}
	\label{sec:odei}
\begin{figure}
	\centering
	\begin{tabular}{cc}
     	\subfigure[$\beta_{0}=0.05$]{\includegraphics[width=0.49\textwidth]{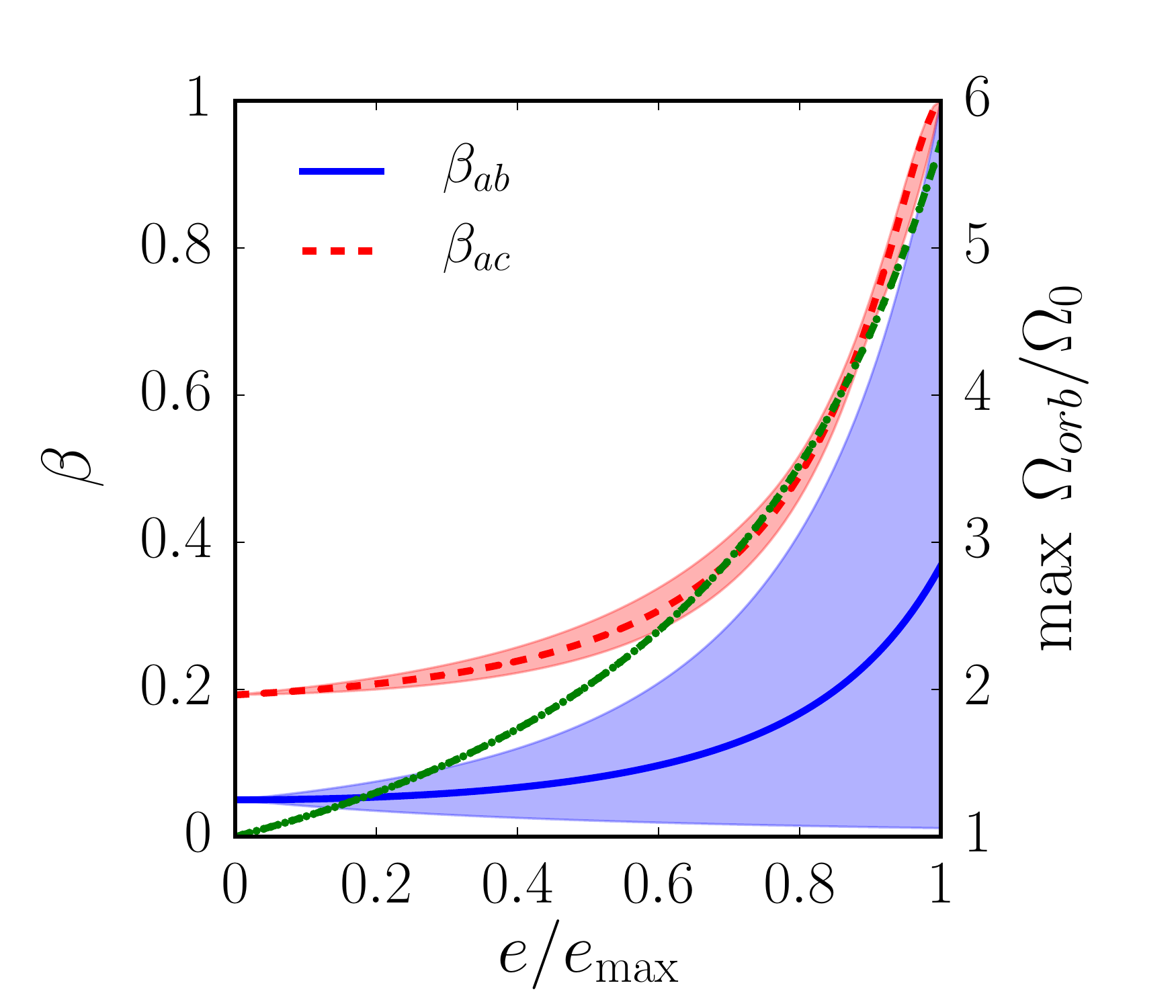}} &
        \subfigure[$\beta_{0}=0.3$]{\includegraphics[width=0.49\textwidth]{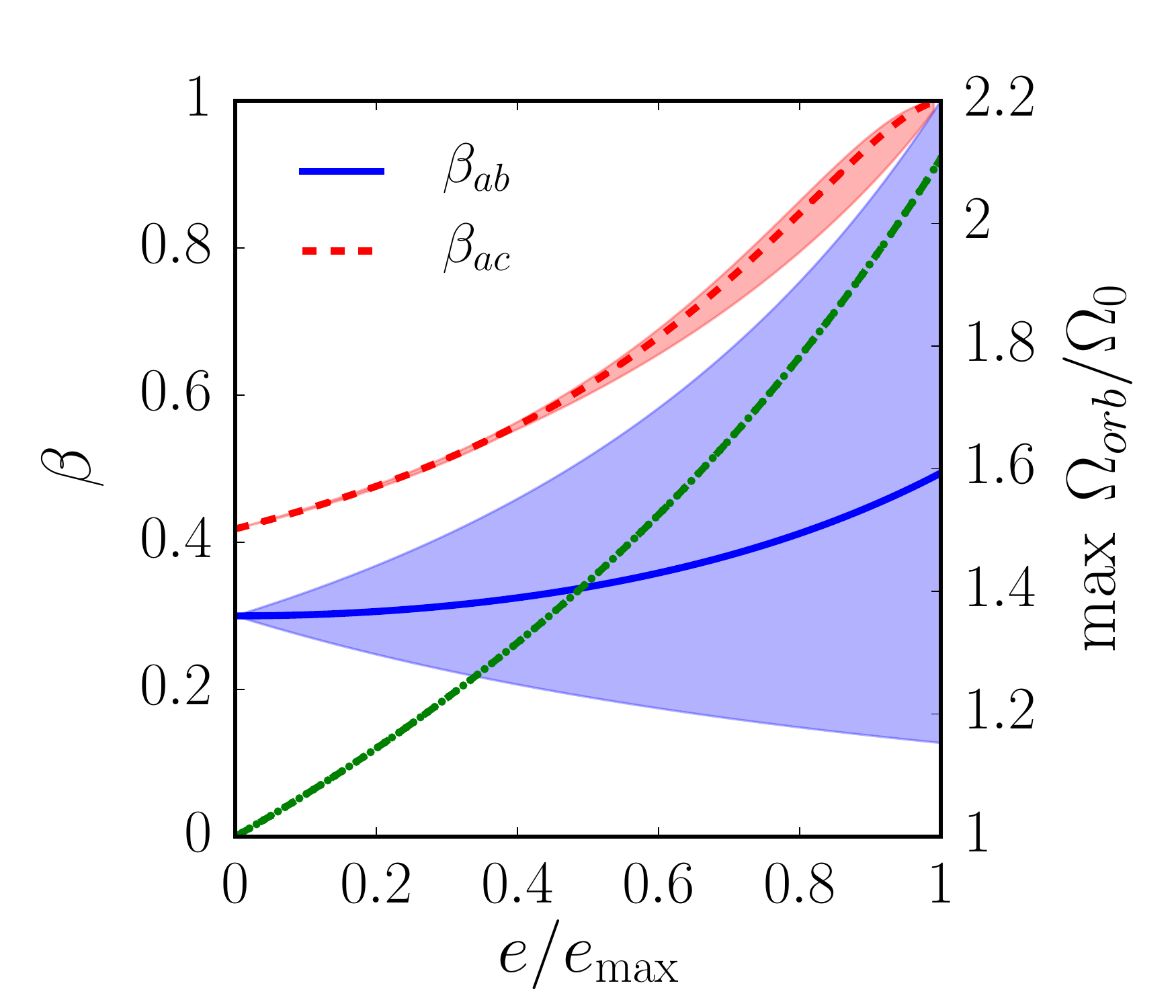}}  \\
		\subfigure[$\beta_{0}=0.05$]{\includegraphics[width=0.49\textwidth]{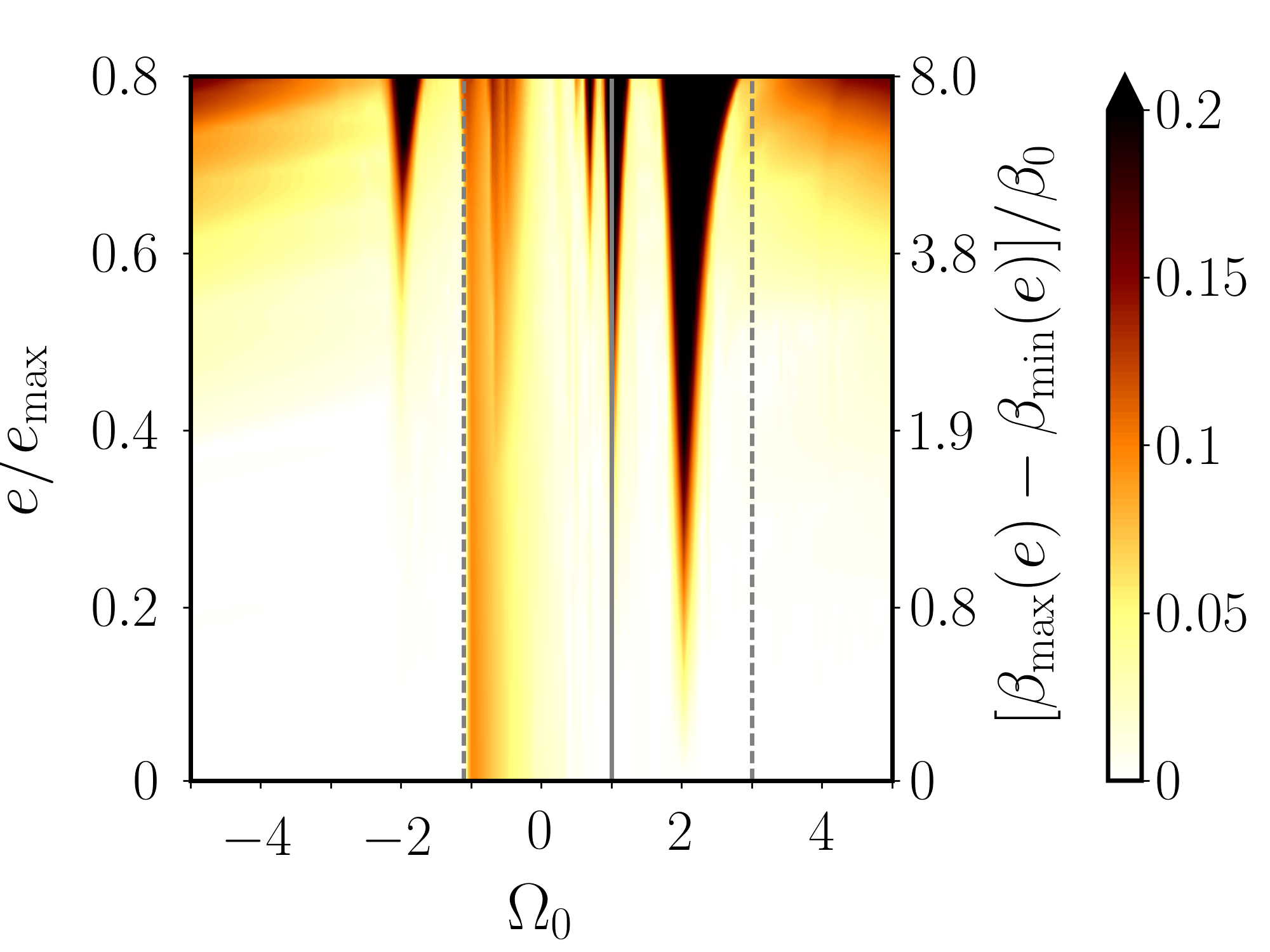}} &
		\subfigure[$\beta_{0}=0.3$]{\includegraphics[width=0.49\textwidth]{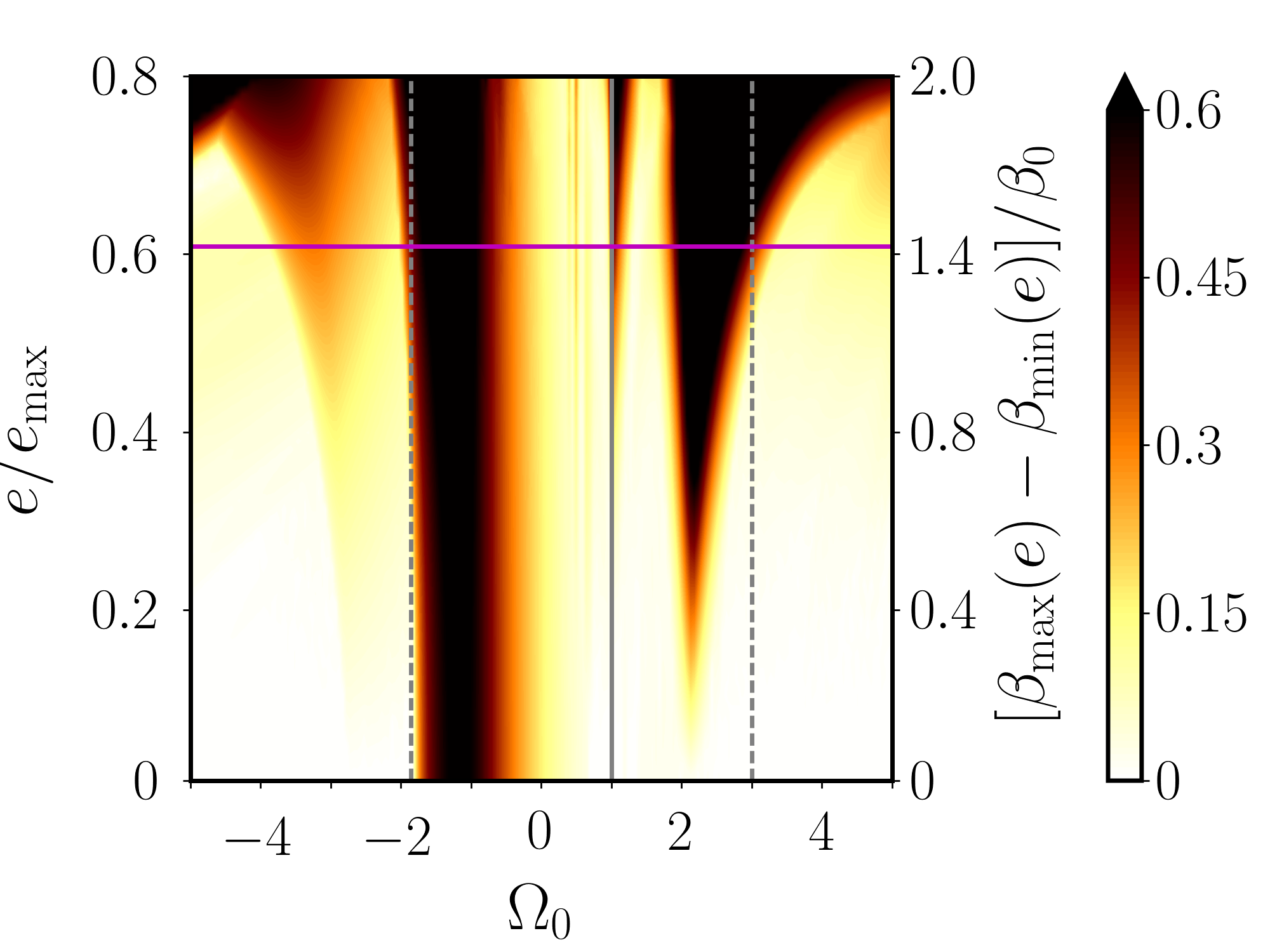}}
	\end{tabular}
	\caption{Survey of the orbitally driven elliptical instability (ODEI). (a) \& (b) Range of values of $\beta_{ab}$ (solid blue line) and $\beta_{ac}$ (dashed red line) for the various eccentric orbits considered. Blue thick (resp. red dashed) line shows the mean value of $\beta_{ab}$ (resp. $\beta_{ac}$) along the orbits. The second vertical axis shows the maximum of ratio $\Omega_{orb}/\Omega_0$ (green dotted line).
	(c) \& (d) Growth rates $\sigma$ of the ODEI in the plane $(e/e_{\max}, \Omega_0)$ for degree $n=10$. The color bar is saturated at $\sigma \geq 0.2$ in (c) and $\sigma = 0.6$ in (d). White areas correspond to marginally stable regions.
	The containers considered are oblate ellipsoids with $R= R_{m} + 0.05$. Vertical black line corresponds to the synchronised case ($\Omega_{0}=1$) driving the LDEI (see \S\ref{sec:ldei}). The horizontal line $e=0$ corresponds to the TDEI (see \S\ref{sec:tdei}). Vertical dashed black lines are the bounds of the forbidden zone FZ$_{\beta_0}$ of the classical TDEI, valid for $e=0$ and $\beta_{ab} = \beta_0$. Horizontal magenta line is $e_R/e_{\max}$, with $e_R$ defined by formula (\ref{eq:eroche}). In (a) the line is outside of the plot ($e_R/e_{\max} = 0.89$).}
	\label{Fig_ODEI_eps_W0}
\end{figure}

\begin{figure}
	\centering
	\begin{tabular}{cc}
    	\subfigure[$\Omega_0 = -2, \, \omega=0.326$]{\includegraphics[width=0.43\textwidth]{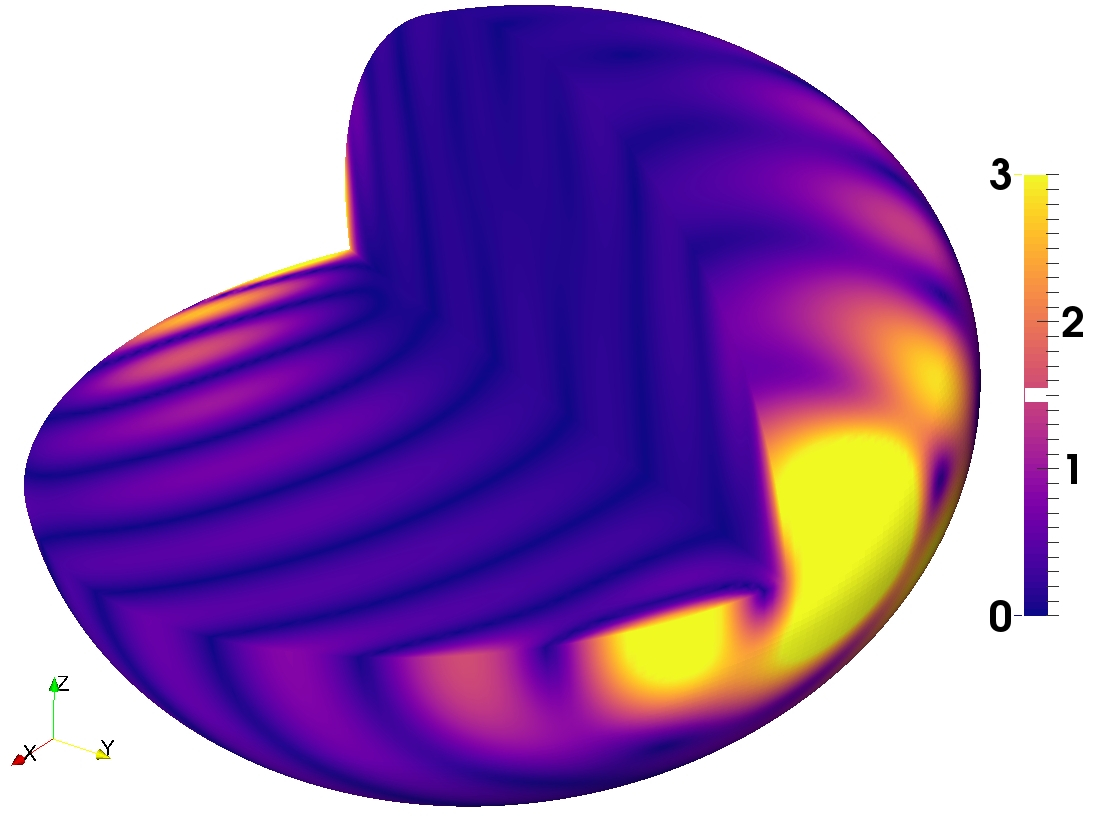}} &
		\subfigure[$\Omega_0 = -1, \, \omega=0$]{\includegraphics[width=0.43\textwidth]{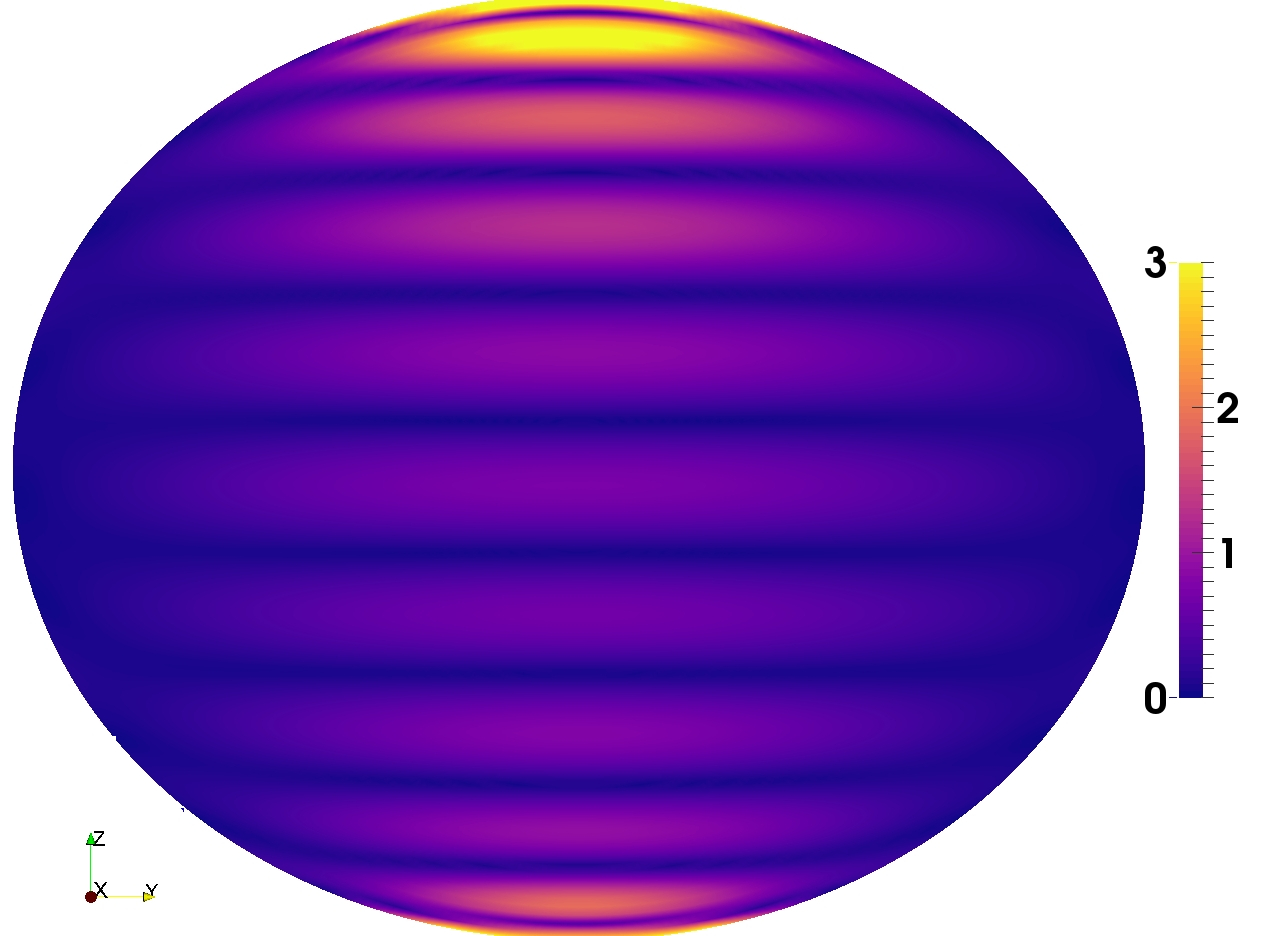}} \\
         \subfigure[$\Omega_0 = -0.49, \, \omega=0.227$]{\includegraphics[width=0.43\textwidth]{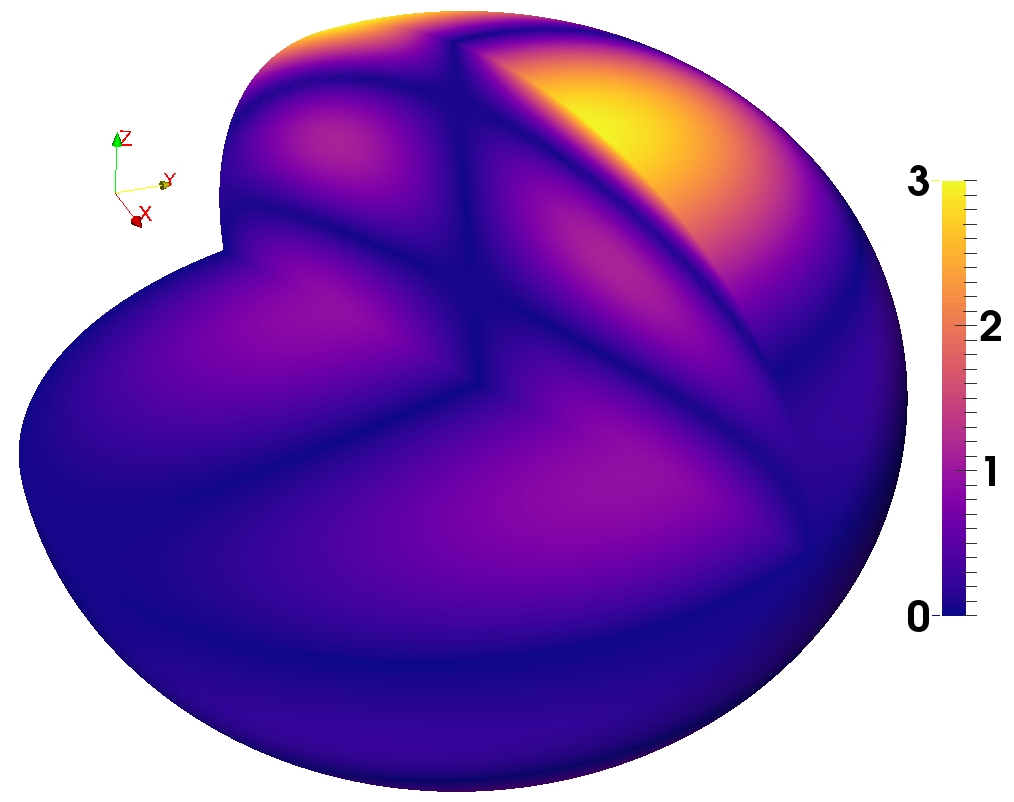}} &
		\subfigure[$\Omega_0 = -0.1, \, \omega=0.03$]{\includegraphics[width=0.43\textwidth]{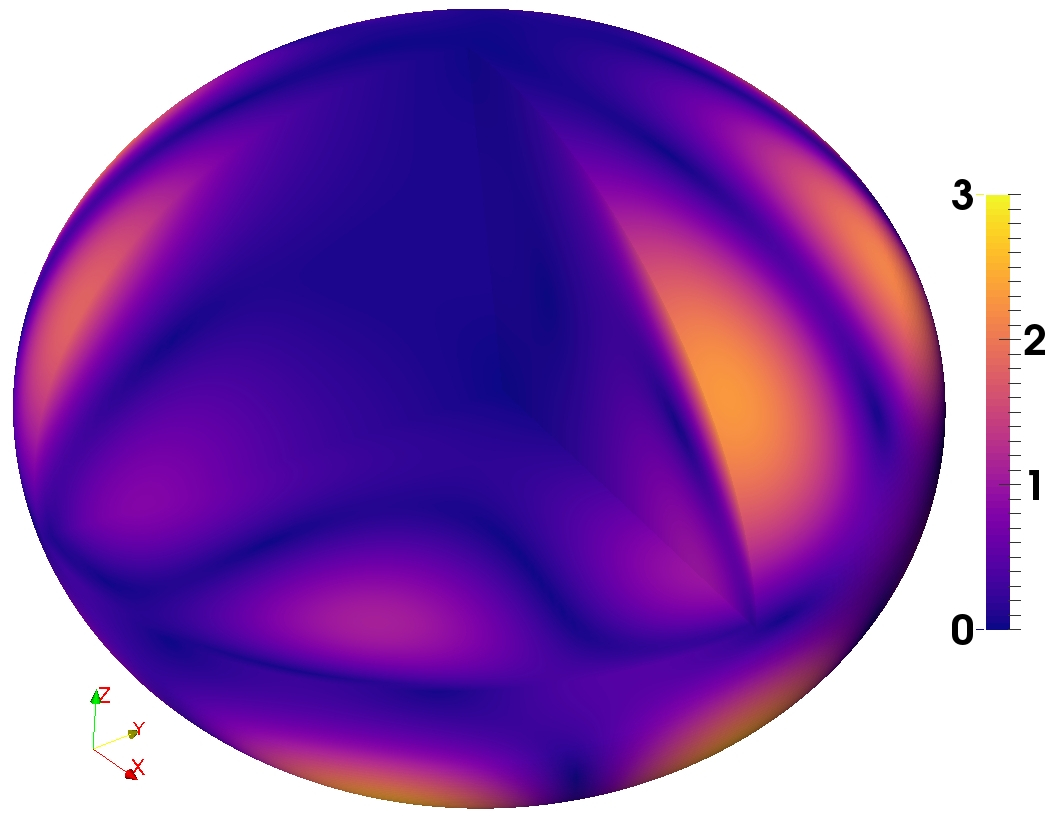}} \\
         \subfigure[$\Omega_0 = 0.5, \, \omega=0,142$]{\includegraphics[width=0.43\textwidth]{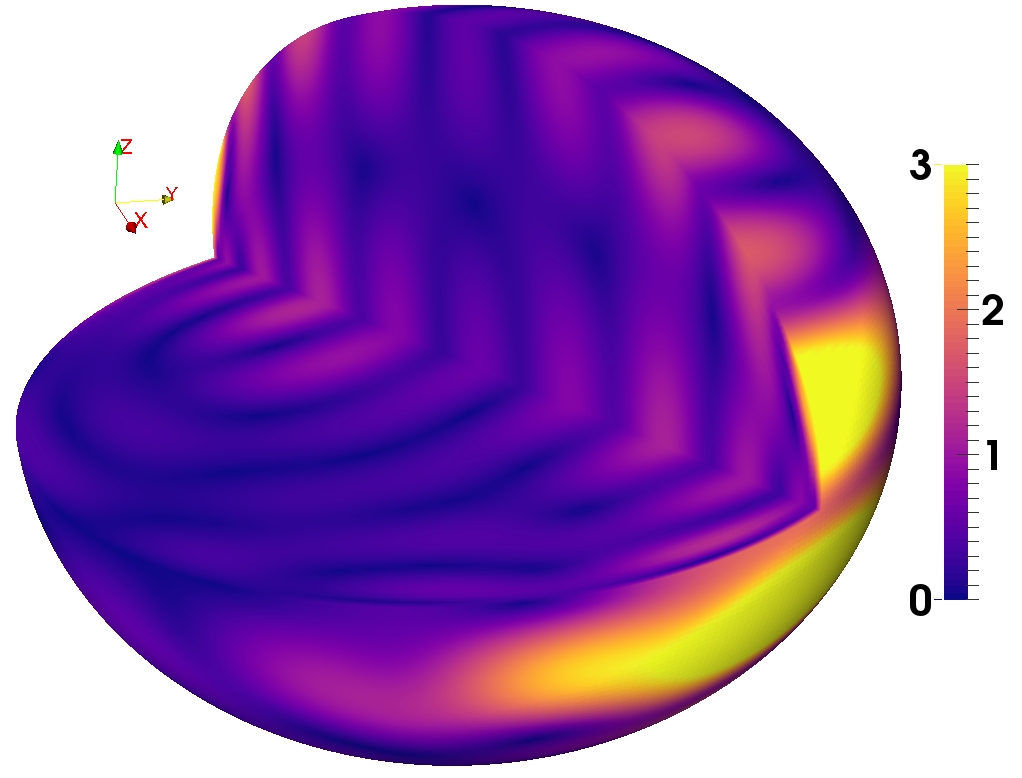}} &
		\subfigure[$\Omega_0 = 2.05, \, \omega=0.347$]{\includegraphics[width=0.43\textwidth]{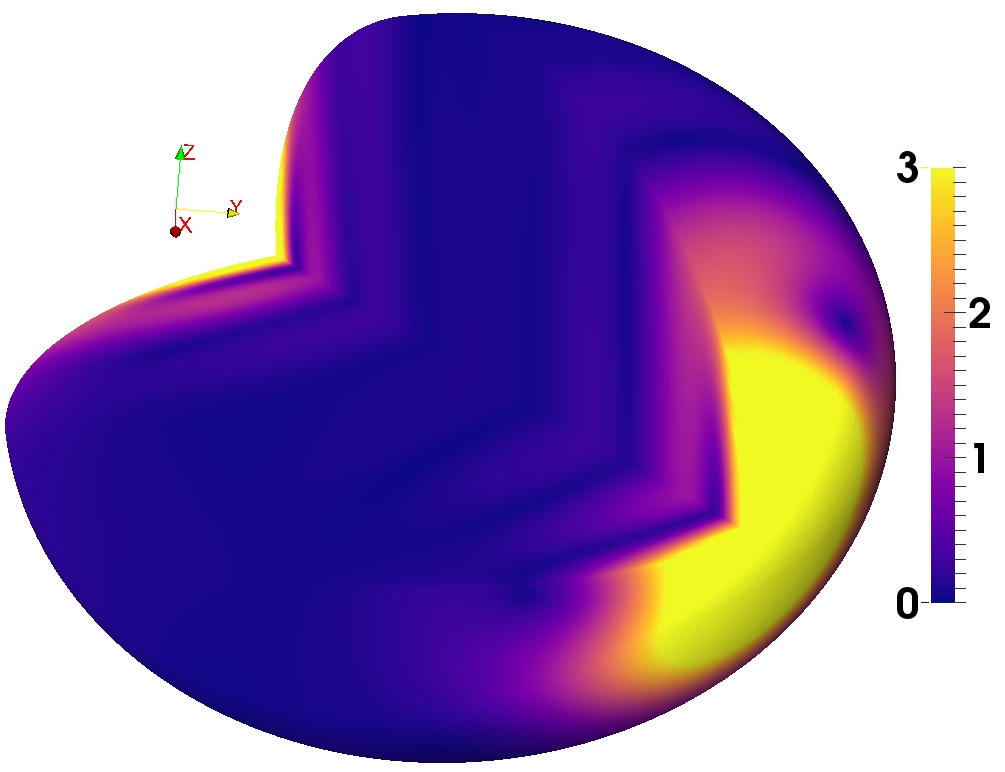}} \\
 \end{tabular}
	\caption{Velocity magnitude $||\boldsymbol{u}||$ of several unstable flows of the figure \ref{Fig_ODEI_cutWb} (c). $n=10$, $e/e_{max}=0.4$ and $\beta_0 = 0.05$. $||\boldsymbol{u}||$ is shown in meridional/equatorial planes and at the outer ellipsoidal surface.
	The color map is saturated for $|| \boldsymbol{u}|| \geq 3$. Flows are computed at $\theta(t) = \pi/2$ on the orbit (see figure \ref{Fig_ODEI_draw}).}
	\label{Fig_ODEI_Flow}
\end{figure}

The general case of a fluid ellipsoid orbiting on a Kepler orbit of eccentricity $0 \leq e < 1$ is now considered. 
Since the Kepler equation (\ref{Eq_Kepler}) is solved at any time step (as in \S\ref{sec:ldei}), the computational cost is more expensive than for computations done in \S\ref{sec:tdei}. As in \S\ref{sec:ldei}, we fix the polynomial degree to the value $n=10$ to survey the whole parameter space. 

We survey the stability of the orbitally driven basic flow in figure \ref{Fig_ODEI_eps_W0} as each
parameter in the set $(\Omega_0, e)$ varies. We arbitrary fix the equilibrium to a small value ($\beta_0 = 0.05$) and to a larger one ($\beta_0 = 0.3$).  In figure \ref{Fig_ODEI_eps_W0} (a) - (b) we show the average, maximum and minimum values of the equatorial and polar ellipticities ($\beta_{ab}(t), \beta_{ac}(t)$) along the orbits. We consider here only oblate containers ($b > c $), which typically describe the shapes of celestial bodies. The maximum value of the orbital angular velocity (normalised by $\Omega_0$) is also shown.

Figure \ref{Fig_ODEI_eps_W0} (c,d) shows the maximum growth rates of the most unstable modes. Some of the associated unstable flows are shown in figure \ref{Fig_ODEI_Flow}. First, the maximum growth rate in each panel tends to increase when $\beta_0$ increases from (c) to (d). Then several aspects of figure \ref{Fig_ODEI_eps_W0} (c,d) are worthy of comment. 

We recover the TDEI considered in \S\ref{sec:tdei}, corresponding to the horizontal line $e=0$.
We also show the bounds of the forbidden zone FZ$_{\beta_0}$, i.e. $\Omega_0 = (1+\beta_0)/(\beta_0-1)$ and $\Omega_0 = 3$ (dashed grey lines). 
The instability with the largest wavelength ($n=1$ basis) is the spin-over mode (not shown). It is similar to the "middle-moment-of-inertia" instability of rigid bodies. The spin-over develops on any circular orbits ($e=0$) for retrograde rotations $-1 \leq \Omega_0 \leq 0$, as expected from previous global analyses \citep[e.g.][]{roberts2011flows,Barker11062016}, but also for any eccentric orbit ($0\leq e$) with an almost constant growth rate (see figure \ref{Fig_DNS} in appendix \ref{app:dns}).
Then an increasing region of the parameter space becomes unstable as the polynomial degree $n$ is increased from $n=1$ to $n=10$, within the expected allowable range $(1+\beta_0)/(\beta_0-1) < \Omega_0 < 3$ where the classical TDEI develops on circular orbits. We observe that the eccentricity has little effect on the growth rates of the TDEI for retrograde orbits (exceptions occur for large eccentricities) within the allowable range $(1+\beta_0)/(\beta_0-1) \leq \Omega_0 \leq 0$. 
For instance we recover the SoP unstable modes at $\Omega_0 = -1$ in figure \ref{Fig_ODEI_Flow} (b). For other values of $\Omega_0$, the exact flow structure of the unstable modes depends on the triaxial shape. So it prevents from directly comparing with flows in figure \ref{Fig_TDEI_Flow} (also obtained at larger $n$), although showing broad common patterns.
This first unexpected result justifies a posteriori the validity of TDEI mechanism in tidally disturbed planets or stars, since it can be extended to eccentric retrograde orbits within the classical allowable range of the TDEI ($(1+\beta_0)/(\beta_0-1) < \Omega_0 \leq 0$).

However, our survey also shows that dynamical tides strongly enhance two instability tongues for prograde eccentric orbits. The growth rates are indeed much larger than the ones predicted on circular orbits. The first one is associated with the LDEI at $\Omega_0=1$ as previously discussed in \S\ref{sec:ldei}. The second tongue seems to be centred on $\Omega_0 = 2 + \beta_0$.
The most unstable flow at $n=10$ is shown in figure \ref{Fig_ODEI_Flow} (f). It exhibits intense motion localised in patches around the equator. It is very different from the TDEI flow at $\Omega_0=2$ in figure \ref{Fig_TDEI_Flow} (d). We expect this localisation to increase as $n$ is increased further.
The enhancing of the growth rate first appears at degree $n=2$, for large enough $e$ (not shown). So it is not associated with the spin-over mode ($n=1$). Then the instability band moves towards smaller eccentricities as $n$ increases (not shown), even when the effects of dynamical tides are \emph{a priori} small ($\Delta \beta_{ab} /\beta_0 \leq 1$). 

\begin{figure}
	\centering
	\begin{tabular}{cc}
		\subfigure[$\beta_{0}=0.05$]{\includegraphics[width=0.49\textwidth]{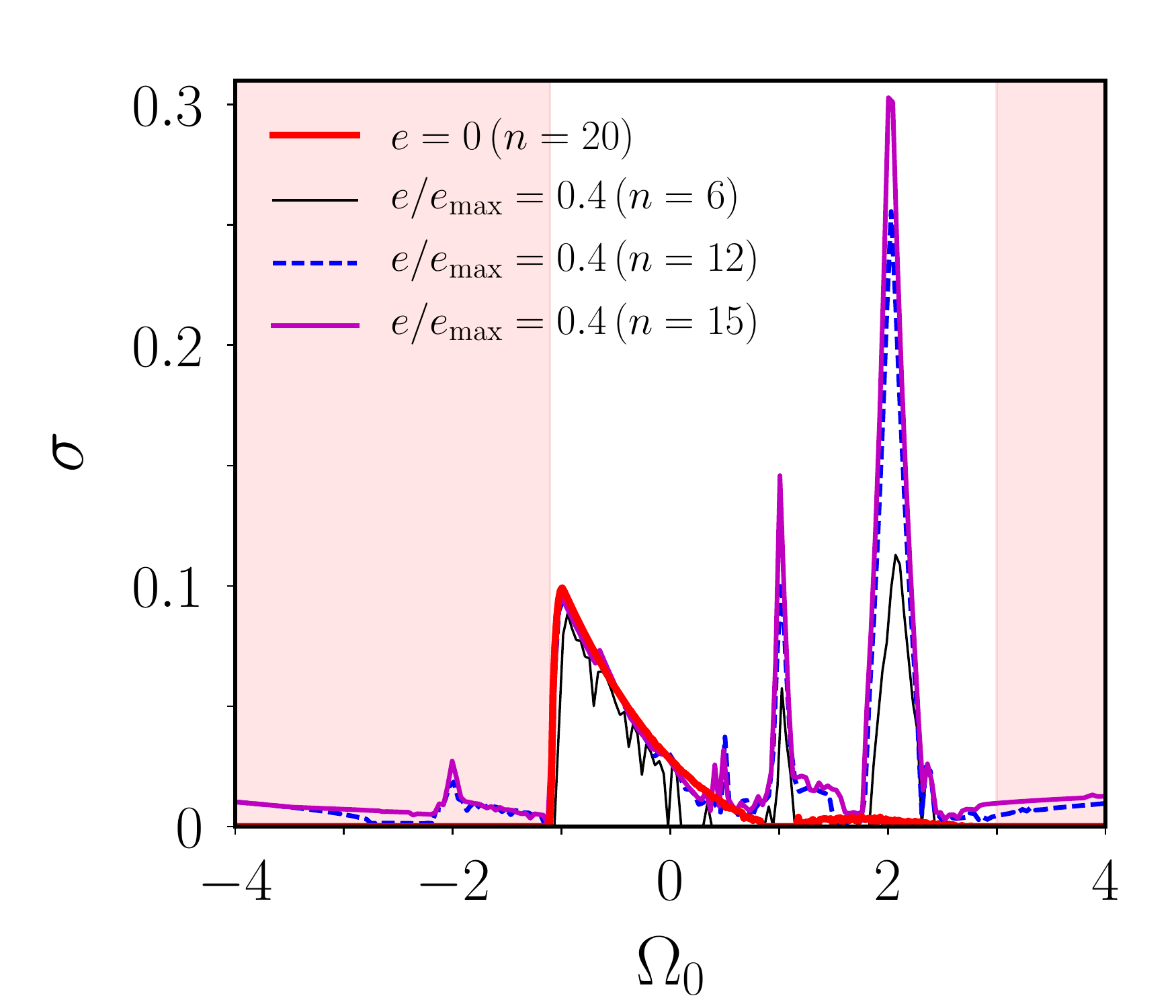}} &
        \subfigure[$\beta_{0}=0.3$]{\includegraphics[width=0.49\textwidth]{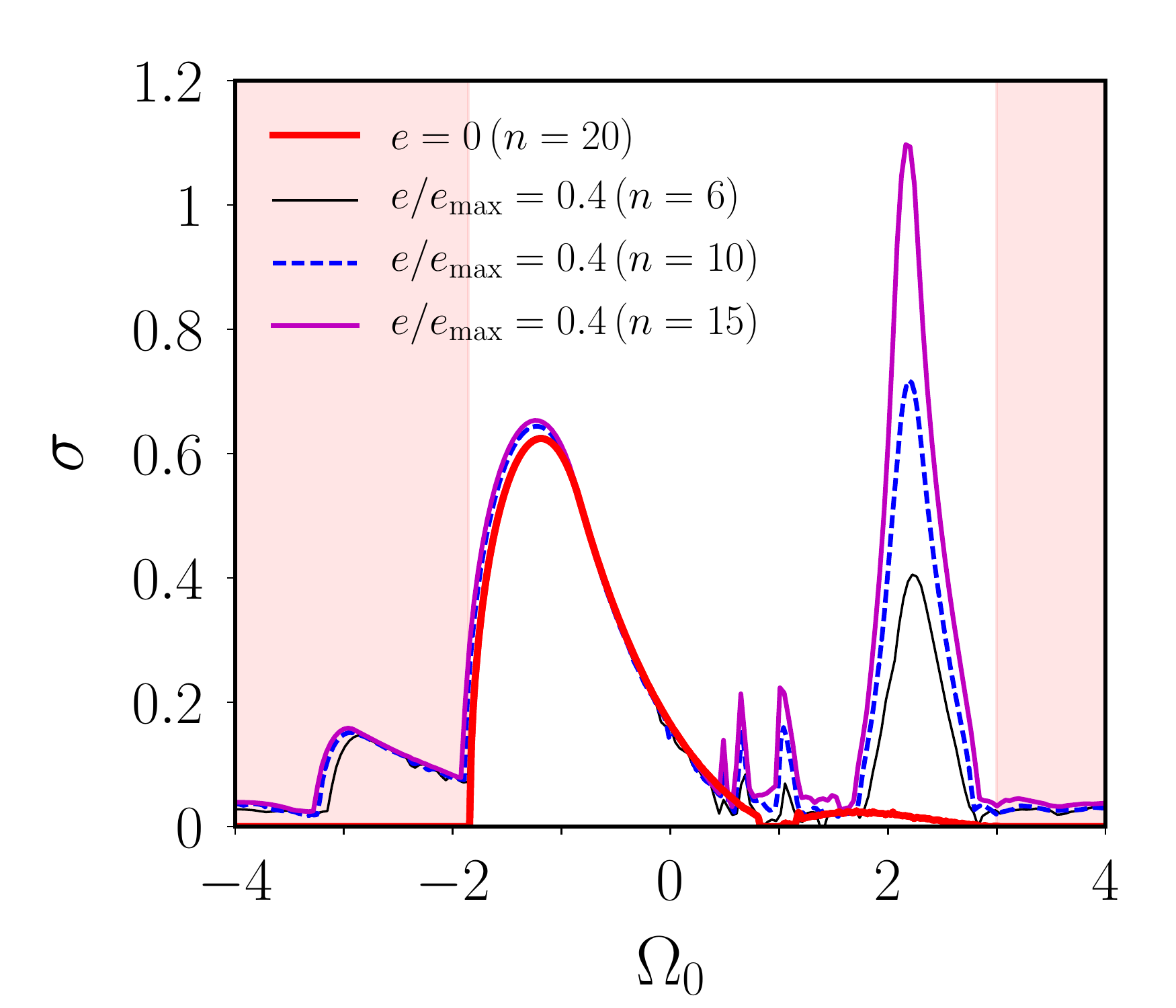}}
	\end{tabular}
	\caption{Growth rates of the ODEI for eccentric Kepler orbits at $e/e_\text{max}=0.4$ compared to the growth rates of the TDEI (red thick solid line). Shaded areas represent the forbidden zone FZ$_{\beta_0}$ for the TDEI ($\Omega_0 \leq (1+\beta_0)/(\beta_0-1)$ and $\Omega_0 \geq 3$).}
	\label{Fig_ODEI_cutWb}
\end{figure}

Another striking result is observed in figure \ref{Fig_ODEI_eps_W0} (c) - (d). We uncovers new violent instabilities within the forbidden zone FZ$_{\beta_0}$ for both retrograde ($\Omega_0 \leq (1+\beta_0)/(\beta_0-1)$) and prograde eccentric orbits ($\Omega_0 \geq 3$). 
The unstable tongue for retrograde orbits appears first at large enough eccentricities and it is not initially associated with the spin-over mode (see figure \ref{Fig_DNS} in appendix \ref{app:dns}). Then increasing the degree $n$ shows that this new tongue is replaced by more unstable tongues which merge with the tongue of the TDEI near $\Omega_0 = (1+\beta_0)/(\beta_0-1)$. The latter tongue also spreads towards more retrograde orbits for large enough eccentricities (when $n$ is large enough). An example of an unstable flow in this tongue at $\Omega_0=-2$ and $\beta_0=0.05$ is shown in figure \ref{Fig_ODEI_Flow} (a). It displays vertical stripes that seem similar to the SoP observed at $\Omega_0=-1$ (but here stacked along an equatorial axis). For $\Omega_0 = -3$ and $\beta_0 = 0.3$, the unstable flow is instead a SoP flow (see the discussion of figure \ref{fig:swan_odei} in \S\ref{sec:physics} below).
For prograde eccentric orbits ($\Omega_0 \geq 3$), these new instabilities are initially associated with an unstable tongue of degree $n=3$. Then, there is also a merging between this tongue and the one appearing near $\Omega_0 = 2 + \beta_0$ at $n=2$, which spreads out towards more and more prograde orbits ($\Omega_0 \geq 3$) for large eccentricities when $n$ increases.
Note that these new unstable tongues exist for orbits of eccentricities $e \leq e_R$ in figure \ref{Fig_ODEI_eps_W0}. So these new instabilities may physically exist in astrophysical fluid bodies. 

We compare now more quantitatively the strength of these instabilities in figure \ref{Fig_ODEI_cutWb} by pushing the degree to $n=15$. We show the growth rates of the TDEI on circular orbits (red solid curve) and the ones of orbitally driven instabilities on eccentric Kepler orbits for a finite value of the eccentricity ($e/e_{\max} = 0.4$).
When $(\beta_0+1)/(\beta_0-1) < \Omega_0 \leq 1$, the growth rate of the ODEI has almost the same value as for the classical TDEI at $e=0$. We note that increasing the degree $n$ yields small variations in $\sigma$. The maximum $\sigma$ is well predicted by the local WKB analysis of the TDEI (see formula \ref{Eq_TDEI_DetuneWKB} in appendix \ref{app:tdei_ledizes}), showing a scaling in $\beta_0$.
Around $\Omega_0 = 1$ we see the peak corresponding to the LDEI, previously computed for various $\beta_0$ in figure \ref{Fig_ODEI_LDEI} (a) at $n=10$. 
The largest growth rates are obtained for the instability located at $\Omega_0 = 2 + \beta_0$ (in the limit $e\ll1$). Its growth rates are approximatively ten times larger than the ones predicted by the classical TDEI circular orbits at the same values of $\Omega_0$. This instability becomes stronger as $n$ is increased from $n=6$ to $n=15$, suggesting a rather small-scale instability. We expect its growth rate $\sigma$ to reach an upper bound for large enough $n$. However we recall that the global method gives only sufficient conditions for instability. So even though the growth rate has not reached yet its asymptotic value, it does not physically rule out the enhanced strength of this instability.
The new instabilities driven by the non-circular Kepler orbits within the classical forbidden zone of the TDEI are also clearly visible. The unstable tongues extend deeply inside the forbidden zone, even at low $\beta_0$. The growth rate is almost insensitive to the chosen $n$ from $n=10$ to $n=15$. It suggests that the asymptotic growth rates have already been (at least at $e/e_{\max} = 0.4$). 

To sum up, we have found new sufficient conditions for inviscid instability (for the values of $n$ considered here). They show that the orbitally driven basic flow (figure \ref{Fig_ODEI_eps_W0}) can be unstable in the allowable region of the classical TDEI, but also inside the classical forbidden zone  for both retrograde and prograde eccentric orbits.
We also show in appendix \ref{app:dns} that these instabilities are recovered in three-dimensional viscous numerical simulations. Therefore we expect this phenomenon to hold in astrophysical bodies.

\section{Physical mechanisms}
\label{sec:physics}
	\subsection{Local approach}
\begin{figure}
	\centering
	\begin{tabular}{cc}
		\subfigure[]{\includegraphics[width=0.49\textwidth]{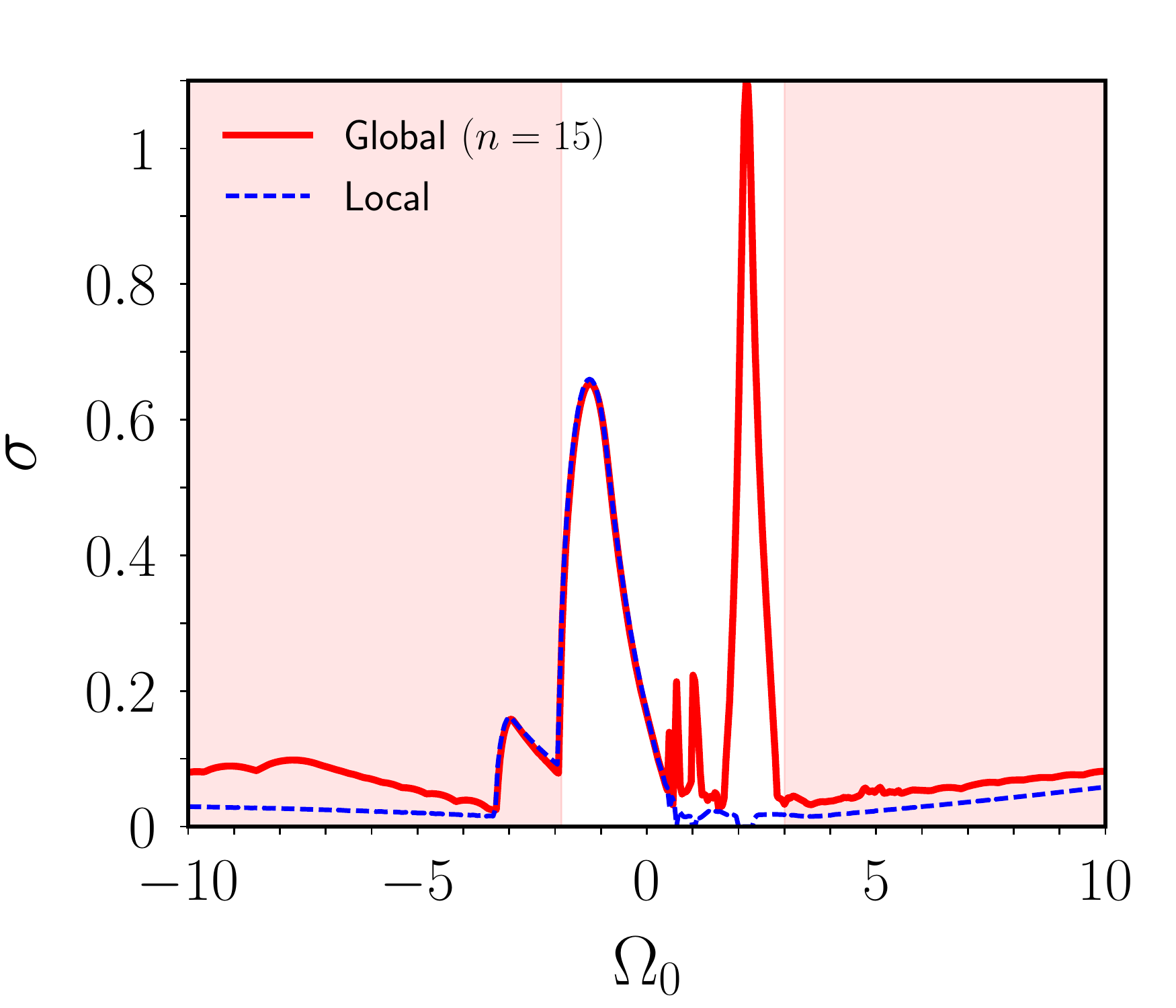}} &
		\subfigure[]{\includegraphics[width=0.49\textwidth]{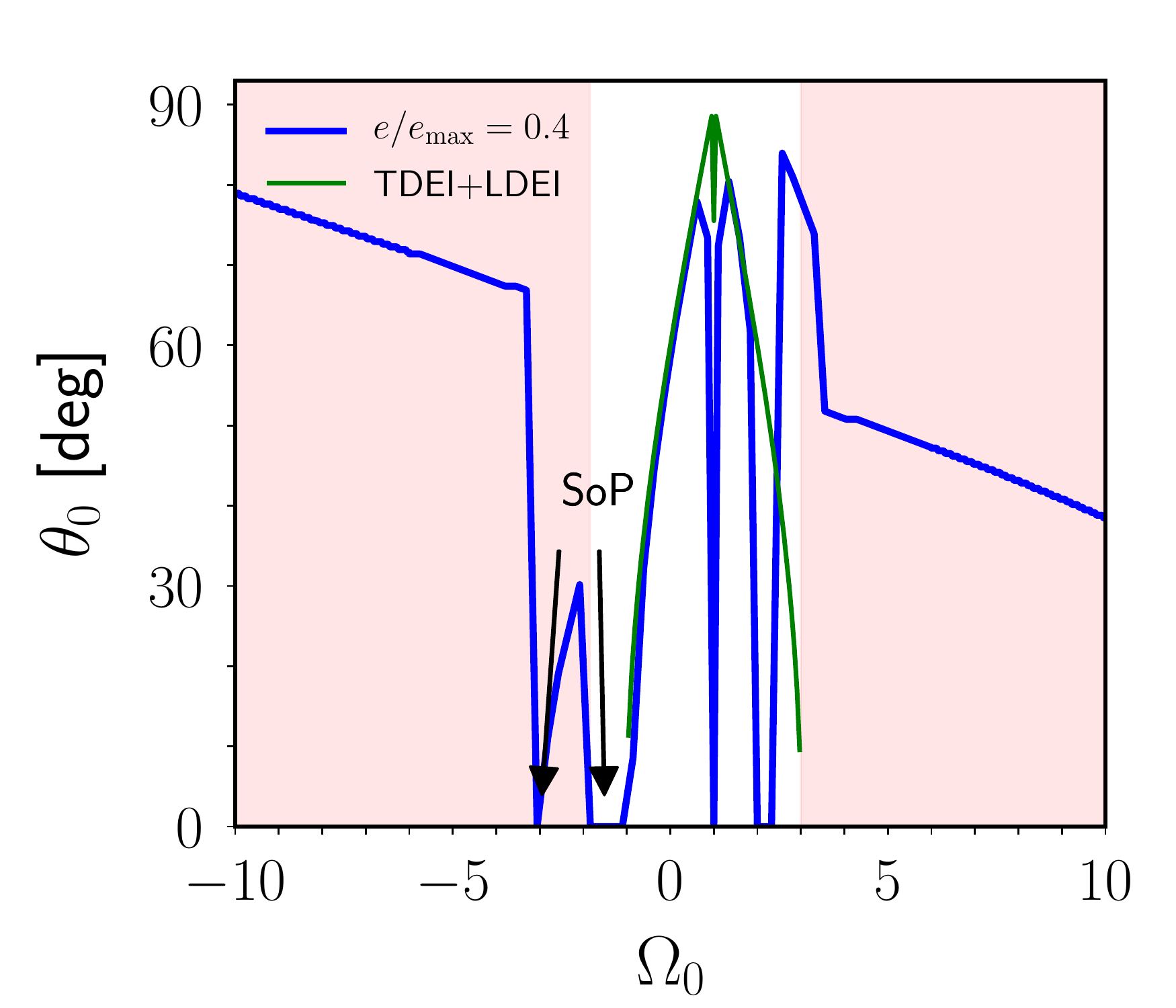}}
	\end{tabular}
    \caption{Comparison between local and global stability analyses of orbitally driven flows (\ref{Eq_BF_Orbit}) for $\beta_0 = 0.3$ and $e/e_{\max} = 0.4$. We compute the local maximum growth rate as the fastest growing solution from a range of initial wave vectors $\boldsymbol{k}_{0} = (\sin\theta \cos \phi, \sin \theta \sin \phi, \cos \theta)$, with 0.5 degree spacing in $\theta \in [0, 90^{\circ}]$. The initial azimuthal angle of the local wave vector $\phi=\pi/4$. The latter maximises the growth rate of the classical TDEI \citep{le2000three}. We have checked that the largest growth rates are insensitive to the value of $\phi$. (a) Local (dashed blue line) and global $n=15$  (solid red line) growth rates $\sigma$ in function of $\Omega_{0}$. (b) Numerical angle $\theta_{0}$ of the initial local wave vector leading to the maximum growth rate as varying $\Omega_0$. Thin green line shows the classical destabilizing angle leading to the TDEI and the LDEI.}
    \label{fig:swan_odei}
\end{figure}

In this section, we discuss the physical mechanism responsible for the orbitally driven instabilities. 
First we perform a local (WKB) stability analysis by solving equations (\ref{Eq_WKB}) in unbounded fluids. 
Indeed the nature of an unstable tongue is be related to the colatitude $\theta_0$ of the initial wave vector $\boldsymbol{k}_{0}$ (see \S\ref{subsec:wkb}) leading to the largest growth rate. 

We aim at solving analytically the stability equations in the limit of weakly eccentric orbits ($e\ll1$). We expand at first order in $e$ the orbital forcing (\ref{Eq_ODEI_Worb}) and (\ref{Eq_ODEI_beta}) to get
\begin{equation}
	\Omega_{orb}(t) = \Omega_{0} \left [ 1 + 2 e \cos (\Omega_{0} t) \right ], \ \, \ \beta_{ab}(t) = \beta_0 \left [ 1 + 3 e \cos(\Omega_{0}t) \right ]. 
	\label{Eq_LongLibOptic_Forcing}
\end{equation}
At this order, the time dependence of $\beta_{ab}(t)$ is monochromatic, in agreement with numerical results of figure \ref{Fig_ODEI_Keplersol} at small $e$.
When $e=0$, the basic state is a pure solid-body rotation  and it admits plane inertial wave perturbations \citep[e.g.][]{greenspan1968}. 
They have periodic wave vectors which are orthogonal to the local velocity vector. 
Plane inertial waves exist when $-1<\Omega_{0}<3$, which is the allowable range of the classical TDEI on circular orbits when $e\to0$.

The basic mechanism of the elliptical instability is a parametric resonance between a pair of inertial waves and the basic flow, provided that certain resonance conditions are met \citep[e.g.][]{le2000three,kerswell2002elliptical}.
This mechanism also applies here.
Indeed when $e \to 0$, two inertial waves can resonate with the orbitally driven basic flow (\ref{Eq_BF_Orbit}) to drive an instability. The latter is governed by an Hill-Schr\"odinger equation, which can be readily obtained following \citet{naing2009local}.
It is not written here for the sake of clarity. 
However the forcing term in the Hill-Schr\"odinger equation is not strictly periodic, as for tidally driven and libration-driven basic flows. Instead it is quasi-periodic with multiple forcing frequencies, such that many resonances are possible.
For a given forcing frequency $f$, the condition of perfect resonance yields
\begin{equation}
	2 \left ( 1 + \widetilde{\Omega}_{0} \right ) \cos \theta_{0} = \frac{f}{2},
	\label{eq:resonaceHill}
\end{equation}
where the left hand side is actually the inertial waves pulsation (Doppler shifted in the rotating frame) and $\widetilde{\Omega}_{0} = \Omega_0/(1-\Omega_0)$.
The forcing frequencies are $f = 1, 2, \widetilde{\Omega}_{0} \dots$ and possible combinations of them (through cosine and sine products). The nature of the unstable tongues, determined by $\theta_{0}$, depends on the considered forcing frequency $f$. The frequency $f=2$ is associated with the classical TDEI on circular orbits \citep{waleffe1990three,le2000three} and $f=1$ with the LDEI on weakly eccentric orbits \citep{herreman2009effects,cebron2012libration,cebron2014libration}. 
For the TDEI, SoP instabilities have for instance wave vectors aligned with the spin axis $\theta_{0} = 0$ \citep{lebovitz1996new,Barker11062016}.

Equation (\ref{eq:resonaceHill}) shows that resonances associated with a given frequency $f$ only occur for values of $\Omega_{0}$ located outside of the forbidden band $|f/(4(1+\widetilde{\Omega}_{0}))| \geq 1$. We recover the existence of the classical TDEI ($f=2$) inside the allowed region $-1<\Omega_{0} <3$. When $f=\widetilde{\Omega}_{0}$, the forbidden band is $|\Omega_{0}| \geq 4$.
However for finite values of $\beta_0$, the unstable tongues have finite widths of order $\mathcal{O}(\beta_0)$ in the limit $0 \leq e\ll 1$. Hence it is possible to excite imperfect resonances by geometric detuning, even though the condition (\ref{eq:resonaceHill}) is not strictly satisfied. A wider range of the parameter space is thus unstable when $\beta_0$ increases, as observed in \S\ref{sec:tdei}, \S\ref{sec:ldei} and \S\ref{sec:odei}. 
For instance the classical TDEI is excited inside the allowable range $(\beta_{0}+1)(\beta_{0}-1)<\Omega_{0} <3$ for finite values of $\beta_0$ (see figure \ref{Fig_TDEI2}).
Therefore considering all the possible frequencies, the resonance condition (\ref{eq:resonaceHill}) shows that the orbitally driven instabilities may \emph{a priori} be triggered well outside the allowed region of the TDEI. 

We further simplify the orbital forcing (\ref{Eq_LongLibOptic_Forcing}) to consider two limiting simplified configurations. Firstly we neglect the dynamical tides (i.e. $\beta(t) = \beta_0)$ to isolate the modulation of the background rotation (i.e. $\Omega_{orb}(t) = \Omega_{0} \left [ 1 + 2 e \cos (\Omega_{0} t) \right ]$). Such a forcing refers to the tidal forcing of a telluric (i.e. rigid) planet moving on an eccentric Kepler orbit. At leading order in $e$, the associated forcing frequency is $f=2$ and the resonance gives the classical growth rate of the TDEI (see appendix \ref{app:tdei_ledizes}). It shows that the time modulation of the background rotation does not destabilise further the tidal basic flow (at leading order in $e$). Higher-order terms in $e$ may be necessary to handle possible new effects. 
Secondly we disable the Coriolis force ($\Omega_{orb}(t) = 0$), but retains the time dependence of the ellipticity along the orbit. In this case, we obtain that the angle $\theta_{0} =\pi/3$ is the most destabilising one for rapidly oscillating tides ($|\Omega_{0}| \gg 1$), leading to $\sigma/\beta_0=9/16$ (as confirmed by solving numerically equations (\ref{Eq_WKB}) in this configuration). The latter formula is identical to the growth rate of the classical TDEI without background rotation (see appendix \ref{app:tdei_ledizes}). It shows that dynamical tides are the key physical mechanism responsible for the instabilities located outside of the allowable range of the classical TDEI, as observed in figures \ref{Fig_ODEI_eps_W0} and \ref{Fig_ODEI_cutWb}. However we note that the associated growth rates are overestimated with respect to the growth rates obtained numerically for the full problem, suggesting that the Coriolis force has a stabilising effect.

\begin{figure}
	\centering
	\begin{tabular}{ccc}
		\subfigure[First order physical librations]{\includegraphics[width=0.49\textwidth]{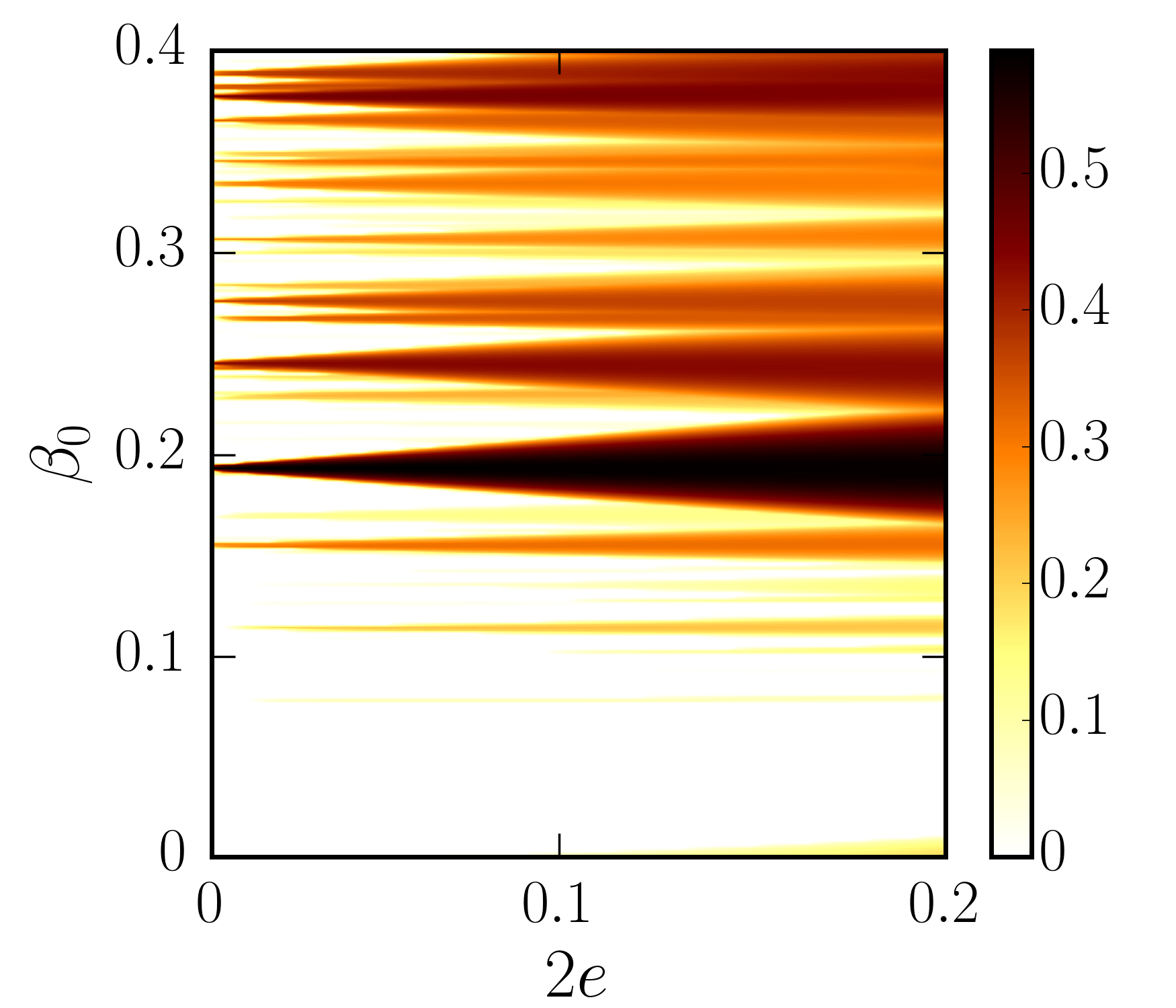}} &
		\subfigure[First order optical librations]{\includegraphics[width=0.49\textwidth]{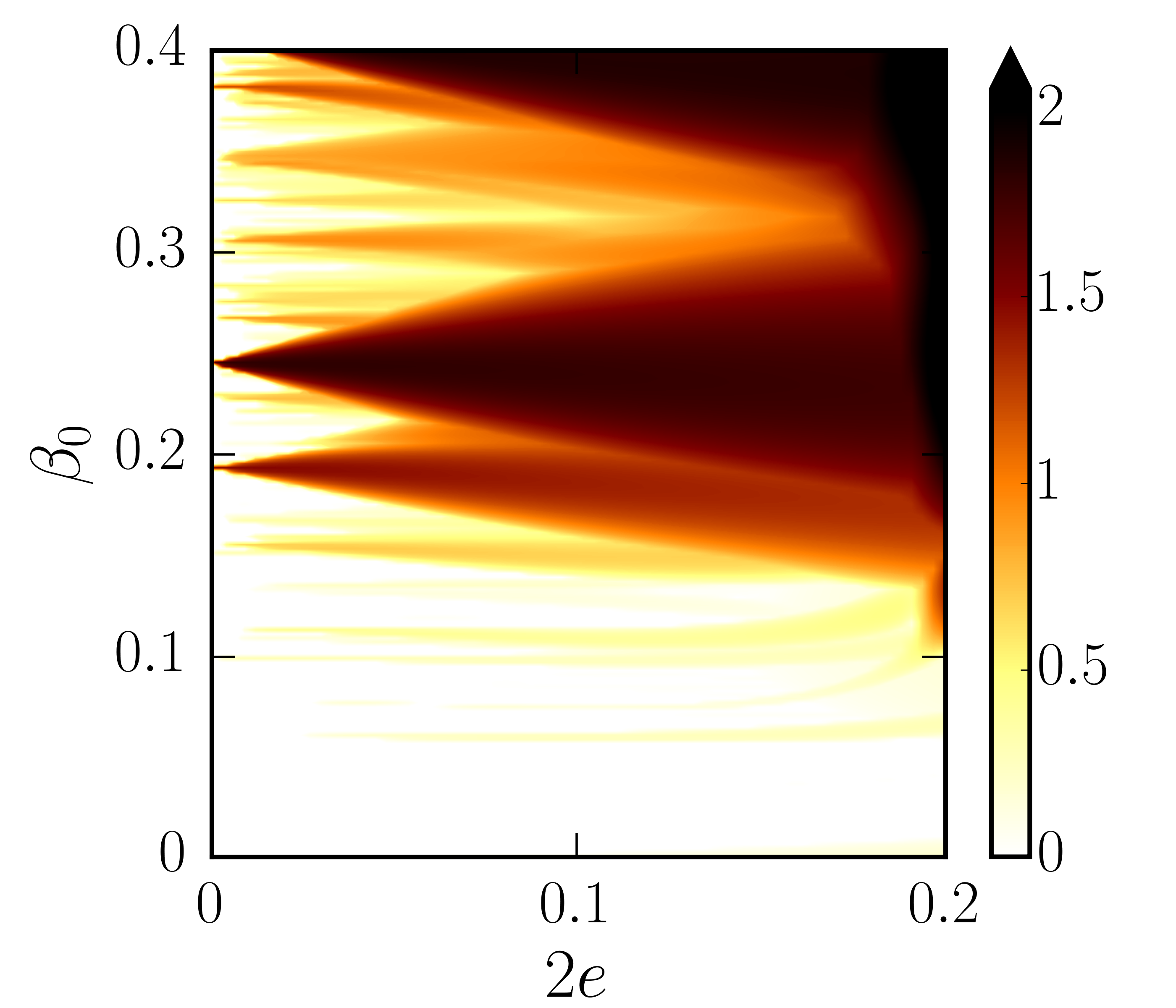}}
	\end{tabular}
	\caption{Survey of the libration-driven elliptical instability for physical librations (\ref{Eq_LongLibPhysic_Forcing}) and first order optical librations in the plane $(\beta_0, 2e)$. Polynomial degree $n=10$. Color map shows the ratio $\sigma/\sigma_\text{wkb}$ with $\sigma_\text{wkb}$ given by formula (\ref{Eq_LDEI_WKB}). White areas correspond to marginally stable regions. Triaxial geometry $a(t) = \sqrt{1 + \beta_{ab}(t)}$, $b(t) = \sqrt{1-\beta_{ab}(t)}$ and $c(t) = [a(t)b(t)]^{-1}$. (a) $\beta_{ab} (t) = \beta_0$ and $\epsilon = 2e$. (b) $\beta_{ab}(t) = \beta_0 (1 + 3e\cos t )$.}
	\label{Fig_LDEI3}
\end{figure}

To get quantitative local predictions in agreement with the global results, we consider the combined effect of rotation and dynamical tides. However obtaining a growth rate formula for any possible resonance, predicted by the equation (\ref{eq:resonaceHill}), is not of practical interest. Indeed resonances may appear or be modified when the full orbital forcing is considered (even at small $e$), possibly leading to more unstable tongues. 
Consequently we solve numerically the local stability equation (\ref{Eq_WKB}) with the SWAN code, taking into account the full orbital forcing (\ref{Eq_ODEI_Worb}) - (\ref{Eq_ODEI_beta}).
We show the comparison between local and global analyses in figure \ref{fig:swan_odei} for $\beta_0=0.3$ and $e/e_{\max}=0.4$. Results obtained at smaller $\beta_0$ are similar and do not change the interpretation.
We first note in figure \ref{fig:swan_odei} (a) a good agreement between local and global growth rates for retrograde orbits ($-4\leq \Omega_0 \leq 0$) and for prograde orbits within the forbidden zone ($\Omega_0 \geq 3$). 
When $(\beta_0+1)/(\beta_0-1) <\Omega_0 \leq 0$, the angle $\cos \theta_0=[2(1+\widetilde{\Omega}_{0})]^{-1}$ in figure \ref{fig:swan_odei} (b) shows that the ODEI reduces to the classical TDEI, which is not modified by the orbital eccentricity. We also recover the SoP instabilities ($\theta_0=0$) when $-2 \leq \Omega_0 \leq -1.5$ for $\beta_0 = 0.3$. Within the forbidden zone of the classical TDEI for retrograde orbits, we find two new tongues of instability not predicted by the TDEI resonance. The modulation of the global rotation is responsible for these instabilities when $-3 \leq \Omega_0 \leq -2$ (not shown). These instabilities were not obtained analytically in the asymptotic limit $e\to0$, because they are due to higher-order terms. 
When $\Omega_0 = -3$ we find a SoP instability, which is also obtained in the global analysis and direct numerical simulations (see appendix \ref{app:dns}). 

Then we find a new tongue of instability within the classical forbidden zone for both rapidly oscillating prograde orbits $\Omega_0 \geq 3$ and retrograde orbits $\Omega_0 \leq -4$. Dynamical tides are responsible for these instabilities. 
The numerical local growth rates are in much closer agreement to the global ones than the analytical growth rate $\sigma/\beta_0=9/16$ obtained without background rotation. We conclude that the Coriolis force has a stabilising effect on these instabilities. Angle $\theta_0$, initially $\theta_0=\pi/3$ without background rotation (i.e. for $\Omega_{orb}(t) = 0$), is modulated by the rotation. It explains the observed linear trend with $\Omega_0$ in figure \ref{fig:swan_odei} (b).

The last striking result in figure \ref{fig:swan_odei} is that the local analysis does not predict the enhancing of growth rates at $\Omega_0 = 1$ (LDEI) and $\Omega_0 = 2+\beta_0$. 
For any finite value of $e$, no local instability is found. A possible explanation is that we find numerical wavenumbers $\boldsymbol{k}(t)$ with secular growths, which challenge the validity of the local analysis. Indeed, local instabilities are generally obtained under the assumption of bounded and asymptotically non-decaying periodic or quasi-periodic wavenumbers \citep[e.g.][]{eckhardt1995local}. We also recall that the local analysis gives only sufficient conditions for instability. Finally it has already been observed that a local analysis can be in disagreement with a global analysis. For instance, global radiative instabilities in compressible Rankine vortex \citep{broadbent1979acoustic} are not predicted by a local WKB analysis \citep{le2001etude}. 

	\subsection{Global approach}
The enhancing of the growth rates when $\Omega_0 = 1, 2+\beta_0$ is left unexplained by the local analysis. Nevertheless, the global analysis provides an explanation for this phenomenon.
In ellipsoids, the elliptical instability is also a parametric resonance between a pair of inertial modes and the basic flow, provided certain resonance conditions are met \citep[e.g.][]{kerswell2002elliptical,lacaze2004elliptical,le2010tidal}. However the present orbital forcing challenges this classical instability mechanism. Indeed, we cannot define properly inertial modes in our time-dependent fluid ellipsoids. 
So identifying the possible resonant couplings is a difficult task. We have to isolate the most unstable modes from the computations and to try to relate them to some inertial modes of a well-chosen ellipsoidal shape (for instance the one associated with the equilibrium tide $\beta_0$). Such an approach relies on an empirical modal decomposition \citep[e.g.][]{schmid2010dynamic,sieber2016spectral}, which is beyond the scope of the present study.

The key phenomenon is the time-dependent ellipticity $\beta(t)$, even in the limit $e \ll 1$. 
To illustrate this point we focus on the first-order optical libration forcing, making use of the forcing (\ref{Eq_LongLibOptic_Forcing}) with $\Omega_0=1$.
In the asymptotic local (WKB) analysis of the LDEI on weakly elliptical orbits ($e\to0$), $\beta_0$ and $e$ are supposed to be of the same order of magnitude in the asymptotic expansion \citep{herreman2009effects,cebron2012libration,cebron2014libration}. So the leading order-effect is the physical libration forcing (\ref{Eq_LongLibPhysic_Forcing}), whereas the time-variable tidal effect (of order $e\beta_0$) is \emph{a priori} of second order.
However the latter effect can become of primary importance if it is large enough, which cannot be probed by the local analysis for the coupled forcing (as explained before).

We compare in figure \ref{Fig_LDEI3} physical librations (a) with optical librations (b), assuming an amplitude $\epsilon = 2e$ for physical librations.
We consider only perturbations of maximum degree $n=10$. Perturbations of higher degrees are not essential for this comparison.
We show the ratio $\sigma/\sigma_\text{wkb}$ to compare the global growth rates $\sigma$ with the local ones $\sigma_\text{wkb}$ predicted by the formula (\ref{Eq_LDEI_WKB}). 
The global growth rates in figure \ref{Fig_LDEI3} (a) do not reach yet the asymptotic local growth rates (\ref{Eq_LDEI_WKB}) for all the values of $\beta_0$, as expected with a global analysis at $n=10$.
In figure \ref{Fig_LDEI3} (b) the unstable tongues generated by physical librations coincide with the unstable tongues generated by optical librations in the limit $e\to0$.
However for finite values of $e$ (even small), we observe that the tongues in figure \ref{Fig_LDEI3} (b) are much wider because of the dynamical tides. Moreover new violent instabilities are triggered, with growth rates much larger than those predicted by formula (\ref{Eq_LDEI_WKB}). It clearly illustrates the enhancing of the LDEI. Note that a similar behaviour is obtained for the unstable tongue appearing at $\Omega_0 = 2 + \beta_0$ (not shown).

\begin{figure}
	\centering
	\includegraphics[width=0.6\textwidth]{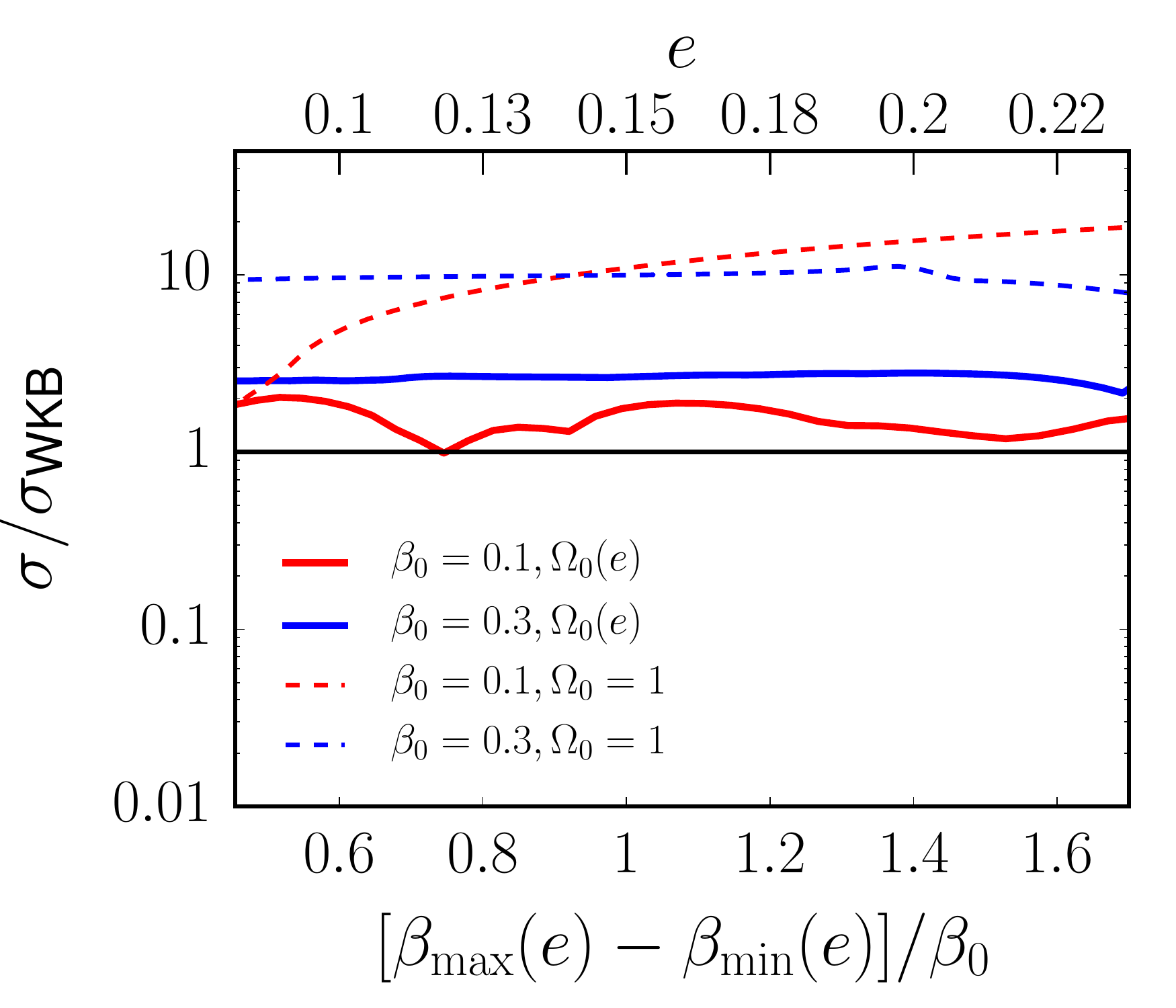}
	\caption{Growth rates of instabilities obtained for synchronised bodies $\Omega_0=1$ and pseudo-synchronised bodies with $\Omega_0(e)$ given by the formula (\ref{eq:ldei_hut81}). Ratio $\sigma/\sigma_\text{wkb}$ with $\sigma_\text{wkb}$ given by formula (\ref{Eq_LDEI_WKB}). Triaxial geometry $a(t) = R \sqrt{1 + \beta_{ab}(t)}$, $b(t) = R\sqrt{1-\beta_{ab}(t)}$ and $c(t) = [a(t)b(t)]^{-1}$ with $R = R_m + 0.05$.}
	\label{Fig_Hut81}
\end{figure}

Finally for the libration forcing it is possible to partially remove the effect of dynamical tides. The tidal torque averaged over weakly eccentric orbits ($e\to0$) vanishes for $\Omega_0=1$. However it no longer holds for eccentric orbits. Following \citet{hut1981tidal}, the tidal torque averaged over an eccentric orbit of eccentricity $e$ vanishes when  
\begin{equation}
	\Omega_0(e) = \frac{(1+3e^2 + \frac{3}{8} e^4) (1-e^2)^{3/2}}{1+\frac{15}{2} e^2 + \frac{45}{8} e^4 + \frac{5}{16} e^6} = 1 - 6e^2 + \mathcal{O}(e^3).
	\label{eq:ldei_hut81}
\end{equation}
When $\Omega_0$ is given by the expression (\ref{eq:ldei_hut81}), the body has reached a pseudo-synchronised state. Pseudo-synchronisation is an important process in the dynamics of binary systems \citep[e.~g][]{hut1981tidal,hut1982tidal}. Indeed pseudo-synchronisation proceeds much faster than circularisation of the orbits \citep{goupil2008tidal}. So the fluid spin rate of a celestial body in an eccentric orbit would tidally evolve towards pseudo-synchronisation (\ref{eq:ldei_hut81}), while the orbit remains eccentric of orbital period $2\pi/\Omega_0$.

We show in figure \ref{Fig_Hut81} the growth rates $\sigma$, normalised by the growth rates of the classical LDEI given by the local formula (\ref{Eq_LDEI_WKB}). We consider both the synchronised case $\Omega_0 = 1$ and the pseudo-synchronised case $\Omega_0(e)$ given by formula (\ref{eq:ldei_hut81}). When the pseudo-synchronisation is reached, the growth rates remain bounded within the range $\sigma/\sigma_{\text{wkb}} \leq 3$ for $\beta_{0} \leq 0.3$. A normalisation with respect to the TDEI (see the formula \ref{Eq_TDEI_sigmaWKB} in appendix \ref{app:tdei_ledizes}) leads to erroneous growth rates for pseudo-synchronised bodies. So we conclude that the LDEI formula (\ref{Eq_LDEI_WKB}) gives a good estimate of the growth rates of orbitally driven instabilities in pseudo-synchronised bodies. Instead for synchronised states ($\Omega_0=1$) the instabilities can be much more vigorous than those predicted by the local formula (\ref{Eq_LDEI_WKB}), with $\sigma/\sigma_{\text{wkb}} \geq 10$ if $\Delta\beta_{ab}/\beta_0 \geq 1$ (as previously discussed in \S\ref{sec:ldei}).

Therefore the net non-zero tidal torque operating along eccentric Kepler orbits is responsible for the enhancement of the instabilities at $\Omega_0=1$. A similar effect also exists at $\Omega_0=2+\beta_0$. So it proves that the dynamical tides are essential to explain the observed enhancing of the growth rates. For the particular case of pseudo-synchronised orbits (\ref{eq:ldei_hut81}), the effects of dynamical tides do not overcome the ones of equilibrium tide on the fluid instabilities, as shown in figure \ref{Fig_Hut81}. The onset of instabilities in pseudo-synchronised bodies is well predicted by the local analysis of the LDEI (\ref{Eq_LDEI_WKB}). So the quantitative predictions of \citet{cebron2013elliptical} for the onset of the elliptical instability in bloated pseudo-synchronised hot Jupiters, based upon the formula \ref{Eq_TDEI_sigmaWKB}), may have to be reassessed.

\section{Conclusion and perspectives}
\label{sec:ccl}
	\subsection{Physical implications}
Rotating fluid ellipsoids have been the subject of many works, going back to \citet{riemann1860untersuchungen}. Their stability is affected by free-surface perturbations, associated with surface gravity modes, and internal hydrodynamic perturbations. Surprisingly, free-surface perturbations weakly affect the stability of fluid ellipsoids \citep{lebovitz1996new,Barker11062016,barker2016nonlinear}. Thus their stability is mainly governed by flow instabilities. Because the viscosity is extremely small in astrophysical bodies, an inviscid analysis is physically relevant. Previous hydrodynamic studies have to be completed, because \citet{lebovitz1996short,lebovitz1996new} consider isolated ellipsoidal fluid masses and \citet{Barker11062016,barker2016nonlinear} ellipsoids moving on circular orbits.

To simplify the problem, we have considered only the case where the mass of the attractor is much larger than the mass of the companion body. It is the simplest framework to model two-body systems such as synchronised moons around planets, gaseous extrasolar planets (Hot Jupiters) around stars or low massive stars orbiting around massive attractors. 
The general two-body problem could also be tackled (e.g. a binary stellar system), solely by changing the hydrostatic estimation of the equatorial ellipticity in formula (\ref{Eq_ODEI_beta}). The radius $r(t)$ has to be replaced by the time-dependent distance between the centres of mass of the two bodies, which are both moving on eccentric orbits. We are confident that our main findings will not change qualitatively in that configuration.

We have revisited the hydrodynamic instabilities of homogeneous, incompressible and rotating ellipsoidal fluid masses subjected to a disturbing tidal potential. 
Several studies are devoted to the stability of fluid ellipsoids subjected to a tidal potential generated by orbital motions on circular orbits \citep{aizenman1968equilibrium,cebron2012elliptical,cebron2013elliptical,Barker11062016}.
Thus our primary purpose was to study how the hydrodynamic stability of fluid ellipsoids is modified by considering a tidal potential generated by orbital motions on eccentric Kepler orbits. 
Our study is complementary to the hydrodynamic stability analysis of \citet{barker2016nonlinear} of Roche-Riemann ellipsoids on circular orbits. 
We recover all the limiting cases of ellipsoidal flow instability \citep{lebovitz1996short,lebovitz1996new,cebron2013elliptical,barker2016nonlinear} and unify them into a global framework.

We may summarise our rather unexpected results in the following way. First, the classical TDEI is unaffected by the dynamical tides for retrograde eccentric orbits for $(1+\beta_0)(\beta_0-1) < \Omega_0 \leq 0$ (outside of the forbidden zone). 
Second, instabilities excited on moderately eccentric orbits can have larger growth rates than those on nearly circular orbits. Indeed dynamical tides are responsible for the enhancing of the vigour of the TDEI near the 2:1 spin-orbit resonance ($\Omega_0 \simeq 2$) and of the LDEI ($\Omega_0=1$). 
Finally, fluid ellipsoids exhibit new fluid instabilities which are triggered within the forbidden zone of the classical TDEI for retrograde ($\Omega_0 \leq (1+\beta_0)(\beta_0-1)$) and prograde ($\Omega_0 \geq 3$) orbits. All these findings show that dynamical tides can drive new instabilities in fluid bodies moving on eccentric Kepler orbits.

We have updated the picture of the linear stability of tidally disturbed fluid ellipsoids. A complete view emerges now. They are prone to various local and global inviscid instabilities. 
On one hand, spheroids are only unstable against free-surface perturbations, associated with surface gravity modes \citep{lebovitz1996new,Barker11062016}. 
Considering tidal effects generated by orbital motions on circular orbits, all ellipsoids (the Roche-Riemann ellipsoids) are unstable against the elliptical instability when $(1+\beta_0)(\beta_0-1) < \Omega_0 < 3$, as predicted by previous analyses \citep{cebron2012elliptical,cebron2013elliptical,barker2016nonlinear}.
Taking into account instabilities of all possible spatial complexity handled by the local and global theories, the parameter space of fluid ellipsoids subjected to a varying tidal torque (eccentric orbits) is unstable against orbitally driven instabilities for both retrograde and prograde eccentric orbits within the range $-10 \leq \Omega_0 \leq 10$ (see figure \ref{fig:swan_odei}). 
Although not considered in our computations, we also expect them to be intrinsically unstable on a wider range of $|\Omega_0|$ (at least for large enough eccentricities).

\begin{table*}
	\centering
	\begin{tabular}{lccccc}
	\\
	\hline
	Stars & $e$ & $P_s$ [d] & $P_{orb}$ [d] & $\Omega_0 = P_s/P_{orb}$ & $\Delta \beta_{ab} / \beta_0$ \\
	\hline
	WASP-17 & 0.028 & 10 & -3.73 & -2.68 & 0.17 \\
	WASP-10 & 0.057 & 11.90 & 3.09 & 3.85 & 0.35 \\
	GJ 674  & 0.07 & 34.80 & 4.69 & 7.42 & 0.43 \\
	HAT-P-1  & 0.067 & 26.60 & 4.46 & 5.96 & 0.41 \\
	WASP-14  & 0.087 & 13.5 & 2.24 & 6.03 & 0.54 \\
	\hline
	\\
	\end{tabular}
	\caption{Orbital parameters of some stars with companions orbiting on eccentric Kepler orbits. $P_s = 2\pi/\Omega_s$ is the spin period (in days) and $P_{orb}=2\pi/\Omega_{orb}$ the orbital period (in days). The last column $\Delta \beta_{ab} / \beta_0$ is defined by formula (\ref{eq:dbetaAB}). The given stars are located within the forbidden zone FZ$_{\beta_0}$ of the classical TDEI. Adapted from \citet{cebron2013elliptical}. Data have been updated from \url{http://exoplanet.eu/}.}
	\label{table:astro}
\end{table*}

Our findings may have important consequences for the tidal dissipation responsible for the circularisation and synchronisation of two-body systems. Some stars, located within the forbidden zone FZ$_{\beta_0}$ of the classical TDEI and with companions orbiting on eccentric Kepler orbits, are reported in table \ref{table:astro} as example. The effects of dynamical tides are not negligible ($\Delta \beta_{ab} / \beta_0 \sim 0.54$ for WASP-14). Consequently we expect these stars to be unstable for the orbitally driven instability (in spite of their presence in the classical forbidden zone).
Two tidal dissipation processes have received most attention, namely tidal friction of the equilibrium tide \citep{zahn1966marees} and tidal friction of eigenmodes forced by dynamical tides \citep{zahn1975dynamical,ogilvie2004tidal,wu2005origin,wu2005origin2,ogilvie2007tidal,goodman2009dynamical,ogilvie2009tidal,rieutord2010viscous}.
The elliptical instability is thus an alternative and promising mechanism \citep{rieutord2004evolution,goupil2008dynamics}. 
 
	\subsection{Perspectives}
The nonlinear outcome of these fluid instabilities remains elusive in astrophysical bodies. Indeed it is not clear whether turbulent flows can develop and sustain an effective mixing in fluid interiors. A parameter survey of their nonlinear behaviours, using efficient numerical simulations, is necessary. For instance \citet{le2017inertial} show that the saturation of the elliptical instability generates turbulence exhibiting many signatures of inertial wave turbulence, a regime possibly expected in planetary interiors. \citet{barker2016nonlinear} also suggests that the elliptical instability may explain the spin synchronisation and circularisation of the shortest period hot Jupiters.

Future work is also required to adopt more realistic interior models. In particular the behaviour of these instabilities when a stable stratification (like a stellar radiative zone) is present is almost unknown, as well as the role of compressibility. 
For instance stars of mass larger than $1.8$ solar mass are stably stratified in their outer layers.
Circularisation and synchronisation are also effective for these stars \citep[e.g.][]{giuricin1984synchronization}. Depending on the considered density profile, the stratification does favour or not the elliptical instability \citep{kerswell1993elliptical,miyazaki1993elliptical,le2006thermo,cebron2010tidal,clausen2014elliptical}. However a unifying theory is still missing and has to be addressed. In addition, the capability of tidal effects to drive self-sustained magnetic field is still controversial. Tidal dynamos were first addressed by \citet{barker2013non,cebron2014tidally}, neglecting density effects.

Finally we have developed two open source numerical codes that may be useful for future linear stability studies of incompressible fluids. They are quite general and can be applied to several other situations. They are freely available for the community at \url{https://bitbucket.org/vidalje/}. On one hand with the SWAN code, local stability analyses of any time-dependent basic flow in unbounded fluids can be performed. On the other hand, global stability analyses of any mechanically driven flow of uniform vorticity in ellipsoids can be carried out with the SIREN code. Unlike previous studies, our code also handles arbitrary (time-dependent) ellipsoidal shapes (not limited to small departures from the sphere). Indeed, it handles ellipsoidal perturbations of unprecedented small wavelengths. We have considered in the present study polynomial degrees as large as $n=25$. It corresponds to more than 6000 basis elements.

A fork of the SIREN code has been used to compute (i) the tilted hydromagnetic eigenmodes of a fluid in a corotating ellipsoid \citep{vidal16diffusionless} and (ii) the viscous decay factors of inertial modes \citep{lemasquerierlibration}. Computing the inertial modes is the first step towards a complete and self-contained viscous stability analysis of inertial instabilities in arbitrary rotating ellipsoids. Indeed, viscous effects can be introduced as a correction of inviscid inertial modes. This viscous correction is required to compare theoretical predictions with simulations or experiments, all performed at finite values of viscosity and deformation.

\section*{Acknowledgements}
JV acknowledges the French {\it Minist\`ere de l'Enseignement Sup\'erieur et de la Recherche} for his PhD grant.
This work was partially funded by the French {\it Agence Nationale de la Recherche} under grant ANR-14-CE33-0012 (MagLune) and by the 2017 TelluS program from CNRS-INSU (PNP) AO2017-1040353.
ISTerre is part of Labex OSUG@2020 (ANR10 LABX56).
Most figures were produced using matplotlib (\url{http://matplotlib.org/}).
The authors strongly acknowledge the four anonymous referees for their comments, which have
strongly improved the quality of the paper.

\appendix{}

\section{Polynomial basis of $\boldsymbol{\mathcal{V}}_{n}$}
\label{Appendix_GP_Vn}
\citet{wu2011high} have proposed an algorithm to build the basis elements of $\boldsymbol{\mathcal{V}}_{n}$. We outline here the method. We consider first a spherical container ($a=b=c$). The vorticity field is decomposed into poloidal $\mathcal{P}_w (\boldsymbol{r})$ and toroidal $\mathcal{T}_w (\boldsymbol{r})$ scalars as
\begin{equation}
	\nabla \times \boldsymbol{u} = \nabla \times \left ( \mathcal{T}_w \boldsymbol{r} \right ) + \nabla \times \nabla \times \left ( \mathcal{P}_w \boldsymbol{r} \right ).
\end{equation}
such that it obeys the solenoidal condition. Vorticity $\nabla \times \boldsymbol{u}$ is then projected onto the finite-dimensional vector space $\boldsymbol{\mathcal{W}}_{n-1}$, made of Cartesian homogeneous monomials $x^iy^jz^k$ of degree $n -1 = i+j+k$ \citep{vantieghem2014inertial}. Note that an element of $\boldsymbol{\mathcal{W}}_n$ is solenoidal but does not necessarily satisfy the impermeability condition. $\mathcal{P}_w (\boldsymbol{r})$ is a homogeneous polynomial of degree $n$ while $\mathcal{T}_w (\boldsymbol{r})$ is a homogeneous polynomial of degree $n-1$. Similarly the velocity field $\boldsymbol{u} (\boldsymbol{r})$ is expanded into poloidal $\mathcal{P}_u (\boldsymbol{r})$ and toroidal $\mathcal{T}_u (\boldsymbol{r})$ scalars as 
\begin{equation}
	\boldsymbol{u} (\boldsymbol{r}) = \nabla \times \left ( \mathcal{T}_u \boldsymbol{r} \right ) + \nabla \times \nabla \times \left ( \mathcal{P}_u \boldsymbol{r} \right ).
\end{equation}
Since there is an isomorphism between vector spaces $\boldsymbol{\mathcal{W}}_{n-1}$ and $\boldsymbol{\mathcal{V}}_{n}$ \citep{vantieghem2014inertial}, we expand $\boldsymbol{u}$ onto $\boldsymbol{\mathcal{V}}_{n}$ such that velocity scalars are related to the vorticity scalars by
\begin{equation}
	\mathcal{T}_u = \mathcal{P}_w
\end{equation}
and
\begin{equation}
	\nabla^2 \mathcal{P}_u = -\mathcal{T}_w \, \ \, \text{with} \, \ \, \mathcal{L}^2 \mathcal{P}_u = 0 \, \ \, \text{at} \, \ \, r=1, \label{Eq_Pol_Wu}
\end{equation}
where $\mathcal{L}^2$ is the angular momentum operator
\begin{equation}
	\mathcal{L}^2 = \left ( y \frac{\partial }{\partial z} - z \frac{\partial }{\partial y} \right )^2 + \left ( z \frac{\partial }{\partial x} - x \frac{\partial }{\partial z} \right )^2 + \left ( x \frac{\partial }{\partial y} - y \frac{\partial }{\partial x} \right )^2.
\end{equation}

The difficult part of the above algorithm is to solve equations (\ref{Eq_Pol_Wu}). However any homogeneous polynomial of degree $p$ can be decomposed into harmonic homogeneous polynomials of maximum degree $p$, which are spherical harmonics \citep{backus1996foundations}. So we project $\mathcal{T}_w (\boldsymbol{r})$ of degree $n-1$ onto spherical harmonics as
\begin{equation}
	\mathcal{T}_w (\boldsymbol{r}) = r^{n-1} \sum \limits_{l=1}^{n-1} \sum \limits_{m=-l}^{l} t_l^m \mathcal{Y}_l^m,
	\label{Eq_Tor_W_SH}
\end{equation}
where $\mathcal{Y}_l^m$ are normalised spherical harmonics of degree $l$ and order $m$ and $\left \{ t_l^m \right \}$ the set of spherical harmonics coefficients. The degree $l=0$ is omitted because of the incompressible condition. Poloidal scalar solution of (\ref{Eq_Pol_Wu}) is of the form
\begin{equation}
	\mathcal{P}_u = \mathcal{P}_{P} + \mathcal{P}_{H},
\end{equation}
with $\mathcal{P}_P$ a particular solution of (\ref{Eq_Pol_Wu}) and the general solution of the homogeneous Laplace equation $\nabla^2 \mathcal{P}_H = 0$. From the expansion (\ref{Eq_Tor_W_SH}), a particular solution of equation (\ref{Eq_Pol_Wu}) in spherical harmonics expansion is
\begin{equation}
	\mathcal{P}_P(\boldsymbol{r}) = r^{n+1} \sum \limits_{l=1}^{n-1} \sum \limits_{m=-l}^{l} \frac{-t_l^m}{(n+2)(n+1) - l(l+1)} \mathcal{Y}_l^m.
	\label{Eq_PolP_U_SH}
\end{equation}
The homogeneous solution has the general form
\begin{equation}
	\mathcal{P}_H(\boldsymbol{r}) = r^{n+2} \sum \limits_{l=1}^{n+2} \sum \limits_{m=-l}^{l} p_l^m \mathcal{Y}_l^m,
	\label{Eq_PolH_U_SH}
\end{equation}
where the set of coefficients $\left \{ p_l^m \right \}$ is determined by the boundary condition
\begin{equation}
	\mathcal{L}^2 \mathcal{P}_P = - \mathcal{L}^2 \mathcal{P}_H.
\end{equation}
Once the coefficients are known we can transform the spherical harmonics expansion back into a Cartesian form. Finally, the Poincar\'e transform \citep{poincare1910precession} 
\begin{equation}
	(x,y,z) \leftarrow \left ( \frac{x}{a}, \frac{y}{b}, \frac{z}{c} \right ) \, \ \, \text{and} \, \ \ (u_x,u_y,u_z) \leftarrow \left ( \frac{u_x}{a}, \frac{u_y}{b}, \frac{u_z}{c} \right )
\end{equation}
is used to convert the solutions in spheres to solutions in ellipsoids of axes $(a,b,c)$.

The implementation of \citet{wu2011high} relies on symbolic computations of (\ref{Eq_Tor_W_SH}), (\ref{Eq_PolP_U_SH}) and (\ref{Eq_PolH_U_SH}). Basis elements up to degrees $n=5$ are explicitly given in their Appendix A. However, their symbolic algorithm breaks down for degrees $n > 6$, because their algorithm seems to fail to compute the spherical harmonic coefficients $t_l^m$ for higher degrees. We have extended their method to build the basis for degrees $n \geq 6$. It is achieved by combining symbolic and numerical calculus in Python. The algorithm is also parallelised to reduce the computation time. With our implementation we can reach degrees $n \geq 6$, because spherical harmonics coefficients are only computed numerically with the open-source library SHTNS \citep{schaeffer2013efficient}. The comparison between (\ref{Eq_Basis_Vn}) and the elements obtained above shows that the two sets are equivalent, changing only by linear combinations of the basis elements. 

In practice, the generation of basis elements is not restricted to a particular degree but we found that the generation of symbolic matrices $\boldsymbol{M}$, $\boldsymbol{N}$ and $\boldsymbol{L}$ becomes impractical for degrees $n>18$ because of high memory usage ($\simeq 200$~GB). Consequently we have adopted the algorithm of \citet{lebovitz1989stability} for high degrees numerical computations, reaching degrees as high as $n=25$ ($\lesssim 20$~GB). The limiting factor is then the CPU time to solve the stability problem.

\section{Precessing flow in spheroidal containers}
\label{app:prec}
\begin{figure}
	\centering
	\begin{tabular}{cc}
		\subfigure[Basis $n=2$ as in \citet{kerswell1993instability}]{\includegraphics[width=0.49\textwidth]{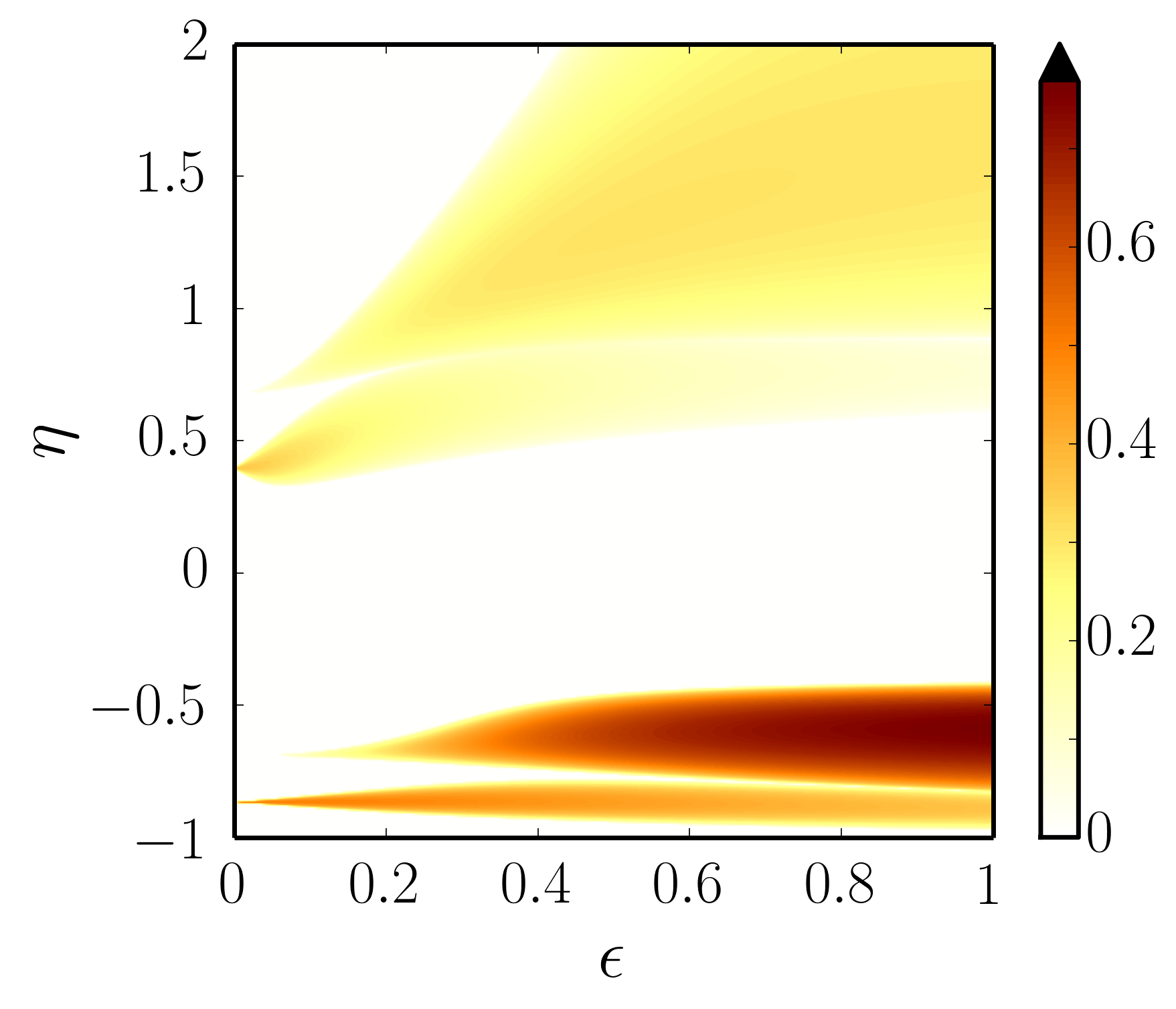}} &
		\subfigure[Basis $n=6$ as in \citet{wu2011high}]{\includegraphics[width=0.49\textwidth]{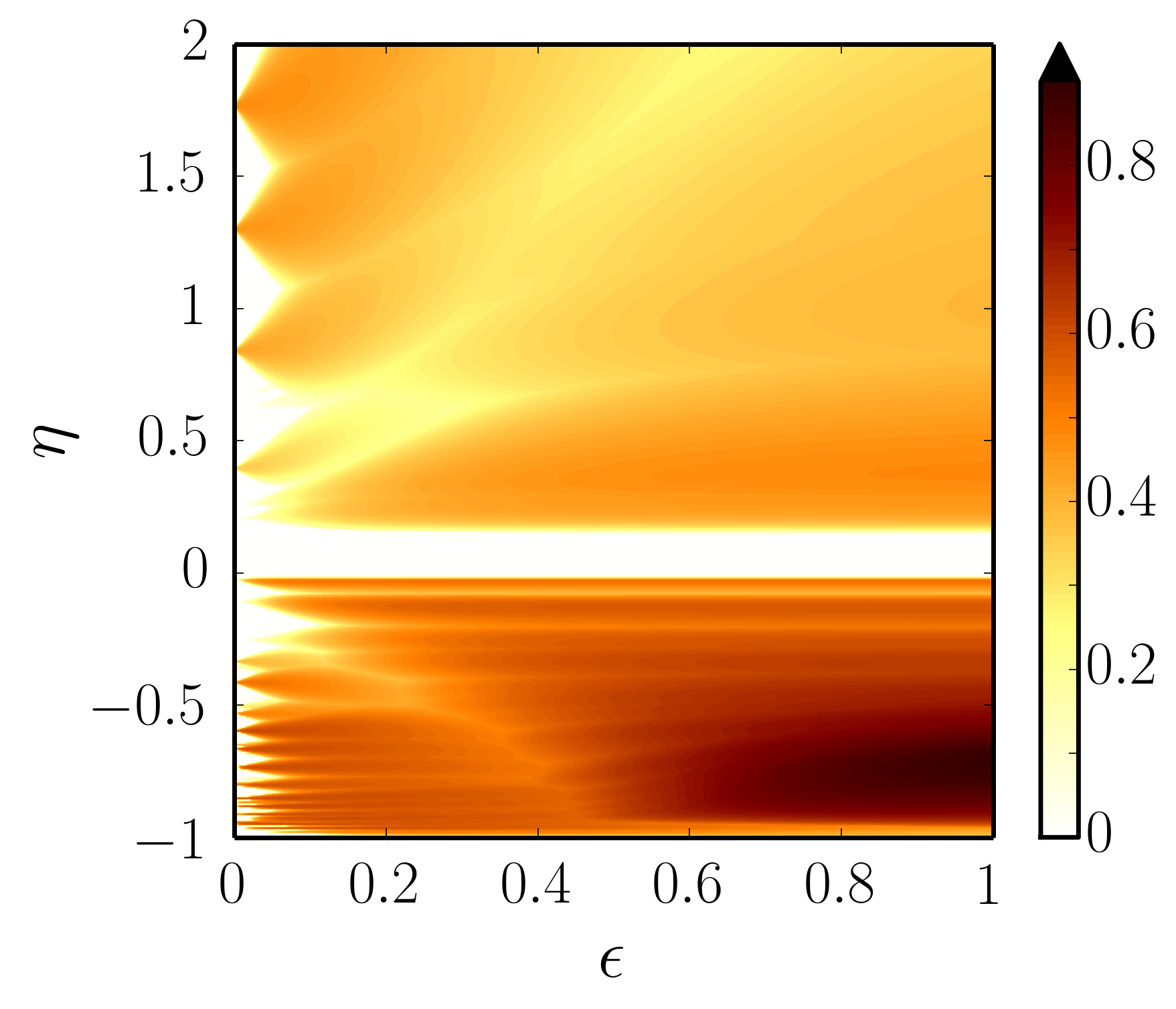}} \\
		\subfigure[Basis $n=15$]{\includegraphics[width=0.49\textwidth]{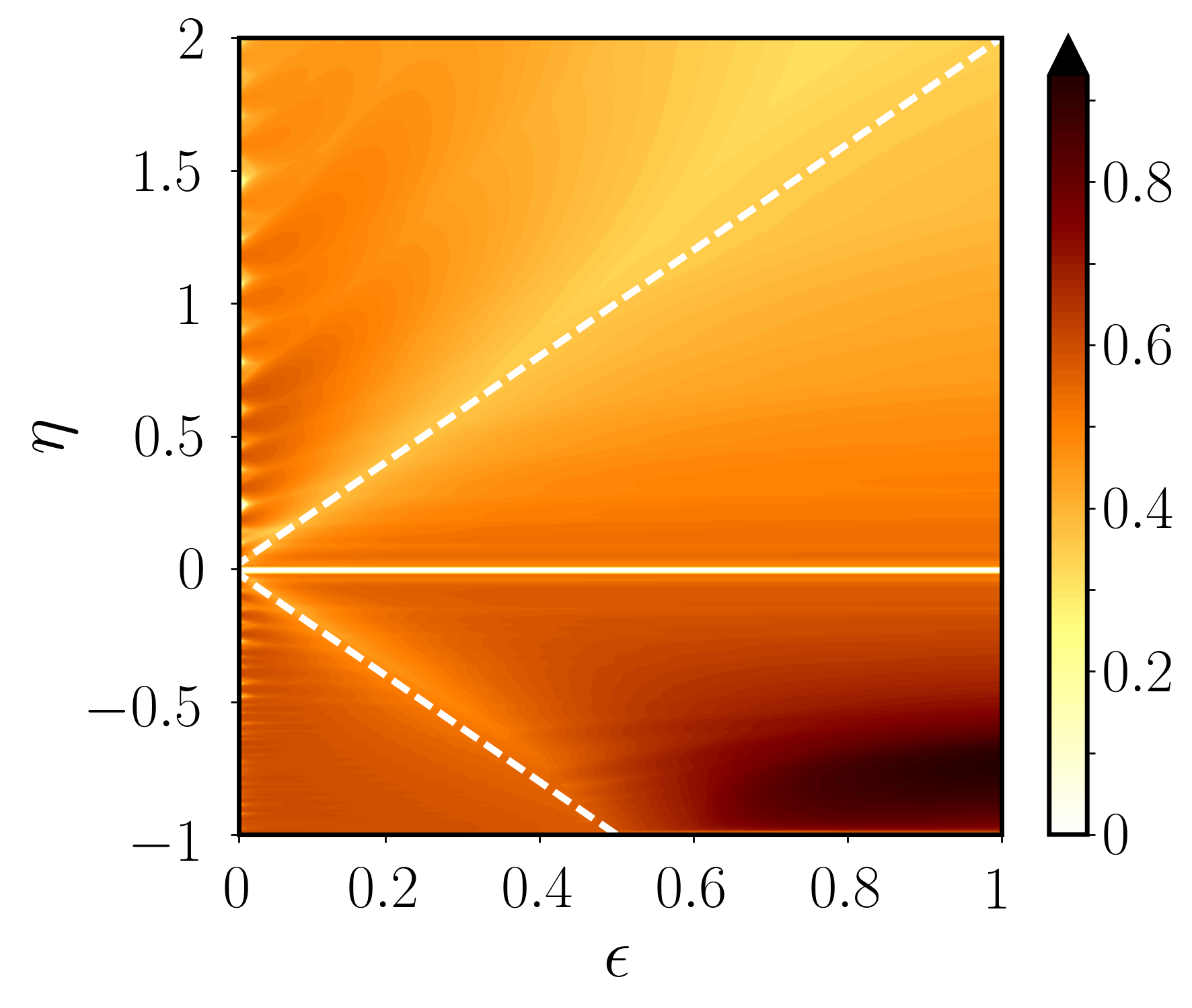}} &
		\subfigure[Numerical local analysis]{\includegraphics[width=0.49\textwidth]{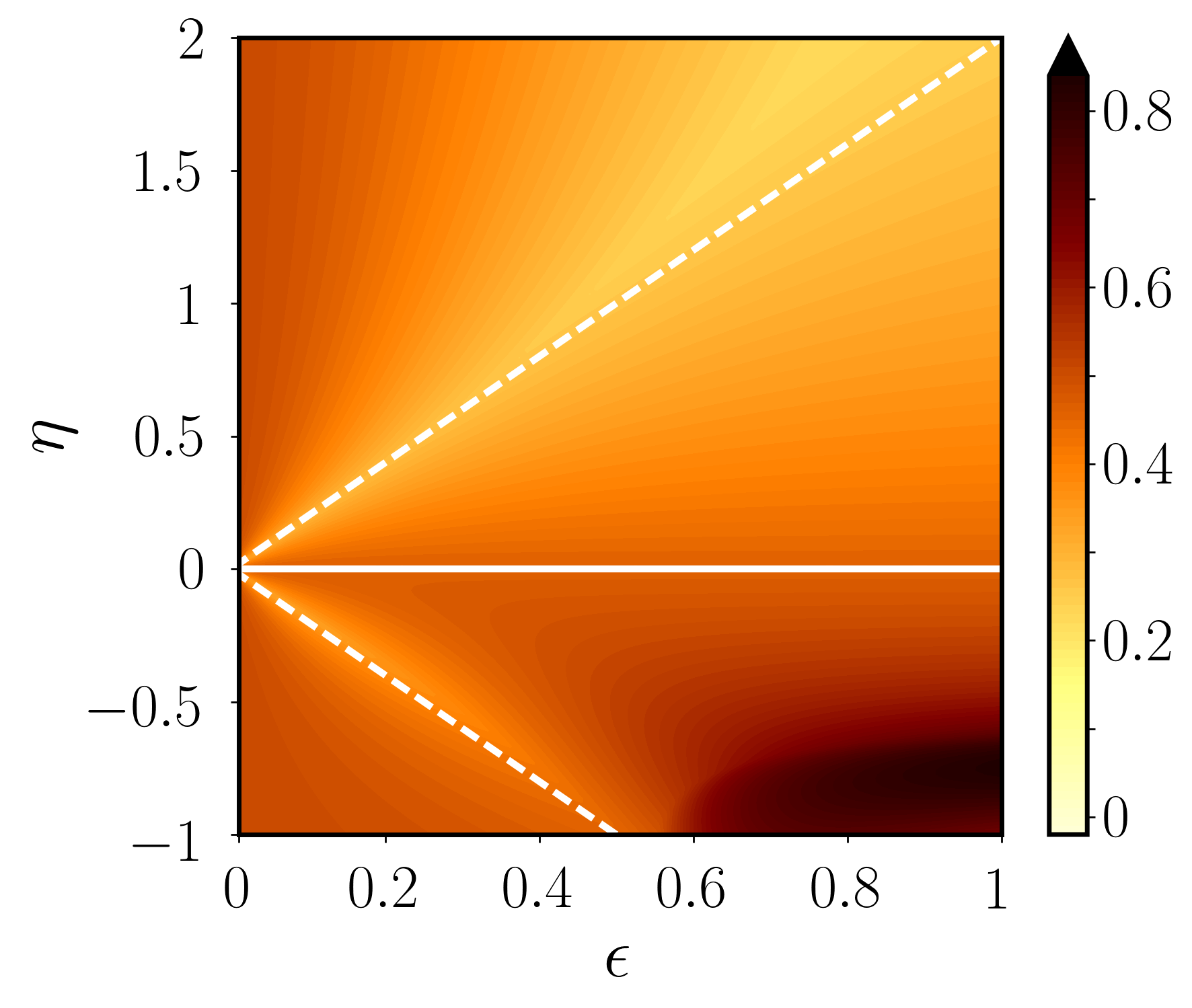}}
	\end{tabular}
	\caption{Survey of the stability of the precessing basic flow (\ref{Eq_BF_PrecS}) in the ($\eta, \epsilon$) plane. Color map shows the ratio $\sigma / \epsilon$. The same color scale is used for the four plots. White areas correspond to marginally stable regions. However the precessing basic flow (\ref{Eq_BF_PrecS}) is divergent for $\eta=0$ (resonance). Hence its stability has not been studied for $\eta=0$ (horizontal white solid line). In (c,d), tilted white dashed lines are given by $\eta = \pm 2 \epsilon$.}
	\label{Fig_Prec_Bench}
\end{figure}

We revisit here the precession-driven instabilities in spheroidal geometry ($a=b \neq c$), studied by \citet{kerswell1993instability} and \citet{wu2011high}. We use this case as a cross-benchmark for our global and local analyses. We work in the precessing frame, where the vertical axis coincides with the spheroidal axis of symmetry. We assume a precession angle of $\pi/2$, such that the body rotation vector is $\boldsymbol{\Omega}^{\mathcal{B}} = \epsilon \, \boldsymbol{\widehat{x}}$, with $\epsilon$ the dimensionless amplitude of the precession forcing (Poincar\'e number). The precession-driven basic flow is \citep{kerswell1993instability,wu2011high}
\begin{equation}
	\boldsymbol{U} = -y \, \widehat{\boldsymbol{x}} + [ x - \mu (1 + \eta) z] \, \widehat{\boldsymbol{y}} + \mu y \, \widehat{\boldsymbol{z}},
	\label{Eq_BF_PrecS}
\end{equation}
with $\eta = 1/c^2 - 1$ the polar flattening and $\mu = 2\epsilon / \eta$ a parameter which measures the ratio of the elliptical distortion over the shearing of the streamlines. 

As shown by \citet{kerswell1993instability}, no instability is associated with the linear basis ($n=1$). \citet{wu2011high} extended the work of \citet{kerswell1993instability} by considering basis up to degree $n=6$. Results for the $n=2$ and $n=6$ bases are shown in figures \ref{Fig_Prec_Bench} (a) and (b), which survey the stability of (\ref{Eq_BF_PrecS}) in the plane ($\eta, \epsilon$). The stability maps are in perfect agreement with the previous studies. Tongues of instabilities emerge from the $\eta$ axis. Tongues are associated with two types of instability, namely elliptical and shear instabilities \citep{kerswell1993instability}. The former have growth rates proportional to $\epsilon^2$ and the latter to $\epsilon$. Showing isocontours of $\sigma / \epsilon$ makes the elliptical tongues thicker than the shearing ones. When $n$ increases, the maximum growth rate of oblate spheroids ($\eta > 0$) first increases quickly. When $n$ is large enough ($n > 10$) the increase slows down and the growth rates reach constant values when $n$ increases further. On the other hand prolate spheroids ($\eta < 0$) have already large growth rates close to 1 for large $\epsilon$ and the maximum values do not really evolve with $n$. 
As noticed by \citet{wu2011high}, the progression of unstable tongues for oblate spheroids ($\eta > 0$) toward the spherical case $\eta = 0$ is quicker than for prolate spheroids ($\eta < 0$) when $n$ increases.

Global analysis at maximum degree $n=15$ is shown in figure \ref{Fig_Prec_Bench} (c). In comparison with $n=6$, new tongues of instabilities appear almost everywhere, filling the map and making the identification of the nature of the unstable tongues difficult. Valleys of less unstable modes are found for prolate and oblate ellipsoids (white dashed lines). 
Global analysis is in excellent agreement with the local analysis shown in figure \ref{Fig_Prec_Bench} (d). This benchmark cross-validates our two numerical codes.

\begin{figure}
	\centering
	\includegraphics[width=0.5\textwidth]{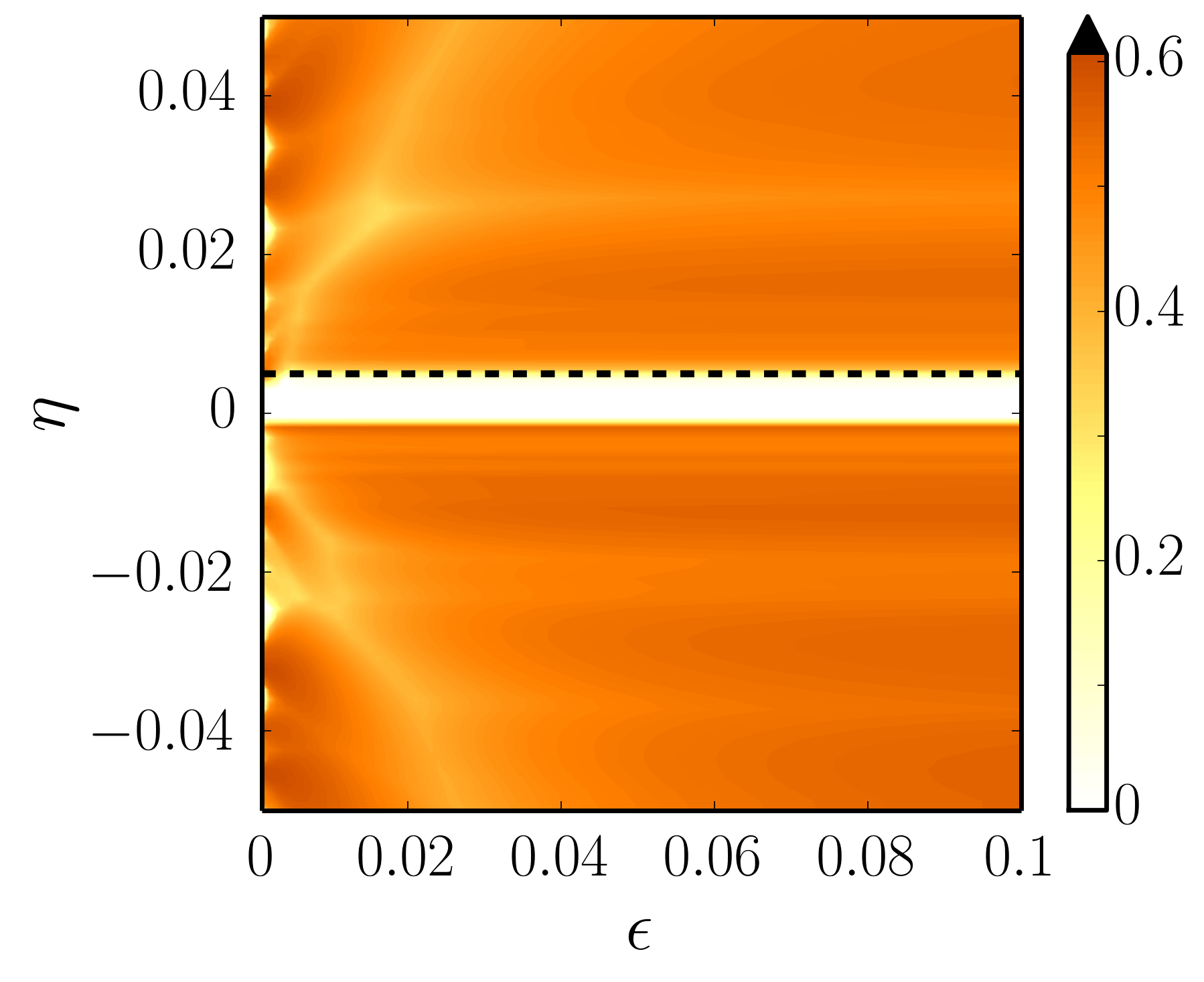}
	\caption{Survey of the stability of the precessing basic flow (\ref{Eq_BF_PrecS}) in the ($\eta, \epsilon$) plane. Basis $n=25$ (zoom in). The Earth oblateness ($\eta=0.005$) is shown by the black dashed line. Colour map shows the ratio $\sigma / \epsilon$. White areas correspond to marginally stable regions. However the precessing basic flow (\ref{Eq_BF_PrecS}) is divergent for $\eta=0$ (resonance). Hence its stability has not been studied for $\eta=0$.}
	\label{Fig_PrecEarth}
\end{figure}

Finally figures \ref{Fig_Prec_Bench} (c,d) draws the possible existence of global instabilities for the limit of very small oblateness relevant in geophysics. For instance the Earth's liquid core has a flattening of $\eta \simeq 0.005$. We push up the maximum degree to $n=25$ in figure \ref{Fig_PrecEarth} and zoom in on the geophysical range of parameter space.
We observe instability for oblate spheroid of oblateness as small as the one of the Earth's core.
The amplitude of precession $\epsilon$ is still rather large to be consistent with geophysical values ($\epsilon \simeq 10^{-7}$), but it is likely that an unstable area appears for smaller $\epsilon$ when $n$ is further increased as predicted by the local analysis in the weak forcing limit \citep{kerswell1993instability}.

\section{Tidally driven elliptical instability}
	\subsection{Asymptotic growth rate}
	\label{app:tdei_ledizes}
\citet{le2000three} uses a multiple-scale analysis in $\beta_{0}$ to solve the local equations (\ref{Eq_WKB}).
The local growth rate $\sigma_\text{wkb}$ is, at leading order \cite[equation 32 of][]{le2000three},
\begin{equation}
 	\frac{\sigma_\text{wkb}}{\left | 1-\Omega_{0} \right | } = \max_{\theta_0} \frac{1}{4} \sqrt{\left ( 1 + \cos \theta_0 \right )^4 \beta_{0}^2 - 4 \left [2 - 4 \left( 1 + \widetilde{\Omega}_{0} \right ) \cos \theta_0 \right ]^2 } + \mathcal{O} \left ( \beta_{0}^2 \right ),
 	\label{Eq_TDEI_DetuneWKB}
\end{equation}
with $\widetilde{\Omega}_{0} = \Omega_0 / (1 - \Omega_0)$ and $\theta_0$ the colatitude angle between the vertical axis $\widehat{\boldsymbol{z}}$ and the initial wave vector $\boldsymbol{k}_0$, ranging in $[0, \pi]$. Angle $\theta_0$ is chosen to maximise $\sigma$.
In the asymptotic limit $\beta_{0} \to 0$, the elliptical instability only exists in the allowable range $-1~<~\Omega_{0}~<~3$ \citep{craik1989stability}.
Outside this range, the flow lies in the forbidden zone of the TDEI for $\beta_0 \to 0$, denoted FZ$_0$. As a result of geometric detuning effects the classical TDEI is excited on a wider allowable range for finite values of $\beta_{0}$ introduced in the main text. Values outside this range define the classical forbidden zone for finite $\beta_0$, denoted FZ$_{\beta_0}$ in the main text.
In the limit $\beta_0 \to 0$, the growth rate (\ref{Eq_TDEI_DetuneWKB}) reduces to the formula originally devised by \citet{craik1989stability} 
\begin{equation}
	\frac{\sigma_\text{wkb}}{ \left | 1-\Omega_{0} \right |} = \frac{(3 + 2 \widetilde{\Omega}_{0} )^2}{16(1+\widetilde{\Omega}_{0} )^2}  \beta_{0},
 	\label{Eq_TDEI_sigmaWKB}
\end{equation}
for $\cos\theta_0 = 1 / [2(1+\widetilde{\Omega}_{0})]$ which maximises $\sigma_\text{wkb}$.
Finally in the allowable range $(\beta_{0}+1)/(\beta_{0}-1)~<\Omega_{0}\leq-1$, the formula (\ref{Eq_TDEI_DetuneWKB}) gives a non-zero growth rate of \cite[equation 44 of][]{le2000three}
\begin{equation}
 	\frac{\sigma_\text{wkb}}{\left | 1-\Omega_{0} \right | } = \sqrt{\beta_{0}^2-4(\widetilde{\Omega}_{0} +1/2)^2}
\end{equation}
for $\theta_0=0$. When $\Omega_0\geq3$, formula (\ref{Eq_TDEI_DetuneWKB}) gives $\sigma_\text{wkb} = 0$, such that there is no TDEI predicted by the local analysis (at this order in $\beta_0$).

Thus the local analysis of \citet{le2000three} given by the general formula (\ref{Eq_TDEI_DetuneWKB}) does predict that the TDEI at finite $\beta_0$ extends well beyond the region that is unstable at $\beta_0 \to 0$, i.e. when $-1~<~\Omega_{0}~<~3$.
Larger values of $\beta_{0}$ lead to more unstable couplings. As explained by \citet{bayly1986three,waleffe1990three,le2000three,kerswell2002elliptical}, the elliptical instability results from a parametric resonance of plane waves with the basic flow. Resonances have finite widths which increase with $\beta_0$. Thus the elliptical instability can be excited even when exact resonance conditions are not satisfied, giving a wider unstable region in parameter space. This coupling effect of $\beta_0$ is highlighted in figure \ref{Fig_TDEI2}, already giving good matching between local and global results at degree $n=15$. 
Note that this effect has recently been put forward by \citet{barker2016nonlinear}, as shown in \S\ref{app:tdei_barker}.

\begin{figure}
	\centering
	\begin{tabular}{cc}
    	\subfigure[$\beta_{0}=0.15$]{\includegraphics[width=0.49\textwidth]{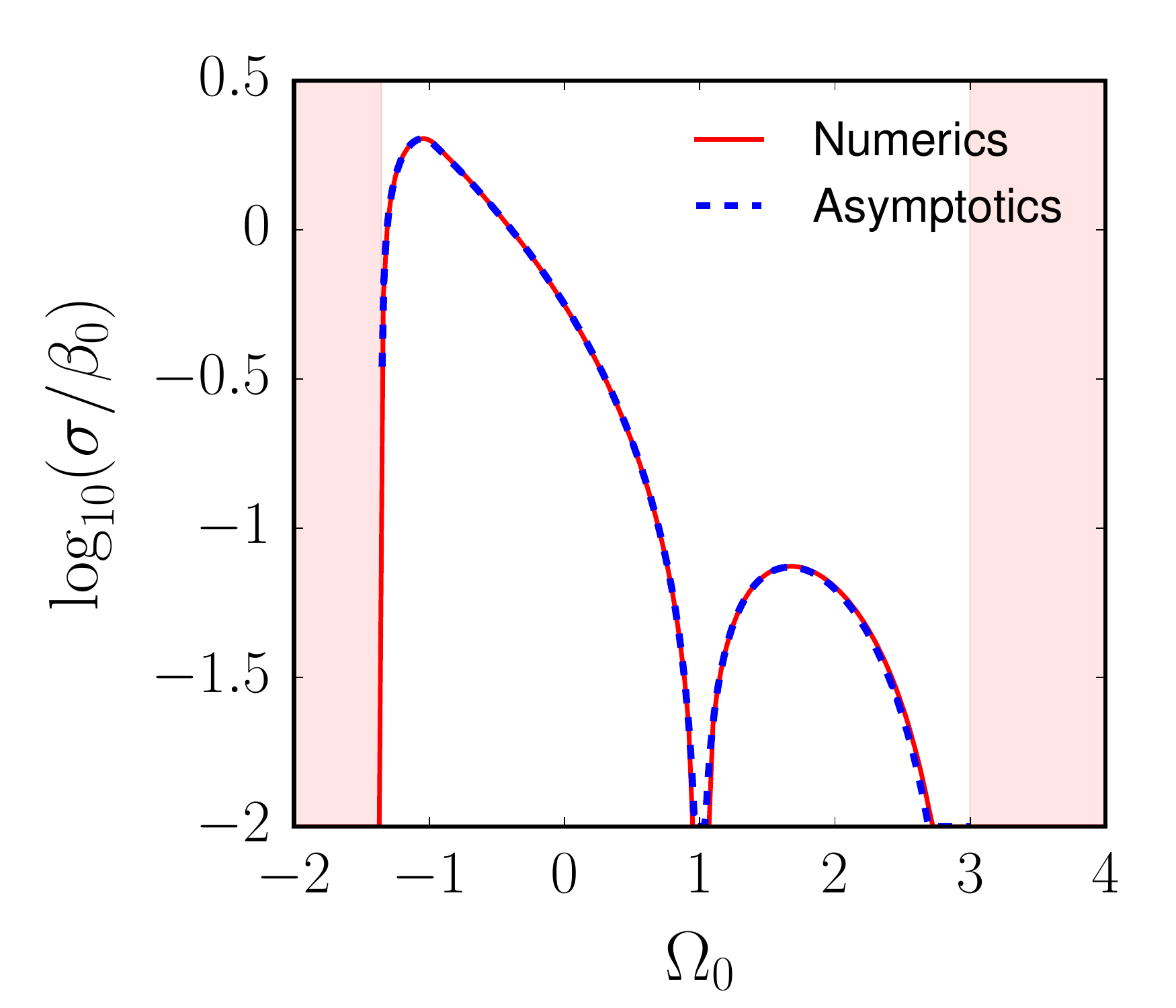}} &
		\subfigure[$\beta_{0}=0.6$]{\includegraphics[width=0.49\textwidth]{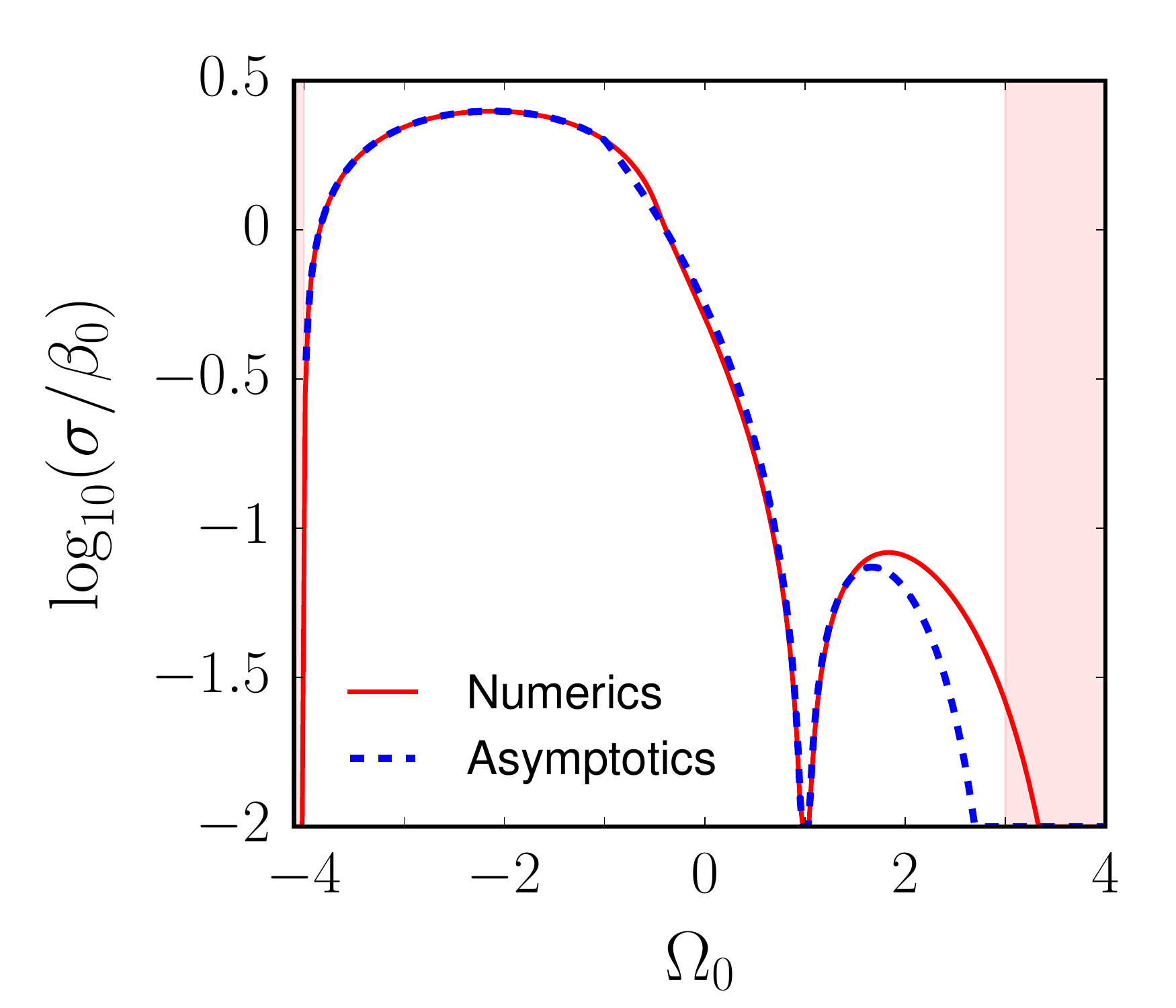}}
    \end{tabular}
	\caption{Comparison of the growth rates of the TDEI between theoretical formula (\ref{Eq_TDEI_DetuneWKB}) and numerical solutions of equations (\ref{Eq_WKB}). Shaded areas are stable regions (forbidden zone FZ$_{\beta_{0}}$). Numerical solutions have been computed with a Floquet analysis.}
	\label{Fig_TDEI_1b}
\end{figure}

Finally note that in figure \ref{Fig_TDEI2} the global analysis slightly predicts instabilities when $\Omega_{0} \geq 3$ while the asymptotic local analysis does not. It is a higher-order effect of finite $\beta_{0}$, which is not taken into account in the asymptotic formula (\ref{Eq_TDEI_DetuneWKB}). Indeed we compare in figure \ref{Fig_TDEI_1b} local formula (\ref{Eq_TDEI_DetuneWKB}) against numerical solutions of local equations (\ref{Eq_WKB}). At small $\beta_{0}$, analytical and numerical solutions are indistinguishable. However for larger $\beta_{0}$, the formula (\ref{Eq_TDEI_DetuneWKB}) under-predicts the upper bound of $\Omega_0$ of the allowable region where the instability occurs (near $\Omega_{0} = 3$), because the assumption of small $\beta_0\ll1$ is no longer satisfied.

	\subsection{Figures of equilibrium}
	\label{app:tdei_barker}
\begin{figure}
	\centering
	\begin{tabular}{cc}
		\subfigure[Local formula (\ref{Eq_Barker_WKB})]{\includegraphics[width=0.49\textwidth]{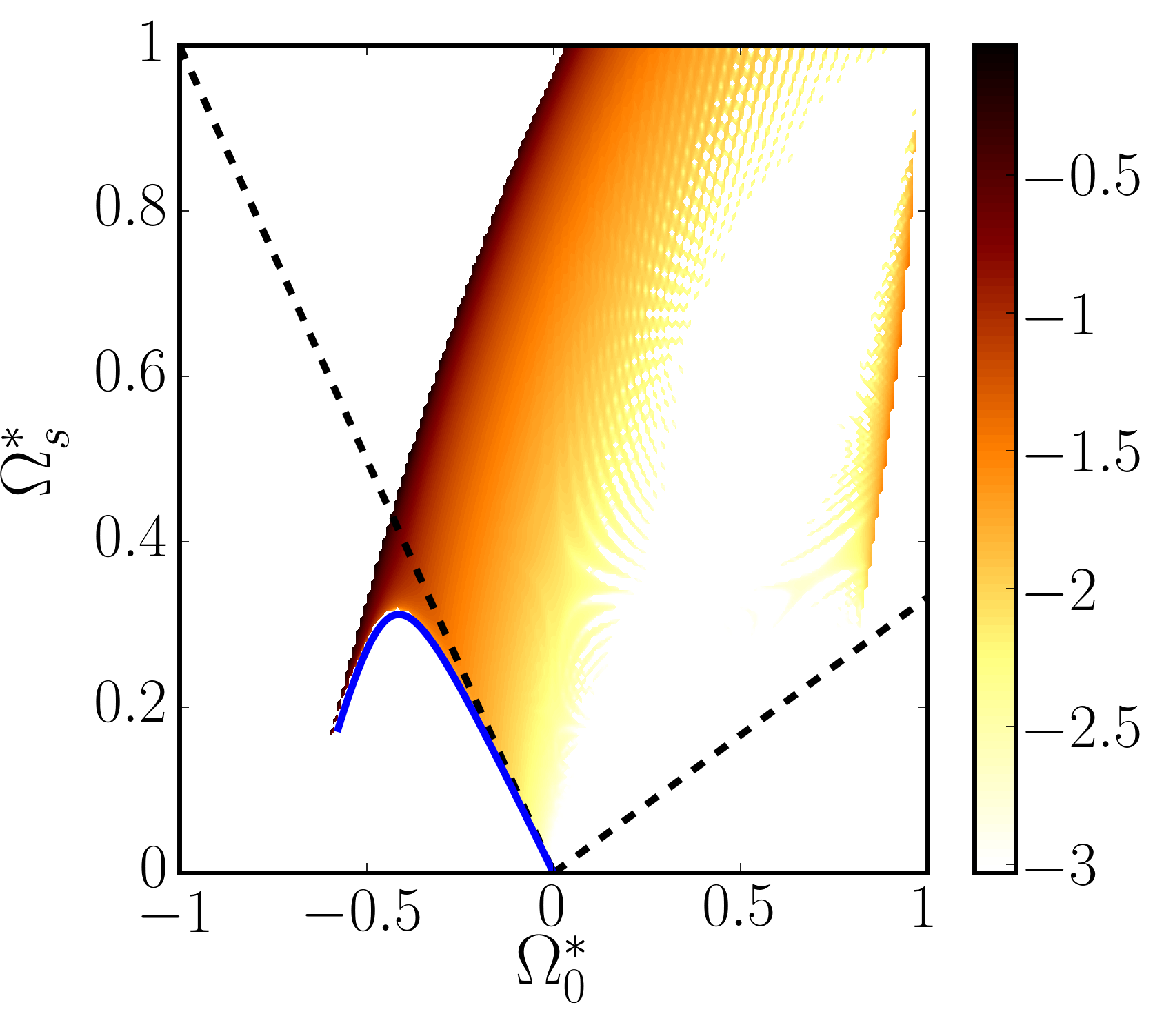}} &
		\subfigure[Global analysis ($n=15$)]{\includegraphics[width=0.49\textwidth]{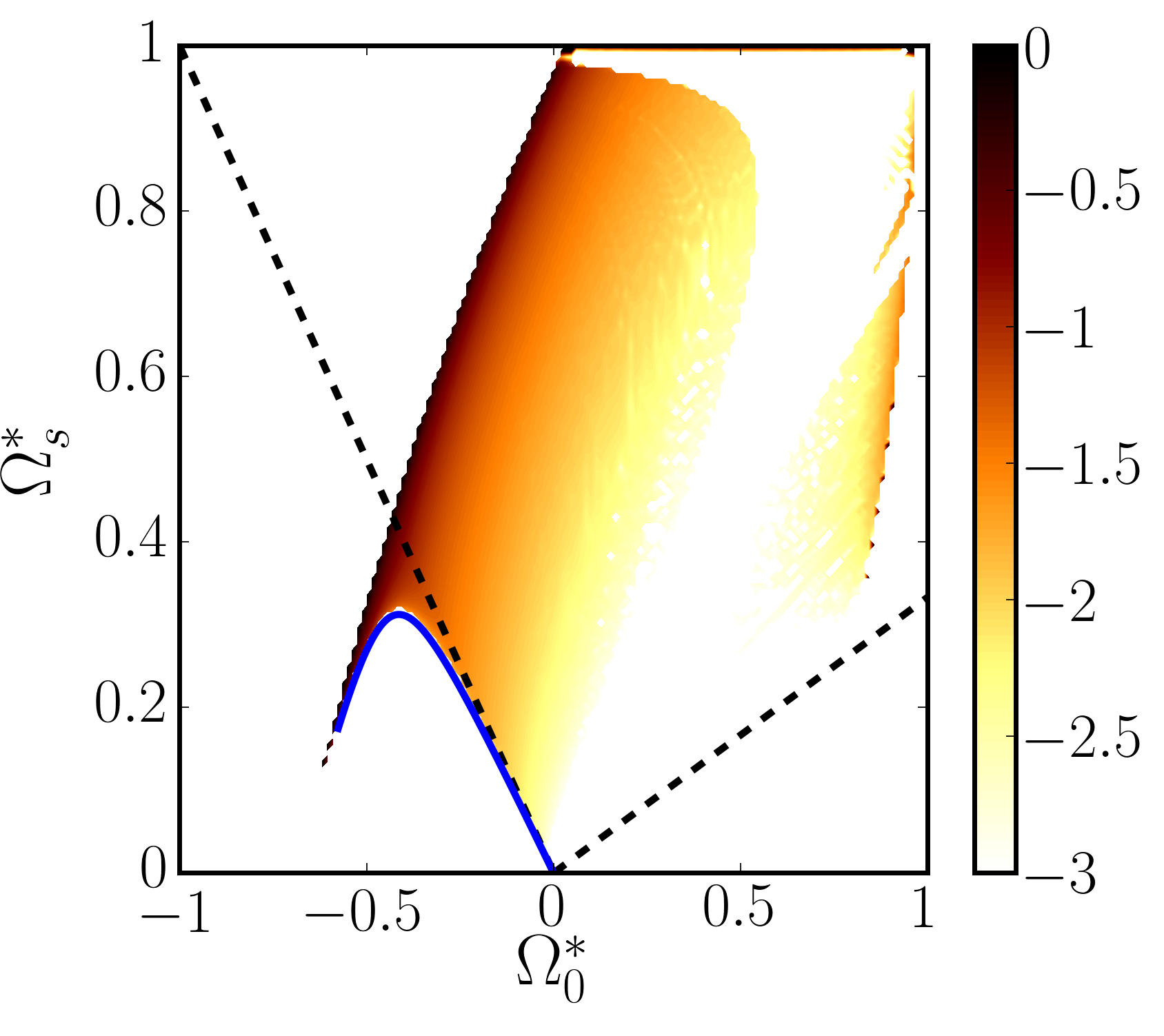}} \\
	\end{tabular}
	\caption{Survey of the parameter space $(\Omega_s^*, \Omega_{0}^*)$ for the elliptical instability as studied by \citet{barker2016nonlinear,Barker11062016}. Color map shows $\log_{10}(\sigma)$ and is saturated for ratio smaller than $10^{-3}$. White areas correspond to undefined ellipsoidal figures of equilibrium such that $\beta_{0} \leq 0$ or $\beta_{0} \geq 1$. Tidal amplitude $A=0.025$. The elliptical instability in the limit $\beta_0 \ll 1$ \citep{craik1989stability} has positive growth rates above the dashed black lines for $\Omega_{0}^* \in \left [ -\Omega_s^*, 3\Omega_s^* \right ]$. Blue solid line represent solutions of the equation  $\Omega_s^* = - \Omega_0^* (2b/a-1)$ devised by \citet{lebovitz1996short,Barker11062016}.}
	\label{Fig_Barker}
\end{figure}

We consider here a self-gravitating fluid domain on a circular orbit ($e=0$) but with semi-axes $(a,b,c)$ which are no longer independent of $\Omega_{0}$ (as opposed to \S\ref{sec:tdei}). \citet{Barker11062016} and \citet{barker2016nonlinear} have recently considered this particular case of TDEI in figures of equilibrium. The equilibrium tide is related to $\Omega_{0}$. To compare our results with theirs, we choose the inverse of the dynamical frequency $\omega_d^{-1}$ as time unit, with $\omega_d=(4 \pi G \rho/3)^{1/2}$ and $G$ the gravitational constant. We introduce two new dimensionless parameters, namely the fluid spin rate $\Omega_s^* =\Omega_{s}/\omega_G$ and the orbital spin rate $\Omega_0^* = \Omega_s^* \Omega_0$ (note that $\Omega_0$ is dimensionless).
The tidal amplitude $A$ is \citep{barker2016nonlinear} 
\begin{equation}
	\beta_{0} = \frac{3A}{2 \left [ 1 - \gamma^2 - (\Omega_0^*)^2 \right ] - A},
\end{equation}
with $\gamma = \Omega_s^* - \Omega_{0}^*$ the differential rotation.
The fluid ellipsoid semi-axes are $a = \sqrt{1 + \beta_{0}}$, $b = \sqrt{1 - \beta_{0}}$ and
\begin{equation}
	c^2 = \frac{2 \left [ (2A+\gamma^2 + (\Omega_{0}^*)^2-1) (A-\gamma^2 - (\Omega_{0}^*)^2+1) +f \right ]}{(A+1)[A+2(\gamma^2+(\Omega_{0}^*)^2-1)]},
\end{equation}
with
\begin{equation}
	f = 2 \gamma \Omega_{0}^* \sqrt{\left [ 1-2A-\gamma^2-(\Omega_{0}^*)^2 \right ] \left [ 1 + A - \gamma^2 - (\Omega_{0}^*)^2 \right ]}.
\end{equation}
\citet{Barker11062016,barker2016nonlinear} find that the hydrodynamic instabilities in ellipsoids with rigid boundaries are quantitatively similar to the ones in ellipsoids with realistic free surface deformations. Consequently the results obtained with rigid boundaries can also be applied to stellar configurations.

\citet{lebovitz1996short,Barker11062016} report a violent instability, called "stack of pancakes"-type instability (SoP), for negative $\Omega_{0}^*$ if the tidal amplitude is sufficiently large. The latter instability, located outside of the unstable range $-\Omega_s^* < \Omega_{0}^* < 3\Omega_s^*$ of the elliptical instability \citep{craik1989stability}, occurs in the interval
\begin{equation}
	-\frac{\Omega_s^*}{2b/a - 1} \leq \Omega_0^* \leq -\frac{\Omega_s^*}{2a/b - 1},
	\label{Eq_SoP_Barker}
\end{equation}
which is centred on $\Omega_s^* = \Omega_0^*$
This instability is already highlighted by \citet{le2000three} as an effect of finite $\beta_0$ (see discussion in \S\ref{sec:tdei}). The local formula (\ref{Eq_TDEI_DetuneWKB}) of \citet{le2000three} can be written in the appropriate dimensionless form
\begin{equation}
	\frac{\sigma}{|\gamma|} = \max_{\theta_0} \frac{1}{4} \sqrt{\left ( 1 + \cos \theta_0 \right )^4 \beta_{0}^2 - 4 \left [ 2 - 4 \left ( 1 + \frac{\Omega_{0}^*}{\gamma} \right ) \cos \theta_0 \right ]^2 } + \mathcal{O}(\beta_0^2).
	\label{Eq_Barker_WKB}
\end{equation}
In figure \ref{Fig_Barker}, we compare the global analysis at degree $n=15$ and the local formula (\ref{Eq_Barker_WKB}).
The agreement between the two approaches is quite good (except near $\Omega_s^* \geq 0.95$). 
However for larger $A$, asymptotic formula (\ref{Eq_Barker_WKB}) under-predicts the boundary of the region in which the instability is present. Numerical solutions of local stability equations (\ref{Eq_WKB}) are in better agreement with global results \citep{Barker11062016}. It is also better explained by formula (\ref{Eq_SoP_Barker}), which does not assume $\beta_{0} \to 0$.

\section{Direct numerical simulations of orbital flows}
\label{app:dns}
The global analysis gives sufficient conditions for inviscid instability, associated with the most unstable inviscid flows. However we could have doubts about their existence in real viscous flows.
So we compare our global results against direct numerical simulations of the stability equation (\ref{Eq_IE}). Unlike the global and local stability methods, we reintroduce the viscous term $E_k \nabla^2 \boldsymbol{u}$ and nonlinear term $(\boldsymbol{u} \cdot \nabla ) \boldsymbol{u}$ in the stability equation (\ref{Eq_IE}). Indeed it is not feasible to carry out three-dimensional numerical simulations in the inviscid linear regime. The impermeable boundary condition $\boldsymbol{u} \cdot \boldsymbol{n} = 0$ is supplemented with the stress-free boundary condition
\begin{equation}
	\boldsymbol{n} \times \left [ \boldsymbol{n} \cdot \left ( \nabla \boldsymbol{u} + (\nabla \boldsymbol{u})^{T} \right ) \right ] = \boldsymbol{0}.
	\label{Eq_StressFree}
\end{equation}
Stress-free boundary condition (\ref{Eq_StressFree}) avoids expensive computations to solve thin viscous boundary layers. This is also the astrophysically relevant viscous boundary condition. We keep the value of the Ekman number fixed at $E_k=[2.10^{-3}, 5.10^{-3}]$. With this condition, we isolate inertial instabilities we are interested in \citep{lorenzani2003inertial,vantieghem2015latitudinal}. Indeed mechanically driven viscous and centrifugal instabilities, which are often triggered in the boundary layers \citep[e.g.][]{lorenzani2001fluid,noir2009experimental,sauret2012fluid}, are ruled out with the stress-free condition (\ref{Eq_StressFree}).

\begin{figure}
	\centering
	\begin{tabular}{cc}
		\subfigure[$n=3$]{\includegraphics[width=0.54\textwidth]{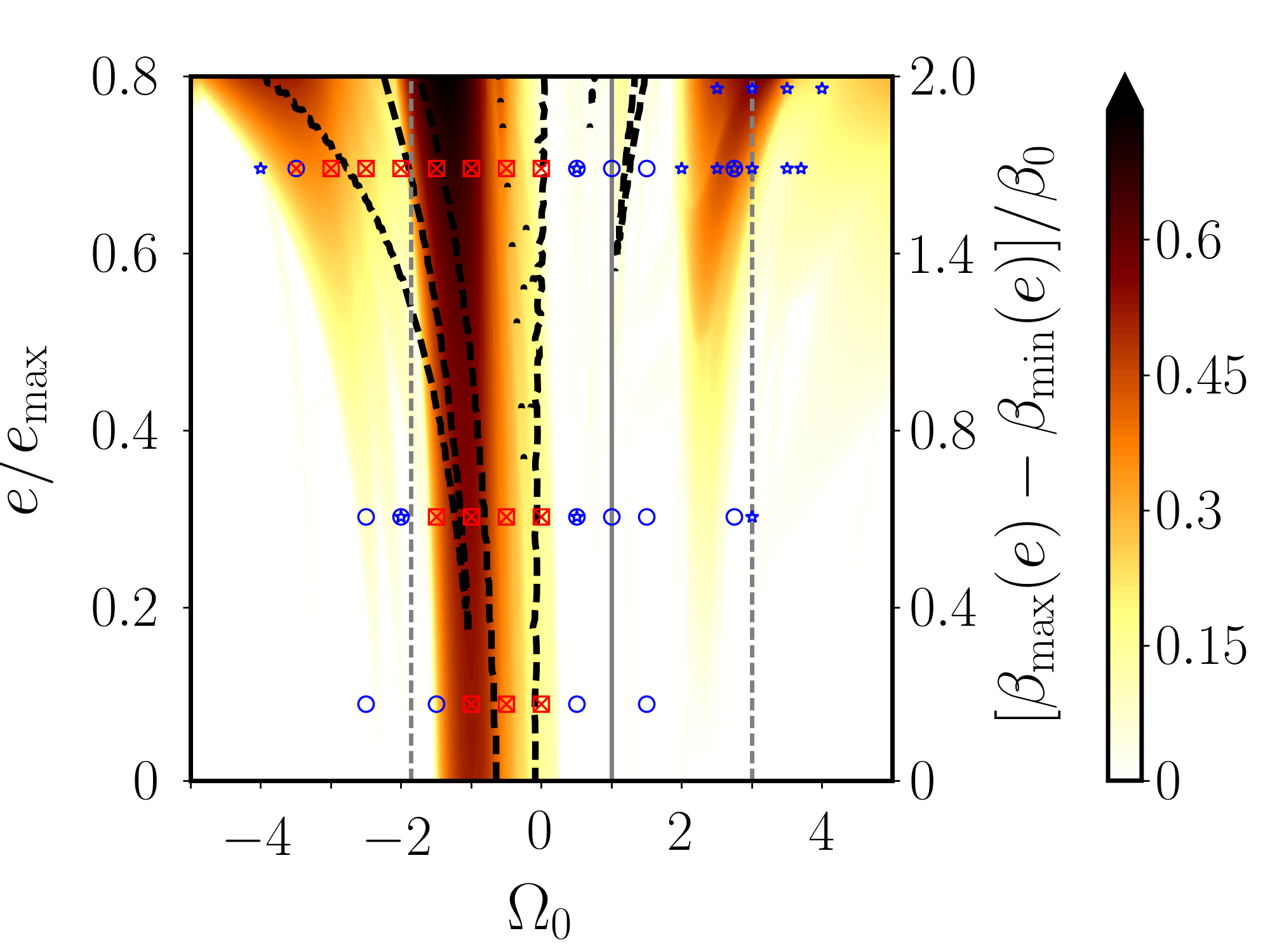}} &
		\subfigure[]{\includegraphics[width=0.35\textwidth]{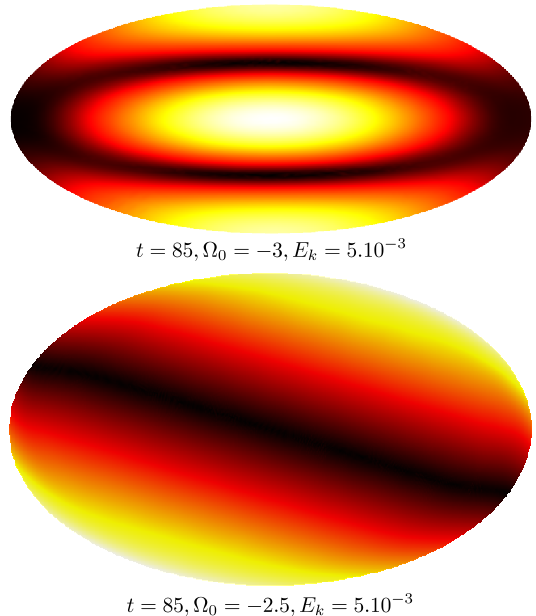}} \\
	\end{tabular}
	\caption{Comparison between the global analysis and direct numerical simulations in COMSOL at $\beta_{0} = 0.3$. (a) Survey of the stability of the orbitally driven flow (\ref{Eq_BF_Orbit}) in the plane $(e/e_{\max}, \Omega_0)$ for degree $n=3$. Isocontours of the growth rate $\sigma$ are shown, saturated at $\sigma \geq 0.6$. White areas correspond to marginally stable regions.
	The containers considered are oblate with $R= R_{m} + 0.05$. Vertical black line corresponds to the synchronised case ($\Omega_{0}=1$) driving the LDEI (see \S\ref{sec:ldei}). The horizontal line $e=0$ corresponds to the TDEI (see \S\ref{sec:tdei}). Vertical dashed black lines are the bounds of the forbidden zone FZ$_{\beta_0}$ of the TDEI valid for $e=0$ and $\beta_{ab} = \beta_0$. Dashed black lines demarcate the two unstable tongues of the spin-over mode $n=1$. Blue circles: stable at $E_k=5.10^{-3}$. Red squares: unstable at $E_k=5.10^{-3}$. Blue stars: stable at $E_k=2.10^{-3}$. Red crosses: unstable at $E_k=2.10^{-3}$. (b) Some unstable flows computed in COMSOL in a meridional plane.}
	\label{Fig_DNS}
\end{figure}

Numerical simulations cannot here benefit from a axisymmetric geometry to use fast and accurate spectral methods usually employed in global astrophysical simulations. We need also to adapt the numerical mesh at each time step. We solve stability equations (\ref{Eq_IE}) - (\ref{Eq_StressFree}) for the perturbation upon the basic flow in their weak variational form with the commercial parallelided finite element code COMSOL, previously used in numerical studies of tidal, librating and precessing flows \citep[e.g.][]{cebron2010systematic,cebron2010tilt,cebron2012libration,noir2013precession}. An unstructured mesh with tetrahedral elements is initially created. The mesh element type employed is the standard Lagrange element P1 - P2, which is linear for the pressure field but quadratic for the velocity field. 
The total number of degrees of freedom ranges between 50~000 and 300~000.
We use the implicit differential algebraic solver (IDA solver), based on backward differentiation formula \citep{hindmarsh2005sundials}. At each time step the system is solved with the sparse direct linear solver PARDISO \citep{schenk2004solving}. 
No stabilisation technique is used.
We solve for the mesh motion using an arbitrary Lagrangian Eulerian (ALE) method, which adapts the numerical mesh at each time step to adjust the ellipsoidal boundary. In practice the ellipsoidal domain is fixed at the origin and the elements are displaced in each Cartesian direction by amounts
\begin{equation}
	\left [ \delta_{x}, \delta_{y}, \delta_{z} \right ] (t) = \left [ x(0) \left ( \frac{a(t)}{a(0)} -1 \right ), y(0) \left ( \frac{b(t)}{b(0)} -1 \right ), z(0) \left ( \frac{c(t)}{c(0)} -1 \right ) \right ]
	\label{Eq_ALE}
\end{equation}
where $(a(0), b(0), c(0))$ are the semi-axes of the ellipsoidal domain at initial time and $(x(0), y(0), z(0))$ the initial position of a given mesh element. The extra computational work per time step makes the code significantly more computationally demanding than a fixed grid version. ALE method has recently been used by \citet{barker2016nonlinear} in nonlinear simulations of tidal flows.
We have checked that our results are not significantly affected by changing the mesh, the size of the domain or the maximum time step.

Figure \ref{Fig_DNS} shows the comparison between the global analysis and direct numerical simulations in COMSOL at $\beta_{0} = 0.3$. Formally, the global method cannot predict accurately neither the viscous growth rate nor the viscous unstable flow of a given spatial complexity. However the viscosity selects the spatial complexity of real viscous flows. It mostly enters as a damping term in the stability problem of tidally driven basic flows \citep[e.g.][]{lacaze2004elliptical,le2010tidal}. So we can heuristically mimic the leading order viscous damping by varying the maximum polynomial degree $n$. That is the reason why in figure \ref{Fig_DNS} (a) we observe that numerical simulations are in good agreement with the global analysis at $n=3$, outside and within the forbidden zone. 
The tongue located around $\Omega_{0}=3$ is not recovered in the simulations, because the Ekman number is too high in the simulations compared to the expected inviscid growth rate. This tongue could be obtained in the simulations by decreasing $E_k$. 

Finally in figure \ref{Fig_DNS} (b) we show some of the most unstable flows in the simulations. The flow $\Omega_0=-3$ has a SoP structure, as predicted by the global and local analyses in \S\ref{sec:physics}. Similarly the flow at $\Omega_0=-2.5$ is a spin-over mode, because the numerical point lies within the spin-over tongue in figure \ref{Fig_DNS} (a).

{
\bibliography{./bib_gp}
\bibliographystyle{jfm}
}

\end{document}